\documentclass[aps,prb,onecolumn,eqsecnum]{revtex4-2}

\usepackage{graphicx}% Include figure files
\graphicspath{ {SingleNodeWeylPaper/} }

\usepackage{bm}% bold math
\usepackage{amsmath}
\usepackage{amssymb}

\newcommand{\bvm}{\begin{verbatim}}
\newcommand{\evm}{\end{verbatim}}

\newcommand{\ocite}{\onlinecite}

\newcommand{\iy}{\infty}

\newcommand{\x}{\text}
\newcommand{\pd}{\partial}

\newcommand{\dg}{\dagger}
\newcommand{\lan}{\langle}
\newcommand{\ran}{\rangle}
\newcommand{\lt}{\left}
\newcommand{\rt}{\right}

\newcommand{\f}{\frac}
\newcommand{\tf}{\tfrac}

\newcommand{\sq}{\sqrt}
\newcommand{\lbl}{\label}

\newcommand{\cd}{\cdot}

\newcommand{\p}{\perp}
\newcommand{\n}{\nabla}

\newcommand{\nm}{\hat{0}}
\newcommand{\um}{\hat{1}}

\newcommand{\tm}{\times}

\newcommand{\eq}[1]{Eq.~(\ref{eq:#1})}
\newcommand{\eqs}[2]{Eqs.~(\ref{eq:#1}) and (\ref{eq:#2})}
\newcommand{\eqss}[3]{Eqs.~(\ref{eq:#1}), (\ref{eq:#2}), and (\ref{eq:#3})}

\newcommand{\eqn}[1]{(\ref{eq:#1})}
\newcommand{\eqsn}[2]{(\ref{eq:#1}) and (\ref{eq:#2})}
\newcommand{\eqssn}[3]{(\ref{eq:#1}), (\ref{eq:#2}), and (\ref{eq:#3})}

\newcommand{\secr}[1]{Sec.~\ref{sec:#1}}
\newcommand{\secsr}[2]{Secs.~\ref{sec:#1} and \ref{sec:#2}}

\newcommand{\figr}[1]{Fig.~\ref{fig:#1}}
\newcommand{\figsr}[2]{Figs.~\ref{fig:#1} and \ref{fig:#2}}
\newcommand{\figssr}[3]{Figs.~\ref{fig:#1}, \ref{fig:#2}, and \ref{fig:#3}}

\newcommand{\spc}{\mbox{ }}

\newcommand{\beq}{\begin{equation}}
\newcommand{\eeq}{\end{equation}}
\newcommand{\beqar}{\begin{eqnarray}}
\newcommand{\eeqar}{\end{eqnarray}}
\newcommand{\beqarn}{\begin{eqnarray*}}
\newcommand{\eeqarn}{\end{eqnarray*}}
\newcommand{\ba}{\begin{array}}
\newcommand{\ea}{\end{array}}
\newcommand{\bwt}{\begin{widetext}}
\newcommand{\ewt}{\end{widetext}}

\newcommand{\Rarr}{\Rightarrow}

\newcommand{\LRarr}{\Leftrightarrow}
\newcommand{\rarr}{\rightarrow}

\newcommand{\dx}{{\text d}}
\newcommand{\ex}{{\text e}}

\newcommand{\ix}{{\text i}}

\newcommand{\Ux}{{\text U}}

\newcommand{\ph}{\hat{p}}

\newcommand{\vh}{\hat{v}}
\newcommand{\wh}{\hat{w}}

\newcommand{\Hh}{\hat{H}}

\newcommand{\Sh}{\hat{S}}

\newcommand{\psih}{\hat{\psi}}
\newcommand{\tauh}{\hat{\tau}}
\newcommand{\chih}{\hat{\chi}}

\newcommand{\dr}{{\bar{d}}}
\newcommand{\er}{{\bar{\epsilon}}}

\renewcommand{\vr}{{\bar{v}}}
\newcommand{\Vr}{{\bar{V}}}

\newcommand{\Ecr}{\bar{\mathcal{E}}}

\newcommand{\psir}{{\bar{\psi}}}

\newcommand{\Cc}{\mathcal{C}}

\newcommand{\Ec}{\mathcal{E}}

\newcommand{\Sc}{\mathcal{S}}
\newcommand{\Tc}{\mathcal{T}}

\newcommand{\Vc}{\mathcal{V}}

\newcommand{\Vcr}{{\bar{\mathcal{V}}}}

\newcommand{\sigb}{{\mbox{\boldmath{$\sigma$}}}}

\newcommand{\taubh}{\hat{\mbox{\boldmath{$\tau$}}}}

\newcommand{\nv}{{\bf 0}}

\newcommand{\db}{{\bf d}}

\newcommand{\hb}{{\bf h}}

\newcommand{\nb}{{\bf n}}
\newcommand{\pb}{{\bf p}}
\newcommand{\qb}{{\bf q}}
\newcommand{\rb}{{\bf r}}
\newcommand{\s}{{\bf s}}

\newcommand{\Ab}{{\bf A}}
\newcommand{\Bb}{{\bf B}}

\newcommand{\pbh}{\hat{\pb}}

\newcommand{\al}{\alpha}
\newcommand{\be}{\beta}
\newcommand{\ga}{\gamma}

\newcommand{\de}{\delta}
\newcommand{\De}{\Delta}

\newcommand{\Sig}{\Sigma}
\newcommand{\tht}{\theta}

\newcommand{\eps}{\varepsilon}
\newcommand{\e}{\epsilon}

\begin{document}
\title{Formalism of general continuum models with boundary conditions,\\
propagation of bound states from nontrivial to trivial topological classes,\\
and the general surface-state structure near one node of a Weyl semimetal}
\date{\today}

\author{Maxim Kharitonov$^{1,2}$}
\address{$^1$Institute for Theoretical Physics and Astrophysics, University of W\"urzburg, 97074 W\"urzburg, Germany\\
$^2$Donostia International Physics Center (DIPC), Manuel de Lardizabal 4, E-20018 San Sebastian, Spain}

\begin{abstract}

We present the {\em (symmetry-incorporating) formalism of general continuum models with boundary conditions} and apply it to the model with the minimal number of degrees of freedom necessary to have a well-defined boundary: a model with a two-component wave function and a linear-in-momentum Hamiltonian. We derive the most general forms (class A) of both the Hamiltonian and boundary condition in 1D (insulator), 2D (quantum anomalous Hall insulator), and 3D (Weyl node) and analytically calculate and explore the corresponding general bound/edge/surface-state structures. In 1D, one bound state exists in the half of the $\text{U}(1)$ parameter space of possible boundary conditions. Considering several dimensions simultaneously ties the models together and uncovers important relations between them. We formulate a version of bulk-boundary correspondence that fully characterizes the vicinity of a Weyl point: the chirality of the surface-state spectrum along a path enclosing the projected Weyl point is equal to the Chern number of the Weyl point. We demonstrate how symmetries are naturally incorporated into the formalism, by deriving the most general form of the model with chiral symmetry (class AIII). We show that the (perhaps unexpected) existence of persistent bound states in the topologically trivial 1D class-A model is not accidental and has at least two topological explanations, by relating it to the topologically nontrivial 2D class-A (by viewing it as an effective 2D quantum anomalous Hall system) and 1D class-AIII (via deviation from the cases of chiral symmetry in the parameter space) models. We identify this as a systematic ``propagation effect'', whereby bound states from topologically nontrivial classes, where they are protected and guaranteed to exist, propagate to the related, ``adjacent'' in dimension or symmetry, topologically trivial classes.
%The main emerging overarching conclusion of our analysis is that bound states appear
%to be more widespread and persistent than the mere strict topological classification suggests and this is a systematic effect.

\end{abstract}
\maketitle

\section{Introduction}

\begin{figure}
\includegraphics[width=.50\textwidth]{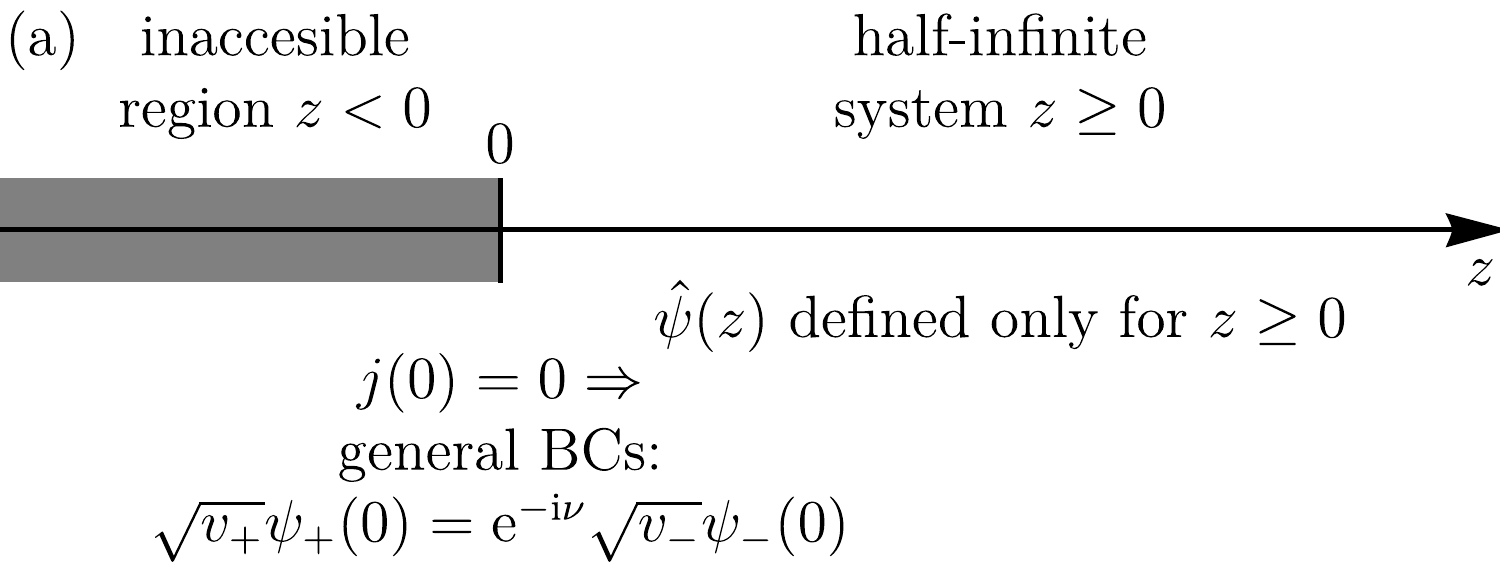}\\
\includegraphics[width=.27\textwidth]{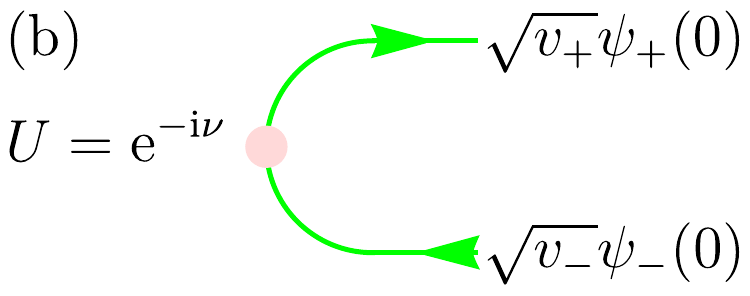}
\caption{(a) 1D half-finite system $z\geq 0$ of the continuum model.
The wave function $\psih(z)$ of the continuum model is defined only in the region $z\geq 0$
and satisfies a boundary condition (BC) at $z=0$. The region $z<0$ is inaccessible region for the low-energy excitations that the model represents.
(b) Illustration of the general BC \eqn{bc} for the considered minimal continuum model [\eqs{H}{psi}],
in the ``standardized'' universal form of the formalism of general BCs~\cite{Ahari,KharitonovGBCs}.
The BC has a natural physical interpretation as a scattering process
between the incident left-moving mode $\psi_-(z)$ and the reflected right-moving mode $\psi_+(z)$ of the wave function.}
\lbl{fig:1D}
\end{figure}

Bound states at boundaries, interfaces, and defects of bulk crystalline electron systems
have by now been firmly established as a widespread feature in a multitude of real materials and theoretical models.
When the bulk has nontrivial topology, certain bound states are guaranteed, as per the concept of bulk-boundary correspondence~\cite{Chiu}.
Moreover, even systems that are (probably) topologically trivial in a rigorous sense
have been theoretically shown to exhibit robust bound states, such as Luttinger-semimetal models~\cite{KharitonovLSM}
(and, as we will show in this work, completely generic minimal low-energy models).

Continuum models (whose Hamiltonians are polynomials in momentum,
typically describing the low-energy expansion of some more complicated, ``more microscopic''
underlying Hamiltonian about the points of interest in the Brillouin zone)
are particularly appealing for theoretical studies of bound states, due to the relative simplicity of their bulk Hamiltonians.
The main challenge on this path is a systematic description of the boundary.
For continuum models, this description comes down to determining proper boundary conditions (BCs),
formulated as linear homogeneous relations between the wave-function components and their derivatives at the boundary.
It has been understood for quite some
time~\cite{AkhiezerGlazman,ReedSimon,Berry,BerezinShubin,Bonneau,Tokatly,McCann,AkhmerovPRL,AkhmerovPRB,Ostaay,Hashimoto2016,Hashimoto2019,Ahari,
KharitonovLSM,KharitonovQAH,Seradjeh,Walter,Shtanko,KharitonovSC,
Enaldiev2015,Volkov2016,Devizorova2017}
that BCs arise as a consequence of the fundamental principle of quantum mechanics:
norm conservation of the wave function upon time evolution, described by the Schr\"odinger equation,
which leads to the hermiticity requirement for the Hamiltonian,
which in turn leads to the conservation of the probability current at the boundary.
This understanding naturally leads to the notion of {\em general BCs},
as a family of all possible BCs (for a given Hamiltonian) that satisfy the current-conservation principle.

Previously, general BCs have been derived from the current-conservation principle
and used to study the bound-, edge-, or surface-state structures for specific continuum models:
for the 1D quadratic-in-momentum one-component model~\cite{Bonneau} (a textbook nonrelativistic Schr\"odinger particle);
for the 3D quadratic-in-momentum  multi-component model~\cite{Tokatly}, in the context of semiconductors;
for the linear-in-momentum two-component model,
in the context of 2D semimetals~\cite{Berry,KharitonovLSM}, 2D quantum anomalous Hall (QAH) systems~\cite{KharitonovQAH},
graphene~\cite{Walter}, 1D insulators~\cite{Ahari}, and 3D Weyl semimetals~\cite{Hashimoto2016,Hashimoto2019};
for the linear-in-momentum four-component model of graphene~\cite{McCann,AkhmerovPRL,AkhmerovPRB,Ostaay}
and 3D Dirac materials~\cite{Shtanko};
for the 3D quadratic-in-momentum two-component model of Weyl semimetals~\cite{Seradjeh};
for the 1D linear-in-momentum four-component model of superconductors with one Fermi surface~\cite{KharitonovSC}.
Also, general BCs with some symmetry constraints have been derived for quadratic- and linear-in-momentum four-component models
of 2D and 3D topological insulators and Dirac materials~\cite{Enaldiev2015,Volkov2016} and two-node 3D Weyl semimetals~\cite{Devizorova2017}.

Recent progress is shaping this topic into the full-fledged {\em formalism of general BCs} for continuum models.
In Ref.~\ocite{Ahari}, Ahari, Ortiz, and Seradjeh formulated a systematic
derivation procedure of the general BCs from the current-conservation principle
for the 1D continuum model with the translation-symmetric Hamiltonian of the most general form,
with any number of wave-function components and any order of momentum.
The procedure is based on the diagonalization of the quadratic form of the current
and leads to the ``standardized'' universal form of the family of the general BCs,
which are parameterized in a nonredundant one-to-one way by sets of unitary matrices.
This formalism was further substantiated and elucidated in Ref.~\ocite{KharitonovGBCs}.
In particular, a natural physical interpretation has been provided in Refs.~\ocite{KharitonovSC,KharitonovGBCs}, where the general BCs can be
regarded as a scattering process at the boundary between the incident (left-moving) and reflected (right-moving) ``modes''
of the wave function and the unitary matrix can be regarded as the scattering matrix of this process [\figr{1D}(b)].

Most of the existing above-cited works focused on the BCs of the {\em most general} form,
but for {\em specific} forms of the Hamiltonians, typically of the mathematically simplest forms,
with fewer or much fewer parameters than the most general possible form allows.
For exhaustive studies of bound states, especially for those in topological systems~\cite{Ryu,Chiu},
two aspects are particularly important.

(i) {\em Generality of the model of a system with a boundary}: that the model does represent the whole class of systems.
By the very definition of topology, topologically stable properties hold not just for particular instances, but for whole {\em classes} of systems,
i.e., families related by all possible continuous deformations that do not violate the central characteristics of the system
(such as the presence of the bulk gap for a gapped system or the presence of nodes for a semimetal).
General BCs exhaust all possible underlying microscopic structures of the boundary,
which is a major advantage of the formalism other methods can hardly offer, see \secr{advantages}.
Even then, the question still remains whether the corresponding bound-state structures are indeed generic to the whole class
until all possible forms of the Hamiltonian have also been considered.
To put it briefly, for a continuum model to represent a whole class of systems with a boundary,
{\em both} its Hamiltonian (describing the bulk) and BCs (describing the boundary) must be of the most general form.

(ii) {\em Symmetries of the model of a system with a boundary}: that the model does respect the symmetries (if present) of the class.
Symmetry constraints are well-known to result in various topological properties~\cite{Ryu,Chiu}.
The standard topological classification scheme is based on chiral $\Sc$ symmetry
and two types $\pm$ of charge-conjugation $\Cc_\pm$ ($\Cc_\pm^2=\pm 1$) and time-reversal $\Tc_\pm$ ($\Tc_\pm^2=\pm 1$) symmetries,
which give rise to ten possible topological symmetry-protected classes.
For such classes, only those continuous deformations of the bulk or boundary are allowed that preserve the symmetries.
Crucially, a system {\em with a boundary} must satisfy the symmetry of a class,
in order for its bound states to reflect the bulk topology of the class.
For continuum models, this means that not only the bulk Hamiltonian, but also the BCs must satisfy the symmetries
(what this means technically has been explained in Refs.~\ocite{KharitonovQAH,KharitonovSC} and will be explained again in \secr{1DAIII}).
Symmetries thus provide constraints on the allowed forms of both the Hamiltonian and BCs
and such models are subsets of the families of general models satisfying only the requirement (i).

When the Hamiltonian of the most general form
(for the chosen ``degrees of freedom'' of the continuum model, see \secr{degrees})
is derived and then the general BCs for this family of Hamiltonians are derived via the formalism of general BCs of Refs.~\ocite{Ahari,KharitonovGBCs},
the generality requirement (i) is satisfied
and such model represents the whole class A of systems with a boundary with no assumed symmetries.
We term such approach as the {\em formalism of general continuum models with BCs}.

When symmetry constraints (if present) are applied to both the Hamiltonian and BCs,
so that the most general form of the Hamiltonian constrained only by symmetries is derived
and the most general form of the BCs constrained only by current conservation principle and symmetries is derived,
both the generality (i) and symmetry (ii) requirements are satisfied
and the such model represents the whole respective symmetry class of systems with a boundary.
In this case, a clarifier can be added to term the approach as the {\em symmetry-incorporating formalism of general continuum models with BCs}.
The formalism is, of course, also applicable to continuum models with any set of symmetries, describing not only topological systems.

\section{Summary of the main results}

In this work, we present such formalism of general continuum models with BCs (\secr{formalism})
and demonstrate its application to the model with the minimal number of degrees of freedom necessary to have a well-defined boundary:
a model with a two-component wave function and a linear-in-momentum Hamiltonian.
(Up to now, applications of this formalism %of general continuum models with BCs
have been carried out for 2D QAH systems in the Supplemental Material of Ref.~\ocite{KharitonovQAH}
and for 1D $\Cc_+$ charge-conjugation-symmetric superconductors with one Fermi surface in Ref.~\cite{KharitonovSC}.)
Such model has one right- and one left-moving mode and there is just one BC relation between them.
The only other model with the same minimal number of degrees of freedom
is a model with a one-component wave function and a quadratic-in-momentum Hamiltonian~\cite{Bonneau,KharitonovGBCs}.

This model provides a rigorous asymptotic low-energy limit of a generic system
in the vicinity of the gap closing (when the gap is much smaller than the typical bandwidth)
and, despite its simplicity, has numerous practical realizations.
Without any assumed symmetries, it describes class-A family of systems with a boundary.
In 1D, it describes a topologically trivial insulator;
in 2D -- a topologically nontrivial quantum anomalous Hall (QAH) system (Chern insulator) in the vicinity of a topological phase transition;
in 3D -- a topologically nontrivial Weyl semimetal in the vicinity of a node.
We derive the most general forms of both the Hamiltonian and BC in 1D (\secr{1D}), 2D (\secr{2D}), and 3D (\secr{3D}).
The above-cited previous studies of the same model with the general BCs but specific, simpler forms of the Hamiltonians
now become special cases or subfamilies of this general model.
The unitary ``matrix'' of the family of all possible BCs is a $\Ux(1)$ scalar $U=\ex^{-\ix\nu}$,
which can be parameterized by a phase-shift angle $\nu\in[0,2\pi)$ on a unit circle $S^1$ (\secr{bc}).
We provide an elegant geometric interpretation of the general BC in terms of the effective pseudospin-$\f12$
of the two-component wave function (\secr{bcgeom}).

We analytically calculate and explore the corresponding general bound/edge/surface-state structures in 1D, 2D, and 3D.
In 1D, one bound state exists in the half of the $\Ux(1)$ parameter space of the general BCs (\secr{bs}).
In 3D, we demonstrate that due to the universal linear momentum scaling the surface-state spectrum near one Weyl node
can be fully characterized by the dimensionless velocity function of the polar angle of the 2D surface momentum.
We also explore the distinction between type-I and type-II Weyl semimetals
and consider the cases of the line-node semimetal and strongly anisotropic Weyl semimetal.
We recover the previously studied~\cite{KharitonovQAH} case of the 2D QAH system, where the most remarkable effect is the significant ``extension''
of the edge-state structure beyond the minimal one required to satisfy the Chern numbers, which is nearly always present.

Further, we demonstrate how symmetries are naturally incorporated into the formalism
by deriving the most general form of the model with chiral symmetry (class AIII) for each dimension,
which describes a topologically nontrivial insulator in 1D, a topologically nontrivial point-node semimetal in 2D, and a line-node semimetal in 3D.

In this work, we purposefully consider the model in several dimensions simultaneously,
as there exists a direct both mathematical and physical connection.
%tying together several important effects and results.
The starting point is the 1D model. In higher dimensions, for preserved translation symmetry along the edge/surface,
the momentum along the edge/surface is conserved and becomes a parameter of the effective 1D Hamiltonian.
Hence, mathematically, the models in higher dimensions reduce to the 1D model,
where the parameters of the latter acquire dependence on the surface momentum.
This way, we demonstrate that there is direct relation between the bound-state structure in 1D,
the surface-state structure of a 3D Weyl semimetal,
and the edge-state structure of the 2D QAH system.

This leads to the physical connection, in particular, of topological properties.
The 1D system can be seen as an effective 2D QAH system upon considering paths in its parameter space.
In 3D, the system on a cylinder passing through a closed path in the surface-momentum plane
can also be regarded as an effective 2D QAH system.
The latter perspective allows us to formulate a version of bulk-boundary correspondence that fully characterizes
the vicinity of a Weyl point:
the chirality of the surface-state spectrum along a path enclosing the projected Weyl point is equal to the Chern number of the Weyl point. 
%(e.g., a circle) 
Within the linear-in-momentum model, this also means that
the chirality of the surface-state velocity as a function of the polar angle of the surface momentum
is equal to the Chern number of the Weyl point.

Perhaps our most important finding concerns the explanation of the existence of the bound states in the general 1D class-A model,
which are present in a half of its parameter space.
This in itself quite remarkable result has previously been obtained~\cite{AkhmerovPRL,AkhmerovPRB,Ostaay,Ahari,Walter}, but remained unexplained.
Since the model describes a topologically trivial class, naively,
one could be tempted to regard these bound states as accidental, with no particular reason for their existence.

This is however not true, as we show that the existence of robust bound states
in the topologically trivial 1D class-A model has at least two topological explanations,
by relating it to two topologically nontrivial classes:
the 2D class-A of QAH systems (as per above, \secr{2DA}) and 1D class-AIII of chiral-symmetric systems,
via deviation from the cases of chiral symmetry in the parameter space. %, the idea expressed already in Ref.~\ocite{KharitonovLSM}).
We argue that this relation could itself be regarded as an explicit manifestation of Bott periodicity~\cite{Ryu,Chiu}.
A mathematically equivalent situation
is the existence of the edge-state band in a 2D semimetals (which are describes the phase-transition point in a 2D QAH system).
This effect has been originally explained in Ref.~\ocite{KharitonovLSM}
in terms of the approximate chiral symmetry of the model, which is equivalent to second explanation above.
%arises from the general edge-state behavior of the 2D point-node semimetal, realized right at the phase transition point.

The existence of bound states in topologically trivial systems (1D class-A insulator and 2D class-A semimetal) is therefore neither rare nor accidental.
We identify this as a systematic ``propagation'' effect,
whereby bound states from the topologically nontrivial classes, where they are protected and guaranteed to exist,
propagate to the related, ``adjacent'' in dimension or symmetry, topologically trivial classes.
This effect, proven here for specific symmetry classes, seems to be generic and could be expected for other symmetry classes as well.
The main emerging overarching conclusion of our analysis is that bound states appear to be more widespread and persistent
than the mere strict topological classification suggests and this is a systematic effect.

\section{Outline of the symmetry-incorporating formalism of general continuum models with boundary conditions\lbl{sec:formalism}}

In this section, we present the key steps of the symmetry-incorporating formalism of general continuum models with BCs
and argue its advantages for the study of bound states.

\subsection{Choosing the relevant degrees of freedom \lbl{sec:degrees}}

The general structure of the continuum model is fixed by the choice of its {\em degrees of freedom},
i.e., the number of wave-function components and the order of momentum in the Hamiltonian.
When symmetries are present, this also includes specifying the transformation properties
of the wave-function components, such as what irreducible representations they belong to.
This choice is made at researcher's discretion, based on the goals: what effect is expected or desired to be captured,
what real physical system or more microscopic model is supposed to be represented, etc.
Once this choice is made, the formalism provides clear steps on how to obtain the continuum model of the most general form
for a system with a boundary for these degrees of freedom (\secsr{deriveH}{deriveBCs}),
as well as an efficient (semi-)analytical method of calculating the bound states (\secr{calcmethod}).

\subsection{Derivation of the most general Hamiltonian \lbl{sec:deriveH}}

The bulk Hamiltonian of the most general form is constructed using the method
known as {\em the method of invariants}, or {\em $k\cd p$ method}~\cite{Hamermesh,Winkler},
well-developed for ``conventional'' spatial and time-reversal $\Tc_\pm$ symmetries.
The method can be straightforwardly adapted to other, more abstract symmetries, such as chiral $\Sc$ and charge-conjugation $\Cc_\pm$.

The Hamiltonian of a continuum model (with translation-symmetric bulk) is a polynomial matrix function of momentum.
One first constructs the needed (for the chosen degrees of freedom) basis matrix functions of momentum.
If present, symmetry constraints narrow down the allowed basis matrices.
The Hamiltonian of the most general form for a given symmetry class is then a linear combination of the allowed basis matrices
with arbitrary coefficients. These coefficients form the parameter space of the Hamiltonian.

\subsection{Derivation of the most general boundary conditions \lbl{sec:deriveBCs}}

The next step is deriving the most general form of the BCs for this family of Hamiltonians.
BCs are a consequence of the fundamental principle of quantum mechanics that the norm of the wave function is conserved upon time evolution.
This leads to the requirement that the Hamiltonian be hermitian.
For a system with a boundary, the latter imposes constraints not only on the form of the Hamiltonian operator,
but also on the Hilbert space of allowed wave functions.
The latter constraint has the form of the probability-current nullification at the boundary.
Only the Hilbert spaces for which the current vanishes identically (for every wave function within it) are admissible.
This current-nullification constraint can be resolved in the form of linear homogeneous relations
between the wave-function components and their derivatives at the boundary,
and these relations are commonly referred to as the {\em boundary conditions} (BCs).
Therefore, BCs are essentially a way of specifying admissible Hilbert spaces
of wave functions, over which the Hamiltonian becomes hermitian.

The general BCs for a given family of Hamiltonians, i.e., all possible BCs satisfying the current-nullification requirement
can be derived using the formalism of Refs.~\cite{Ahari,KharitonovGBCs},
which provides a universal ``standardized'' form, parameterized by unitary matrices.
Together with the Hamiltonian of the most general form,
these general BCs provide the continuum model of the most general form for a class-A system with a boundary.
If present, symmetries can also be naturally incorporated~\cite{KharitonovLSM,KharitonovQAH,KharitonovSC} into the formalism of general BCs.
Symmetries typically provide additional constraints, which lead to subfamilies of general BCs that satisfy these symmetries.
These symmetry-constrained general BCs,
together with symmetry-constrained general Hamiltonian provide the continuum model of the most general form for the respective symmetry class.

\subsection{(Semi-)Analytical method of calculating bound states \lbl{sec:calcmethod}}

One can then calculate the bound states of the so-derived general model.
The (semi-)analytical method, explained in Ref.~\cite{KharitonovGBCs} and used in some previous works, seems the most efficient.
The method follows directly from the theory of linear differential equations.
It amounts to constructing the general solution to the stationary Schr\"odinger equation at a given energy that decays into the bulk.
Applying the BCs, one obtains the equation for the energy of the bound states (and the corresponding wave functions).
For the simplest models, such as the model we consider in this work, the general bound-state structure can be found entirely analytically.
For more complicated models, one can still carry out this procedure ``semi-analytically'', with minimal computer aid.
The latter may be required for (i) finding particular solutions; (ii) solving the final equation for the bound-state energy.

\subsection{Advantages of the formalism \lbl{sec:advantages}}

The main fundamental value of the formalism is that since the so-constructed models are of the most general forms,
they represent {\em whole classes} of systems, with all possible structures of the bulk and boundary.
As such, these models deliver general bound-state structures descriptive of whole classes.
For models with moderate numbers of degrees of freedom,
the parameter spaces of both the Hamiltonian and BCs are typically small enough that they can be fully explored.
This level of general and exhaustive analysis that the formalism offers,
particularly in the description of the boundary, is something that is hardly possible within other methods:
in lattice-model or ab-initio calculations,
one typically resorts to a demonstration of the bound states for a particular (oftentimes, simplest) boundary,
and the question whether this represents generic bound-state properties of the whole class remains open.

The semi-analytical method of calculating bound states also offers important technical advantages:
bound states can be found for a truly half-infinite system and, as a result,
the low-energy features (such as the vicinity of the nodes of semimetals) can be resolved with any desired accuracy.
This is in contrast to the commonly employed finite-size numerical calculations,
where resolution is limited by the spatial quantization effects.

We demonstrate all these points in this work for the minimal continuum model.

\section{General 1D two-component linear-in-momentum continuum model \lbl{sec:1D}}

%don't do, at least now:
%\begin{figure}
%\caption{The wave-function components $\psi_\pm(z)$ [\eq{psi}].., according to the sign of their contribution to the
%probability current $j(z)$ [\eq{j}]. The velocities $\pm v_\pm$ determine the sign of the contribution.}
%\lbl{fig:1Dmodes}
%\end{figure}

\subsection{General Hamiltonian \lbl{sec:H}}

\begin{figure}
\includegraphics[width=.27\textwidth]{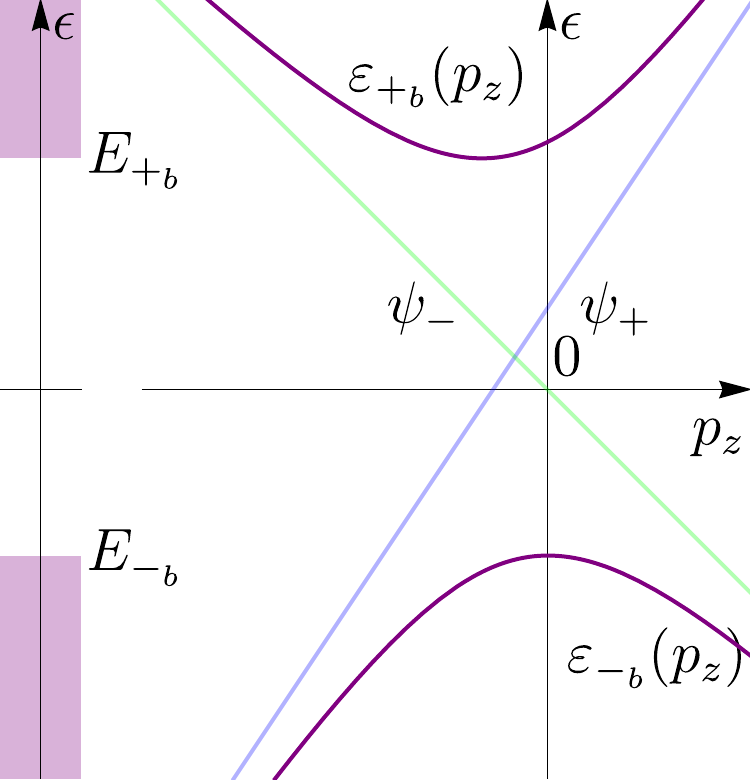}
\caption{Bulk-band spectrum $\eps_{\pm_b}(p_z)$ [\eq{e}] of the general 1D linear-in-momentum two-component continuum model
with the Hamiltonian $\Hh(\ph_z)$ [\eq{H}].
The light blue and green linear bands is the decoupled spectrum \eqn{edp=0} of the right- and left-moving modes $\psi_\pm$ [\eq{psi}], respectively,
in the absence of the gap terms, for $d_\p=0$.
}
\lbl{fig:e}
\end{figure}

In this section, we demonstrate the application of the formalism of general continuum models with BCs
to the 1D model with the minimal number of degrees of freedom necessary to have a well-defined boundary.
The most general form of the linear-in-momentum Hamiltonian for the two-component wave function $\psih(z)$ reads
\beq
    \Hh(\ph_z)=\hat{d}+\vh_z\ph_z,
\lbl{eq:Hinit}
\eeq
where $\hat{d}$ and $\vh_z$ are arbitrary hermitian ($\hat{d}^\dg=\hat{d}$, $\vh_z^\dg=\vh_z$) energy and velocity matrices, respectively, and
\[
	\ph_z=-\ix\pd_z
\]
is the momentum operator.
Throughout, $\dg$ denotes hermitian conjugation of a matrix of any size. We use units in which the Planck constant $\hbar=1$.

We first exploit the available basis degrees of freedom to recast the model in the form most suitable for making analytical progress
in deriving the general BCs and subsequent solution of the bound-state problem.
(We stress that this is always possible for the model of the most general form; no specific assumptions about its form are made.)
Such degrees of freedom are:
(i) the SU(2) rotation (three real parameters) of the wave-function basis;
(ii) the coordinate and momentum basis; the latter becomes particularly important in higher dimensions.
Such procedure has already been carried out in the 2D case in Ref.~\cite{KharitonovQAH}.

The central technical advancement in the formalism of general BCs made in Ref.~\cite{Ahari} was to present the probability current in the diagonal form.
For the linear-in-momentum Hamiltonian \eqn{Hinit}, the current is determined by the velocity matrix $\vh_z$.
To achieve that, we use two (out of three) real parameters of the SU(2) wave-function basis.
In the wave-function basis that diagonalizes the velocity matrix, the Hamiltonian of the most general form \eqn{Hinit} reads
\beq
    \Hh(\ph_z)
%    =\tauh_0 h_0(\ph_z) +\tauh_x h_x(\ph_z)+ \tauh_y h_y(\ph_z) + \tauh_z h_z(\ph_z)
%	=\tauh_0 h_0(\ph_z)+(\taubh\cd\hb(\ph_z))
    =\tauh_0 d_0 +\tauh_x d_x+ \tauh_y d_y + \tauh_z d_z +(\tauh_0 v_{0z}+\tauh_z v_{zz})\ph_z.
\lbl{eq:H}
\eeq
Here, $\tauh_0=\um$ and $\tauh_{x,y,z}$ are the unity and Pauli matrices, respectively
(note that, in general, the $x,y,z$ labels of the Pauli matrices have no meaning of or relation to the real-space coordinate system).
We also assume $v_{zz}>0$, which is always possible to achieve by properly ordering the basis states
(the term $\tauh_z v_{zz} \ph_z$ {\em must} be present for a system with a boundary, see the next \secr{bc}).
The wave-function components in this basis will be denoted as
\beq
    \psih(z)=\lt(\ba{c}\psi_+(z)\\ \psi_-(z)\ea\rt);
\lbl{eq:psi}
\eeq
the meaning of the $\pm$ labels will become clear in the next \secr{bc}.

The energy matrix $\hat{d}$ is still of the most general form, represented by the first four terms in \eq{H} with the real parameters $d_{0,x,y,z}$.
For the subsequent analysis and pseudospin interpretation, it is convenient to treat the coefficients as components of the 3D vector
\beq
    \db=(d_x,d_y,d_z)=(d_\p\cos\de,d_\p\sin\de,d_z),
\lbl{eq:d}
\eeq
where $d_\p=\sq{d_x^2+d_y^2}$ is the absolute value and $\de$ is the polar angle of its projection $(d_x,d_y)$ onto the $xy$ pseudospin plane.

Further simplifications of the form \eqn{H} are, in principle, possible.
The term $\tauh_0 d_0$ describes a trivial energy shift, which can be eliminated by changing the energy reference point.
The term $\tauh_z d_z$ can be eliminated by an overall momentum shift, realized in real space by the wave-function change
$\psih(z)=\ex^{-\ix \f{d_z}{v_{zz}} z}\psih'(z)$.
The angle $\de$ can be set to $\de=0$ by adjusting the relative phase factor of the basis states,
which is the remaining third real parameter of the SU(2) rotation of the wave-function basis.
As a result, in 1D, one can always reduce the Hamiltonian of the most general form \eqn{Hinit}
to the form $d_x\tauh_x+(\tauh_0 v_{0z}+\tauh_z v_{zz})\ph_z$ by utilizing the available basis degrees of freedom.
However, we choose not to exploit these possible further simplifications for two reasons.
First, the general dependence of the bound-state solution on these parameters (especially on $\de$, \secr{2DA})
in 1D is instrumental in uncovering the fundamental properties of the model.
Second, in 2D and 3D, $d_{0,x,y,z}$ become functions of momentum along the edge/surface and cannot be eliminated.
They provide part of the dependence of the edge/surface-state spectra and can affect them in a qualitative way;
in particular, $d_0$ in 3D determines whether the Weyl semimetal is type-I or type-II (\secr{Wtype}).

\subsection{Derivation of the general boundary condition \lbl{sec:bc}}

Although the general BC for the two-component linear-in-moment model has been derived many times
before~\cite{Berry,McCann,AkhmerovPRL,AkhmerovPRB,Ahari,KharitonovLSM,KharitonovQAH,Hashimoto2016,Hashimoto2019,Walter},
here, we rederive it following the ``standardized'' procedure and form of Ref.~\ocite{Ahari}
and provide details to emphasize the important points of the formalism.
Also, for pedagogical purposes, we demonstrate the derivation of the current-conservation principle itself from the norm-conservation principle,
and emphasize the point that BCs are essentially a way of specifying admissible Hilbert spaces over which the Hamiltonian becomes hermitian.

We consider a half-infinite sample occupying the region $z\geq 0$, \figr{1D}(a).
The wave function $\psih(z)$ is defined only in this region and is undefined in $z<0$. Physically, this means the following.
As already mentioned above, typically, a continuum model arises as a low-energy limit of some (``more'') microscopic model.
The wave function of this underlying microscopic model (which is at least implied, even when not specified) is defined for all $z$.
In the region $z\geq0$, this model has low-energy excitations, which are described  by the continuum model of interest,
with the wave function $\psih(z)$ [\eq{psi}] and the Hamiltonian $\Hh(\ph_z)$ [\eq{H}].
Whereas in the region $z<0$, the microscopic model has a gap in the spectrum that is much larger than the relevant low-energy scale.
As a result, the microscopic wave function decays rapidly into the region $z<0$.
The half-space $z<0$ is therefore the region ``inaccessible'' for the low-energy excitations that the wave function $\psih(z)$ describes.

BCs for continuum models have their origin in the very fundamentals of quantum mechanics,
namely, the norm conservation of the wave function $\psih(t)=\psih(z,t)$ upon time evolution, described by the Schr\"odinger equation
$\ix\pd_t\psih(z,t)=\Hh(\ph_z)\psih(z,t)$:
\[
	\pd_t\lan\psih(t),\psih(t)\ran=0.
\]
This leads to the requirement that the Hamiltonian be hermitian,
\beq
    0=\lan\psih,\Hh\psih\ran-\lan\Hh\psih,\psih\ran
    =\int_0^{+\iy}\dx z\,\psih^\dg(z)[\hat{d}-\hat{d}^\dg+(\vh_z-\vh_z^\dg)\ph_z]\psih(z)+\ix j(0).
\lbl{eq:herm}
\eeq
Here,
\[
    \lan \psih_1,\psih_2\ran=\int_0^{+\iy}\dx z\, \psih_1^\dg(z)\psih_2(z)
\]
is the scalar product, given by the integral over the region $z\geq 0$ of the half-infinite system.

There are two contributions in \eq{herm}: the integral bulk contribution and the boundary contribution $\ix j(0)$,
both of which have to vanish individually for any wave function $\psih(z)$ in the Hilbert space.
The bulk contribution leads to the usual requirement that $\Hh(p_z)$
is a hermitian matrix for any real momentum $p_z$: $\Hh^\dg(p_z)=\Hh(p_z)$ $\LRarr$ $\hat{d}^\dg=\hat{d}$, $\vh_z^\dg=\vh_z$.
This, however, is a necessary, but not a sufficient condition for the Hamiltonian $\Hh(\ph_z)$
to be a hermitian operator for a system with a boundary.
The other requirement is that the probability current at the boundary vanishes
\beq
    j(0)=0.
\lbl{eq:j=0}
\eeq
The quadratic form of the probability current reads
\beq
    j(z)=\psih^\dg(z)\vh_z\psih(z)
    =v_+\psi_+^*(z)\psi_+(z)-v_-\psi_-^*(z)\psi_-(z),
\lbl{eq:j}
\eeq
where
\[
    \vh_z=\tauh_0 v_{0z}+\tauh_z v_{zz}=\lt(\ba{cc} v_+ & 0 \\ 0 & -v_-\ea\rt),
\]
is the diagonal velocity matrix of the Hamiltonian \eqn{H} with
\beq
	v_\pm=v_{zz}\pm v_{0z};
\lbl{eq:vpm}
\eeq
$\pm v_\pm$ are the velocities of $\psi_\pm(z)$ components in the absence of the coupling terms, when $d_\p=0$.
The current does not depend at all on the energy matrix $\hat{d}$, constant in momentum.

Unlike the bulk contribution to \eq{herm}, which provides a constraint on the form of the Hamiltonian operator $\Hh(\ph_z)$,
the boundary contribution \eqn{j=0} is ultimately a constraint on the Hilbert space.
The current at the boundary has to be nullified identically for any wave function in the Hilbert space.
Resolving this constraint \eqn{j=0} in the form of linear homogeneous relations for the wave function at the boundary
is essentially a way of specifying the admissible Hilbert spaces over which the Hamiltonian becomes hermitian.
And these relations are commonly referred to as the {\em boundary conditions} (BCs).

Inspecting \eq{j}, it is clear that the current at the boundary can be nullified for a nonvanishing wave function
only if the velocities $\pm v_\pm \gtrless 0$ are of opposite signs (and which is why $v_{zz}>0$ necessarily has to be nonzero);
when they are of the same sign, the quadratic form of the current is either positive- or negative-definite.
For $v_{zz}<v_{0z}$, one has $\pm v_\pm >0$ and both wave-function components $\psi_\pm(z)$
propagate in the positive $z$ direction [both are ``right movers'' in the geometry of \figr{1D}(a)].
For $v_{0z}<-v_{zz}$, one has $\pm v_\pm <0$ and both $\psi_\pm(z)$
propagate in the negative direction (both are ``left movers'').
In these cases, there are no BCs that would nullify the current,
and hence, for a system with a boundary, no Hilbert space exists over which the Hamiltonian could be hermitian.
Physically, this means that in 1D a system with such Hamiltonian cannot have a boundary, it cannot be terminated;
such Hamiltonian can only describe a system with no boundary (infinite or looped).
One physical realization of such situation is the edge of a 2D quantum Hall system with two edge modes of the same chirality.
This also has a similar implication for type-II Weyl semimetals in 3D, which we discuss in \secr{Wtype}.

Only for
\beq	
	-v_{zz}<v_{0z}<v_{zz},
\lbl{eq:vrange}
\eeq
$\psi_\pm(z)$ propagate in the positive and negative directions (hence the labeling), $\pm v_\pm\gtrless 0$,
and are right- and left-movers, respectively (due to the convention $v_{zz}>0$), \figr{e}.
Only in this regime, the current can be nullified by proper BCs and the boundary can be introduced.
In the remainder of the paper, we assume the regime \eqn{vrange}.

In this regime, it is instructive~{Ahari} to further introduce the ``normalized'' right- and left-moving modes
\beq
    \psir_\pm(z)=\sq{v_\pm} \psi_\pm(z),
\lbl{eq:psir}
\eeq
in terms of which the current reads
\beq
    j(z)=\psir_+^*(z)\psir_+(z)-\psir_-^*(z)\psir_-(z).
\lbl{eq:jpsir}
\eeq
Finding the BC that identically nullifies the current becomes particularly transparent in terms of these variables.
Considering the linear homogeneous relation
\beq
    \psir_+(0)=U\psir_-(0)
\lbl{eq:bcU}
\eeq
and inserting it into the current nullification constraint \eqn{j=0}, one obtains
\[
    j(0)=(U^*U-1)\psir_-^*(0)\psir_-(0)=0,
\]
which is true when $|U|=1$.

Hence, \eq{bcU} is the sought BC for any
\beq
	U=\ex^{-\ix\nu}\in\Ux(1), \spc \nu\in[0,2\pi)\sim S^1,
\lbl{eq:U}
\eeq
parameterized by the angle $\nu$ covering the full unit circle $S^1$.
We emphasize that this is, in fact, a {\em family} of all possible BCs.
Each value of $\nu$ on the circle corresponds to one distinct BC, which specifies one distinct admissible Hilbert space,
over which the Hamiltonian is hermitian.
Physically, different $\nu$ describe different boundaries with, e.g., different microscopic structures in the underlying microscopic models.
In terms of the wave-function components \eqn{psi}, the general BC [\eqs{bcU}{U}]
for the linear-in-momentum Hamiltonian \eqn{H} of the most general form reads
\beq
    \sq{v_+}\psi_+(0)=\ex^{-\ix\nu}\sq{v_-}\psi_-(0), \spc \nu\in[0,2\pi).
\lbl{eq:bc}
\eeq

This is the case with the minimal number, one, of the right- and left-moving modes, for which the boundary can be defined.
As demonstrated in Ref.~\cite{Ahari}, the same structure of the general BCs holds for any 1D continuum Hamiltonian
with any (equal) number right- and left-moving modes, and they are parameterized by a unitary matrix.
Here $U\in\Ux(1)$ is such scalar unitary matrix.
The general BCs can be given~\cite{KharitonovGBCs} a natural physical interpretation, illustrated in \figr{1D}(b),
as a scattering processing between the incident $\psi_-(z)$ (left-movers for $z\geq 0$)
and reflected $\psi_+(z)$ (right-movers) modes, where $U$ can be seen as a scattering matrix.
According to this interpretation, in the above minimal case,
the angle $\nu$ parameterizing all possible general BCs \eqn{bc} acquires the meaning of the phase shift.

\subsection{Geometric pseudospin interpretation of the general boundary condition \lbl{sec:bcgeom}}

\begin{figure}
\includegraphics[width=.33\textwidth]{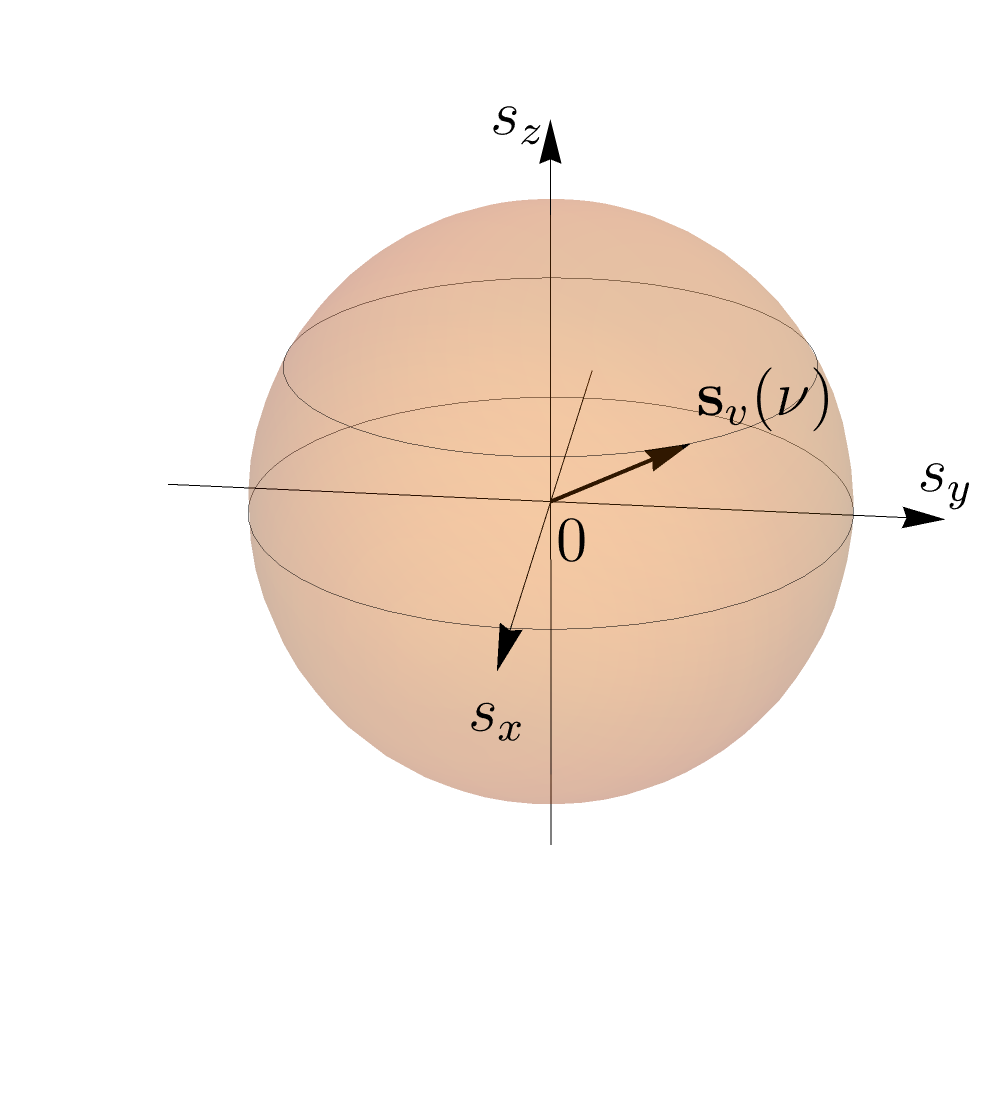}
\caption{Geometric interpretation of the general BC condition \eqn{bc} for the 1D two-component linear-in-momentum continuum model
in terms of the effective pseudospin-$\f12$. The pseudospin $\s_v(\nu)$ [\eq{sv}] of the wave function $\psih(0)$ [\eq{psi(0)}] at the boundary must 
be on circle of the Bloch unit sphere, obtained as its intersection with the plane $s_z=\vr_{0z}$ [\eq{svz}], fixed by the velocity ratio \eqn{vr0z}.
The position of the pseudospin $\s_v(\nu)$ on this circle can be arbitrary and specifies the BC [\eq{bcps}] via the angle parameter $\nu\in[0,2\pi)$.}
\lbl{fig:bcgeom}
\end{figure}

We notice that the BC \eqn{bc} for the linear-in-momentum Hamiltonian \eqn{H} has an elegant and insightful geometrical interpretation (\figr{bcgeom}),
if one regards the two-component wave function $\psih(z)$ [\eq{psi}] as that of pseudospin-$\f12$.

As is well-known~\cite{LLIII}, for any two-component (coordinate-independent) spinor
\beq
	\wh(\s)=\lt(\ba{c}w_+(\s)\\w_-(\s)\ea\rt),
\lbl{eq:w}
\eeq
there exists a direction in the 3D space, which can be characterized by a 3D unit vector $\s=(s_x,s_y,s_z)$, $\s^2=1$,
onto which the spinor has a definite $+\f12$ pseudospin projection,
meaning that it is an eigenvector of the operator $(\taubh\cd\s)=\tauh_x s_x+\tauh_y s_y+\tauh_z s_z$ with the eigenvalue $+1$:
\[
    (\taubh\cd\s)\wh(\s)=\wh(\s).
\]
We will refer to this property simply as to ``the spinor $\wh(\s)$ having the pseudospin $\s$'' and indicate this by the argument $\s$.
The pseudospin can be parameterized by the spherical angles as
\beq
	\s=(\sin\tht_s\cos\phi_s,\sin\tht_s\sin\phi_s,\cos\tht_s).
\lbl{eq:s}
\eeq
Below we assume that the spinor $\wh(\s)$ is normalized, $\wh^\dg(\s)\wh(\s)=1$, and its overall phase factor is fixed;
after this, $\wh(\s)$ is fully determined by $\s$.
We emphasize that the pseudospin-$\f12$ and its {\em effective} 3D space
may in general have nothing to do with the physical degrees of freedom that the wave function $\psih(z)$ represents.

Let us present the wave function at the boundary in the form
\beq
    \psih(0)=\psi_0\wh(\s),
\lbl{eq:psi(0)init}
\eeq
where $\psi_0$ is a complex number that fixes the amplitude and phase of $\psih(0)$, and $\wh(\s)$ fixes the pseudospin $\s$.
This way, the two complex $\psi_\pm(0)$ numbers are parameterized by one complex number $\psi_0$ and two real parameters of $\s$.
Inserting this form into the current-nullification constraint \eqn{j=0},
\[
    j(0)=|\psi_0|^2(v_{0z}+v_{zz} s_z)=0,
\]
we see that it fixes the $s_z=s_{zv}$ component (equivalently, the $\tht_s=\tht_v$ angle) of the pseudospin at the boundary to the value
\beq
    s_{vz}=\cos\tht_v
    =-\vr_{0z},\spc \sin\tht_v
    =\sq{1-\vr_{0z}^2},
\lbl{eq:svz}
\eeq
determined by the dimensionless velocity [\eq{vrange}]
\beq
	\vr_{0z}=\f{v_{0z}}{v_{zz}}\in(-1,1).
\lbl{eq:vr0z}
\eeq
At the same time, the angle $\phi_s=\nu\in[0,2\pi)$ can be arbitrary.
Hence, the wave function
\beq
    \psih(0)=\psi_0\wh(\s_v(\nu)),
    \spc
    \wh(\s_v(\nu))=\lt(\ba{c} \ex^{-\ix\f{\nu}2}\cos\f{\tht_v}2 \\ \ex^{+\ix\f{\nu}2}\sin\f{\tht_v}2 \ea\rt),
\lbl{eq:psi(0)}
\eeq
at the boundary with the pseudospin
\beq
    \s_v(\nu)=(s_{vx},s_{vy},s_{vz})=(\sin\tht_v\cos\nu,\sin\tht_v\sin\nu,\cos\tht_v),
\lbl{eq:sv}
\eeq
in which $\tht_v$ is fixed by \eq{svz} and $\nu\in[0,2\pi)$ is arbitrary, satisfies the current-nullification principle.

The condition \eqn{psi(0)} of having $\s_v(\nu)$ pseudospin at the boundary can be equivalently recast
as $\psih(0)$ being orthogonal to the spinor
\[
    \wh(-\s_v(\nu))=\lt(\ba{c} -\ex^{-\ix\f{\nu}2}\sin\f{\tht_v}2 \\ +\ex^{+\ix\f{\nu}2}\cos\f{\tht_v}2 \ea\rt)
\]
with the pseudospin $-\s_v(\nu)$, as expressed by the scalar product
\beq
    \wh^\dg(-\s_v(\nu))\psih(0)=0.
\lbl{eq:bcpsinit} %ps=pseudospin,init=initial
\eeq
Spelling out the latter,
\beq
    \ex^{\ix\f{\nu}2}\sin\tf{\tht_v}2\psi_+(0)-\ex^{-\ix\f{\nu}2}\cos\tf{\tht_v}2\psi_-(0)=0,
\lbl{eq:bcps}
\eeq
we confirm that this BC is indeed identical to \eq{bc} obtained from the formalism.
But this way, the general BC receives an elegant geometric interpretation:
the allowed pseudospin $\s_v(\nu)$ [\eq{sv}] of the wave function \eqn{psi(0)} at the boundary
must be on the circle of the Bloch unit sphere, as depicted in \figr{bcgeom}.
The position of the pseudospin on this circle can be arbitrary and specifies the BC [\eq{bcps}] via the angle parameter $\nu\in[0,2\pi)$.

\subsection{Bulk spectrum \lbl{sec:e}}

The bulk spectrum of the Hamiltonian \eqn{H} consists of the two bands
\beq
    \eps_{\pm_b}(p_z)=d_0+v_{0z} p_z\pm_b\sq{d_\p^2+(d_z+v_{zz} p_z)^2},
\lbl{eq:e}
\eeq
shown in \figr{e}.
The subscript of $\pm_b$ is meant to designate ``bulk'', to distinguish these signs
from those used to denote the wave-function \eqn{psi} components $\psi_\pm(z)$.
For absent $d_\p=0$, the spectrum is gapless with the linearly dispersing bands
\beq
	\eps_{\pm_b}(p_z)|_{d_\p=0}=d_0 + v_{0z}p_z \pm_b |d_z+v_{zz}p_z|=d_0+v_{0z} p_z \pm (d_z + v_{zz} p_z).
\lbl{eq:edp=0}
\eeq
In the last formula, the bands correspond to decoupled $\psi_\pm(z)$.
The bands cross at energy and momentum
\beq
	(\e,p_z)=\lt(d_0-\vr_{0z} d_z,-\f{d_z}{v_{zz}}\rt).
\lbl{eq:epzcross}
\eeq
The effect of the diagonal energy term $\tauh_z d_z$ is therefore to shift the bands in momentum, and also in energy when $v_{0z}\neq0$ is nonzero.

The effect of the off-diagonal energy terms $\tauh_xd_x+\tauh_yd_y$ is to open the gap around this crossing point.
For nonzero $d_\p$ [\eq{d}], the boundaries
\beq
	E_{+_b}=\min_{p_z}\eps_{+_b}(p_z), \spc
	E_{-_b}=\max_{p_z}\eps_{-_b}(p_z)
\lbl{eq:Edef}
\eeq
of the upper (``conduction'') $+_b$ and lower (``valence'') $-_b$ bulk bands \eqn{e} are
\beq
	E_{\pm_b}
	=d_0-\vr_{0z}d_z \pm_b \sq{1-\vr_{0z}^2} d_\p
	=d_0+d_z \cos\tht_v \pm_b d_\p \sin\tht_v .
\lbl{eq:E}
\eeq
The gap in the spectrum equals
\[
	E_{+_b}-E_{-_b}=2d_\p\sq{1-\vr_{0z}^2}=2d_\p\sin\tht_v.
\]

The bands $\eps_{\pm_b}(p_z)$ of the gapped spectrum are symmetric with respect to the crossing point \eq{epzcross}.

\subsection{Bound state \lbl{sec:bs}}

\begin{figure}
\includegraphics[width=.40\textwidth]{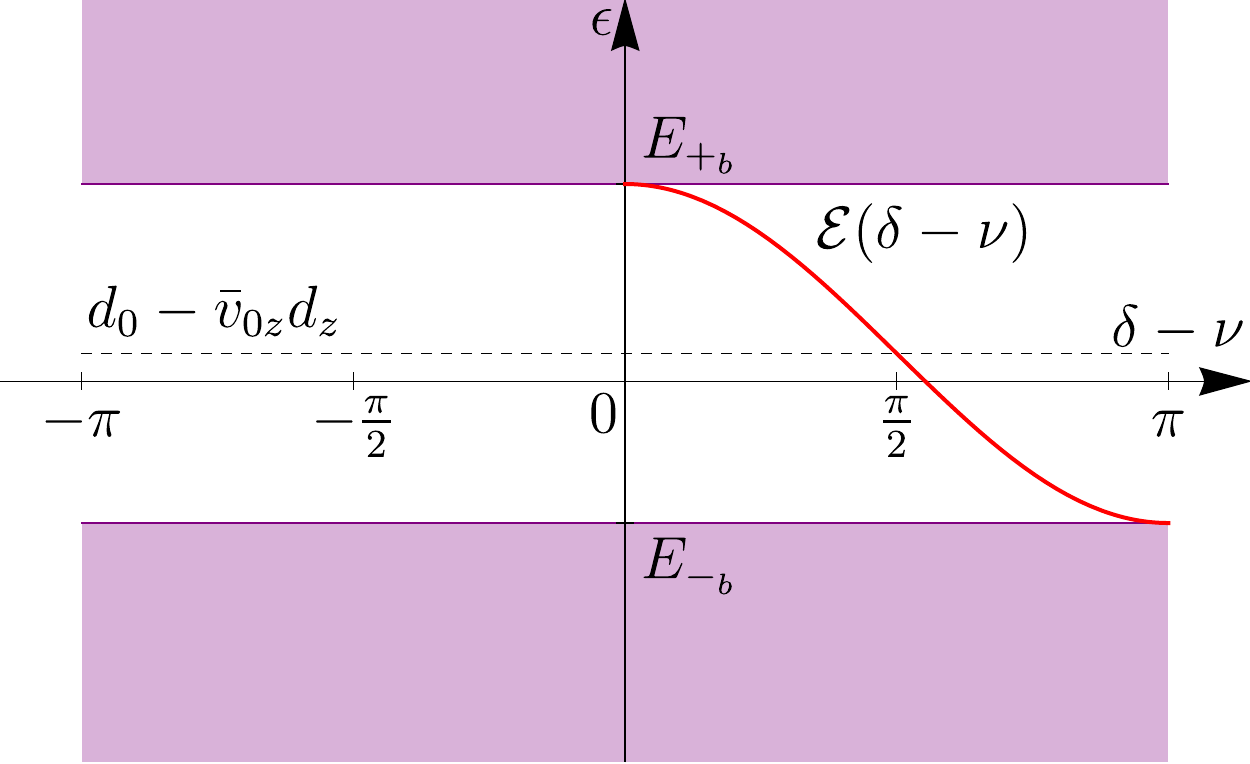}
\caption{The bound state $\Ec(\de-\nu)$ [red, \eq{Ec}] of the 1D linear-in-momentum two-component model of the most general form,
with the Hamiltonian $\Hh(\ph_z)$ [\eq{H}] and BC \eqn{bc}. The purple regions depict the continuum of the bulk states.
The model describes class A of the general topological classification~\cite{Ryu,Chiu}, which is topologically trivial in 1D.
Nonetheless, the bound state exists generically in the half of the parameter space of either the angle $\de$ of the gap terms [\eq{d}]
or the phase-shift angle $\nu$ of the general BC \eqn{bc}.
We provide two explanations for the existence of this bound-state band in \secr{2reasons}, which is one of the key results of this work.
}
\lbl{fig:Ec}
\end{figure}

We now calculate the bound state for the model for the two-component wave function
with the most general forms of the linear-in-momentum Hamiltonian [\eq{H}] and BC [\eqs{bc}{bcps}]
using the method outlined in \secr{calcmethod}, which for this model is fully analytical.
For a half-infinite sample occupying the region $z\geq0$,
the wave function \eqn{psi} of the bound state must decay into the bulk: $\psih(z)\rarr \nm$ as $z\rarr+\iy$.
Due to this, the energy
\beq
	\e\in(E_{-_b},E_{+_b})
\lbl{eq:egap}
\eeq
of the bound state can belong only to the gap of the bulk spectrum \eqn{e}
between the bulk-band boundaries \eqn{E}, which is present when $d_\p>0$ is nonzero.
We first construct the general solution to the stationary Schr\"odinger equation
\[
	\Hh(\ph_z)\psih(z)=\e\psih(z)
\]
at a given energy $\e$ that decays into the bulk.
Such general solution is a linear combination of particular solutions of the form $\chih(\e)\ex^{\ix p_z(\e) z}$
with complex momenta $p_z(\e)$ satisfying the characteristic equation
\[
	\det[\Hh(p_z(\e))-\e\um]=0
\]
and the corresponding eigenvector $\chi(\e)$ satisfying
\[
	[\Hh(p_z(\e))-\e\um]\chih(\e)=\nm.
\]
For the $z>0$ sample, there is just one momentum solution
\beq
	p_z(\e)=-\f{d_z}{v_{zz}}+\f{d_\p}{v_{zz}\sq{1-\vr_{0z}^2}}\lt[-\vr_{0z}\er'+\ix\sq{1-\er'^2}\rt]
\lbl{eq:pz(e)}
\eeq
with the positive imaginary part, which corresponds to a particular solution that decays as $z\rarr+\iy$.
Here, we introduce the dimensionless shifted energy
\beq
	\er'=\f{\e'}{d_\p\sq{1-\vr_{0z}^2}},
\lbl{eq:er'}
\eeq
where $\e'$ is the energy relative to the center \eqn{epzcross} of the bulk spectrum:
\beq
    \e=d_0- \vr_{0z} d_z+\e'.
\lbl{eq:ee'}
\eeq

The eigenvector of this particular solution reads
\beq
	\chih(\e)
	=\lt(\ba{c} \f1{\sq{v_+}}\ex^{\ix[\Phi(\e)-\de]} \\ \f1{\sq{v_-}}\ea\rt),
\lbl{eq:chi(e)}
\eeq
where we introduce the phase function $\Phi(\e)$ as
\beq
    \ex^{\ix\Phi(\e)}=\er'+\ix\sq{1-\er'^2}.
\lbl{eq:Phidef}
\eeq
Importantly, due to the square root in the imaginary part, the phase function takes values in the range
\beq
    \Phi(\e)\in[0,\pi].
\lbl{eq:Phirange}
\eeq

The general solution to stationary the Schr\"odinger equation at energy within the bulk gap [\eq{egap}],
decaying into the bulk, consists of just this one particular solution \eqn{chi(e)} and reads
\[
    \psih(z)=c\chih(\e)\ex^{\ix p_z(\e)z},
\]
where $c$ is a free constant fixed by normalization.
Inserting this form into the general BC \eqn{bc}, we obtain the equation for the bound-state energy
\[
	\ex^{\ix \Phi(\e)}=\ex^{\ix(\de-\nu)}.
\]
Due to the range constraint \eqn{Phirange} of the phase function,
there is one bound-state solution, which exists only for $\de-\nu\in(0,\pi)$, when $\sin(\de-\nu)>0$.
The dimensionless shifted energy \eqn{er'} of the bound-state solution reads
\[
	\Ecr'(\de-\nu)=\cos(\de-\nu), \spc
	\de-\nu\in (0,\pi),
\]
and its actual energy reads [\eq{ee'}]
%need both? the first one shown  $\de-\nu$, the second one shows linearity in $d$
\beq
	\Ec(\de-\nu)
	=d_0-\vr_{0z}d_z+\sq{1-\vr_{0z}^2}d_\p\cos(\de-\nu)
	=d_0-\vr_{0z}d_z+\sq{1-\vr_{0z}^2}(d_x\cos\nu+d_y\sin\nu),
	\spc \de-\nu\in(0,\pi).
\lbl{eq:Ec}
\eeq

The most important dependence of the bound-state energy $\Ec(\de-\nu)$
is that on the angle $\de$ of the gap terms [\eq{d}] and the angle $\nu$ in the general BC \eqn{bc}.
We observe that it depends only on their {\em difference} $\de-\nu$. The dependence is plotted in \figr{Ec}.
As $\de-\nu$ spans the interval $(0,\pi)$, the bound-state energy $\Ec(\de-\nu)$ spans the gap $(E_{-_b},E_{+_b})$ of the bulk spectrum,
merging with the boundaries \eqn{E} of the upper $+_b$ and lower $-_b$ bulk bands at the ends of the interval:
\beq
    \Ec(0)=E_{+_b},\spc \Ec(\pi)=E_{-_b}.
\lbl{eq:Ecmerge}
\eeq
The bound state exists in the {\em half} of the $\Ux(1)\sim S^1\sim[0,2\pi)$ parameter spaces of either $\de$ or $\nu$ and is absent in the other half.
For a fixed $\nu$, as a function of $\de$, the bound-state ``band'' $\Ec(\de-\nu)$ merges with the bulk boundaries at
\beq
	\de_{+_b}=\nu,\spc \de_{-_b}=\nu+\pi
\lbl{eq:demerge}
\eeq
and occupies the region
\[
	%\de_{+_b}<\de<\de_{-_b},\spc
	\de\in(\de_{+_b},\de_{-_b}).
\]
When $\Ec(\de-\nu)$ is considered as a function of one of $\de$ or $\nu$, the other determines the position of the band.

We also notice that, owing to the geometric pseudospin interpretation of the general BC introduced in \secr{bcgeom} and \figr{bcgeom},
the bound-state energy \eqn{Ec} can be presented in the elegant form
\beq
    \Ec(\de-\nu) =d_0+(\db\cd\s_v(\nu)), \spc \de-\nu\in(0,\pi),
\lbl{eq:Ecd}
\eeq
as a scalar product of the ``energy vector'' $\db$ [\eq{d}] in the Hamiltonian \eqn{H}
and the pseudospin unit vector $\s_v(\nu)$ [\eq{sv}] characterizing the BC \eqn{bcps}.

Inserting the found bound-state energy \eqn{Ec} into \eqs{pz(e)}{chi(e)}, we obtain the complex momentum
\[
	p_z(\Ec(\de-\nu))=-\f{d_z}{v_{zz}}+\f{d_\p}{v_{zz}\sq{1-\vr_{0z}^2}}\lt[-\vr_{0z}\cos(\de-\nu)+\ix\sin(\de-\nu)\rt],
\]
and the wave function
\beq
	\psih_\x{bs}(z)=\psi_0\wh(\s_v(\nu))\ex^{\ix p_z(\Ec(\de-\nu))z}.
\lbl{eq:psibs}
\eeq
of the bound state. In accord with \eq{psi(0)}, the eigenvector $\chih(\Ec(\de-\nu))$ of the bound-state solution
is collinear with the pseudospin state $\wh(\s_v(\nu))$ at the boundary, fixed by the general BC [\eqs{bcpsinit}{bcps}].

\section{Topology and bulk-boundary correspondence of a generalized quantum anomalous Hall system \lbl{sec:topo}}
%no transition whatsoever, but ok?

Consider a closed path in the parameter space $(d_0,d_x,d_y,d_z,v_{0z},v_{zz})$ of the general 1D Hamiltonian \eqn{H}.
Let $\al\in[0,2\pi)\sim S^1$ be the variable parameterizing this closed path, defined on a unit circle $S^1$.
This {\em one-parameter subfamily} of 1D Hamiltonians $\Hh(\ph_z,\al)$ is periodic in $\al$,
and as such, it can be viewed as {\em one} effective 2D Hamiltonian, in the 2D space of $(p_z,\al)$, which is effectively a cylinder.
It belongs to class A, topologically nontrivial in 2D, and represents an effective 2D quantum anomalous Hall (QAH) system
(here, in accord with the common terminology, ``anomalous'' just means that real magnetic field is not involved).
This construction is also known as the 1D {\em adiabatic charge pump}~\cite{RiceMele1982,FuKane2006}.

The bulk topology of a QAH system is characterized by the Chern number topological invariant.
For the two-band model we consider, the Chern number $C$ is the skyrmion charge of the mapping
from the $(p_z,\al)$ space onto the unit Bloch sphere $S^2$ realized by the pseudospin
\beq
	\s(p_z,\al)=-\f{\hb(p_z,\al)}{|\hb(p_z,\al)|}, \spc \hb=(h_x,h_y,h_z),\spc |\hb|=\sq{h_x^2+h_y^2+h_z^2},
\lbl{eq:stopo}
\eeq
of the valence band $-_b$.
Here,
\[
	\hb(p_z,\al)=\db(\al)+(0,0,v_{zz}(\al))p_z
\]
is the effective ``Zeeman field'' of the Hamiltonian \eqn{H} presented in the form
\beq
	\Hh(p_z,\al)
	=\tauh_0 h_0(p_z,\al)+(\taubh\cd\hb(p_z,\al)),
\spc
	h_0(p_z,\al)=d_0(\al)+v_{0z}(\al)p_z,
\lbl{eq:HZeeman}
\eeq
The explicit expression for the Chern number reads
%{\bf[sort out the sign, other than that correct?]}
\beq
	C=\f1{4\pi}\int \dx p_z \dx \al \, \sin\tht_s(p_z,\al)\pd_{p_z} \tht_s \pd_{\al} \phi_s
	=\f1{4\pi}\int \sin\tht_s \dx\tht_s \dx\phi_s,
\lbl{eq:C}
\eeq
where $\tht_s$, $\phi_s$ are the spherical angles [\eq{s}] of the pseudospin \eqn{stopo}.

The key consequence of the so-introduced bulk QAH topological structure
is that the bound-state energy $\Ec(\al)$ [\eq{Ec}] of the 1D system
can now be viewed as an effective ``edge-state'' band as a function of the parameter $\al$ for the 2D QAH system.
As a result, bulk-boundary correspondence~\cite{Chiu} holds,
which states that the signed number $N[\Ec(\al)]$ of the chiral bands is determined by the bulk Chern number:
\beq
	N[\Ec(\al)]=C.
\lbl{eq:bbc}
\eeq
An edge-state band is called chiral if it connects lower $-_b$ (``valence'') and upper $+_b$ (``conduction'') bulk bands.
The {\em chirality}, defined as the individual signed number $\pm1$ of a chiral band,
is determined by the order of the merging points $\al_{\pm_b}$ with these bulk bands on the $\al$ axis.
I.e., if there is just one chiral band for the introduced effective 2D QAH system,
the chirality is equal to the sign of the difference:
\[
	N[\Ec(\al)]=\x{sgn}(\al_{+_b}-\al_{-b}).
\]
If there are several chiral bands, then the total signed number $N[\Ec(\al)]$ in \eq{bbc}
is given by the sum of such individual chiralities.

Therefore, as per the bulk-boundary correspondence \eqn{bbc},
in a topologically nontrivial phase of an effective 2D QAH system with nonzero Chern number,
the corresponding signed net number $N$ of chiral bands is guaranteed to exist.
In \secr{2DA}, we will make use of this construction
(considering closed paths in the parameter space of the 1D system, which represent effective 2D QAH systems)
to provide one explanation for existence of the bound state \eqn{Ec}.

\section{Chiral symmetry in the formalism of general continuum models with boundary conditions \lbl{sec:S}}

\begin{figure}
\includegraphics[width=.40\textwidth]{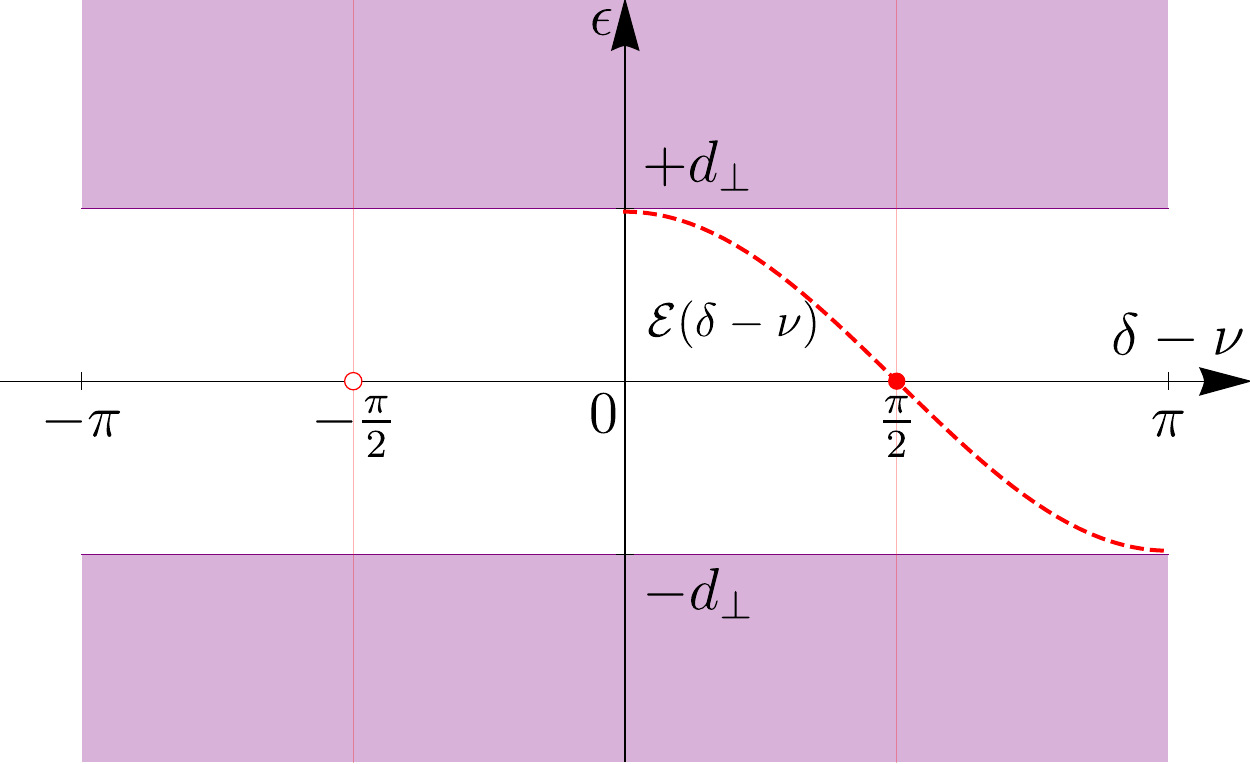}
\spc
\includegraphics[width=.32\textwidth]{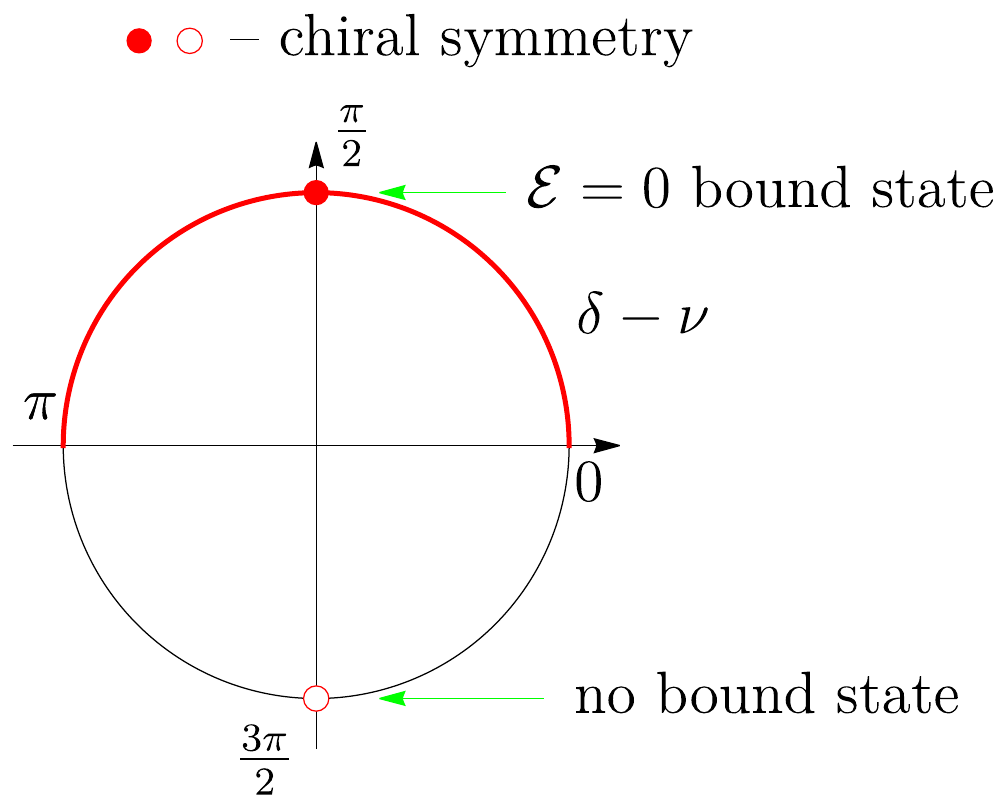}
\caption{Bound-state structure of the 1D chiral-symmetric linear-in-momentum two-component model of the most general form,
with the Hamiltonian $\Hh^\Sc(\ph_z)$ [\eq{HS}] and BC \eqn{bcgenS}.
Only two values $\de-\nu=\pm\f{\pi}2$ [\eq{Sdenu}] (marked with red vertical lines) correspond to chiral symmetry
of the system with the boundary, when both the Hamiltonian and BC are symmetric under one chiral-symmetry operation.
There are no bound states at $\de-\nu=-\f{\pi}2$ and there is a zero-energy $\Ec=0$ bound state at $\de-\nu=+\f{\pi}2$.
The emergence of the %[chiral {\bf[-in a different sense!]}
bound-state band $\Ec(\de-\nu)$ [dashed red, \eq{Ec}, \figr{Ec}]
upon the deviation from the chiral-symmetry point $\de-\nu=+\f{\pi}2$
is one of the explanations for its existence in a topologically trivial class A, which we provide in \secr{1DAIII}.
}
\lbl{fig:EcS}
\end{figure}

In this section, we demonstrate how symmetries can be naturally incorporated into the formalism of general continuum models with BCs.
We demonstrate that for chiral symmetry~\cite{Ryu,Chiu}.

Chiral symmetry is a unitary operation ($\Sh\Sh^\dg=\tauh_0$)
\beq
    \Sh\psih(z),
\lbl{eq:Sdef}
\eeq
that squares to unity:
\beq
    \Sh^2=\tauh_0.
\lbl{eq:S2=1}
\eeq
A Hamiltonian $\Hh^\Sc(\ph_z)$ is said to obey chiral symmetry $\Sc$
if there exists such an operation $\Sh$, under which the Hamiltonian changes its sign:
\beq
    \Sh \Hh^\Sc(\ph_z)\Sh^\dg=-\Hh^\Sc(\ph_z).
\lbl{eq:SH}
\eeq

We would like to derive the most general forms of {\em both} the chiral symmetry operation $\Sh$
and the Hamiltonian $\Hh^\Sc(\ph_z)$ that obeys that symmetry.
However, there is an obstacle here, as one can be found from \eq{SH} only once the other is specified:
if the Hamiltonian is given, then the most general form of the symmetry operation can be found;
if the symmetry operation is given, then the most general form of the Hamiltonian can be found.
However, if neither are specified from the start, the problem is not well-posed.
A systematic way to resolve this obstacle is as follows.
As discussed in \secr{bc}, among the terms in the general Hamiltonian $\Hh(\ph_z)$ [\eq{H}],
the term $\tauh_z v_{zz} \ph_z$ is essential in order for the boundary and BC to be well-defined.
Hence, this term must be present also in the chiral-symmetric Hamiltonian.
Therefore, we first find the most general form of $\Sh$ under which the term $\tauh_z v_{zz} \ph_z$ is chiral-symmetric.
The most general form of such operation is the linear combination
\beq
    \Sh(s)=\tauh_x \cos s +\tauh_y \sin s = \lt(\ba{cc} 0&\ex^{-\ix s}\\\ex^{+\ix s}&0\ea\rt)
\lbl{eq:S}
\eeq
with an arbitrary angle $s$. Note that the angle $s$ is defined modulo $\pi$, since $\Sh(s+\pi)=-\Sh(s)$ is effectively the same operation.

Such form with any $s$ prohibits all $\tauh_0$ terms in the Hamiltonian; hence, $d_0=0$ and $v_{0z}=0$ in the general $\Hh^\Sc(\ph_z)$.
Checking the remaining gap terms $d_{x,y}$ in $\Hh(\ph_z)$, we find that they are chiral-symmetric for
\beq
	s=\de+\tf\pi2 \x{ mod }\pi,
\lbl{eq:Ssde}
\eeq
\beq
    \Sh(\de+\tf\pi2)
    =\tauh_x \cos(\de+\tf\pi2) +\tauh_y \sin(\de+\tf\pi2)
    =-\tauh_x \sin\de +\tauh_y \cos\de.
\lbl{eq:Sde}
\eeq
Hence,
\beq
    \Hh^\Sc(\ph_z)=(\tauh_x\cos\de+\tauh_y\sin\de)d_\p + \tauh_z(d_z+v_{zz}\ph_z)
\lbl{eq:HS}
\eeq
is the most general form of the Hamiltonian [\eq{SH}], chiral-symmetric under the operation $\Sh(\de+\tf\pi2)$.

As per above, \eq{Ssde} is a two-way relation.
If $\de$ in the Hamiltonian \eqn{HS} is given,
then there always exists a chiral-symmetry operation of the form \eqn{Sde}, under which the Hamiltonian \eqn{HS} is chiral-symmetric.
If $s$ in the chiral-symmetric operation \eqn{S} is given,
then there always exists a Hamiltonian \eqn{HS} with $\de$ satisfying \eq{Ssde}, which is chiral-symmetric under such operation.
Importantly, note that since $\de$ is defined modulo $2\pi$, while $s$ is defined modulo $\pi$,
there are two solutions $s\pm\f\pi2$ for $\de$ for a given $s$.

Next, we want to find all BCs that satisfy chiral symmetry.
As already explained in Refs.~\ocite{KharitonovLSM,KharitonovQAH,KharitonovSC},
for the bound states to reflect the topological properties of a certain symmetry class, the {\em system with a boundary} must respect that symmetry.
For continuum models, this means that not only the bulk Hamiltonian [\eq{SH}], but also the BC has to satisfy that symmetry.
Symmetry of the BC means the following.
As explained in \secr{bc}, BCs are essentially a way of specifying admissible Hilbert spaces, over which the Hamiltonian becomes hermitian.
Since the transformed wave function must also belong to the same Hilbert space,
a BC is chiral-symmetric if the wave-function $\Sh\psih(z)$ [\eq{Sdef}], transformed by the chiral-symmetry operation, also satisfies this BC.
This provides constraints on the parameter space of BCs, i.e., BCs under certain symmetry are subsets of the general BCs.

Applying the above procedure,
the {\em general} BC \eqn{bc} (i.e., not yet chiral-symmetric) for the chiral-symmetric Hamiltonian under $\Sh(s)$ [\eq{S}] reads
\beq
    \sq{v_{zz}}	\psi_+(0)=\ex^{-\ix\nu}\sq{v_{zz}}\psi_-(0),
\lbl{eq:bcgenS}
\eeq
with yet arbitrary $\nu\in[0,2\pi)$.
Inserting $\Sh(s)\psih(0)$ into this general BC, we obtain
%\[
%    \sq{v_{zz}}	\ex^{-\ix s}\psi_-(0)=\ex^{-\ix\nu}\sq{v_{zz}}\ex^{+\ix s}\psi_+(0).
%\]
\[
    \sq{v_{zz}}	\psi_+(0)=\ex^{\ix(-2s+\nu)}\sq{v_{zz}}\psi_-(0).
\]
The latter BC is equivalent to \eq{bcgenS} when
\beq
	s=\nu \x{ mod }\pi.
\lbl{eq:Ssnu}
\eeq
We obtain that the BC \eqn{bcgenS} is chiral symmetric under the operation $\Sh(s)$ when the relation \eqn{Ssnu} holds.

Further, since for a system with the boundary, both the bulk Hamiltonian and the BC have to satisfy the same symmetry, both \eqs{Ssde}{Ssnu} have to hold.
This happens when $\de$ and $\nu$ are related as
\beq
	\de-\nu=\pm\f{\pi}2.
\lbl{eq:Sdenu}
\eeq
Since both $\de$ and $\nu$ are defined modulo $2\pi$, while $s$ is defined modulo $\pi$, two options arise.
This relation is to be understood as follows:
when either $\de$ or $\nu$ is fixed, there are only two discrete values of the other, when the system with the boundary is chiral-symmetric.

To summarize, the Hamiltonian \eqn{HS} and BC \eqn{bcgenS}, when one of the relations \eqn{Sdenu} holds,
describe the most general form of the model of the chiral-symmetric system with a boundary,
which realizes a topologically nontrivial in 1D class AIII.

Inspecting the bound-state solution \eqn{Ec} for these two cases \eqn{Sdenu},
we find that there is a zero-energy bound state $\Ec=0$ for $\de-\nu=+\f{\pi}2$ and an absent bound state for $\de-\nu=-\f{\pi}2$ (\figr{EcS}).
In both of these chiral-symmetric cases, the bound-state structure is symmetry-protected:
(i) when the $\Ec=0$ bound state is present, it cannot be removed without breaking chiral symmetry (or closing the bulk gap);
(ii) when the bound state is absent, a bound state cannot be introduced without breaking chiral symmetry.

\section{Two topological reasons for robust bound states in a topologically trivial 1D class-A system\lbl{sec:2reasons}}

The model studied in \secr{1D} represents a 1D class-A gapped system,
which, according to the general topological classification~\cite{Ryu,Chiu}, is topologically trivial.
Based on this, one could naively be tempted to conclude that the bound states contained in this model [\secr{bs}, \eq{Ec}],
present in half of the parameter space, are of completely accidental nature, with no particular reason for their existence.
This is, however, not the case, as a more elaborate analysis shows, which we now present.
In fact, we provide not one, but two topological explanations
for the robust bound states in a topologically trivial 1D class-A system
and argue that this relation could itself be regarded as an explicit manifestation of Bott periodicity.

\subsection{Connection of 1D class-A system to 2D class-A quantum anomalous Hall system \lbl{sec:2DA}}

The first explanation is based on the construction presented in \secr{topo},
which relates the 1D class-A model of interest to the 2D class-A model, with the same symmetry but different dimension:
if a closed path, parameterized by $\al\in S^1$ on a unit circle,
in the parameter space of a 1D class-A system is considered, it can be regarded as an effective 2D class-A QAH system.
The latter is a topologically nontrivial class, and if the system is in a topologically nontrivial phase with nonzero Chern number,
it is guaranteed to have chiral edge-state band(s) via bulk-boundary correspondence \eqn{bbc}.
This has a (perhaps somewhat unanticipated) consequence that the bound states in a topologically trivial 1D class-A system
are nonetheless guaranteed to exist in a substantial (i.e., not negligible) part of the parameter space of $\al$.
Even though they are not topologically protected, since they will merge with the bulk-state bands as $\al$ spans its range $[0,2\pi)$,
the bound states are nonetheless in this sense robust.

This is precisely what happens for the 1D model studied in \secr{1D}.
We notice that for the circular path of the fixed radius $d_\p$ in the space $(d_x,d_y)$ of the gap terms,
parameterized by the angle $\al=\de \in[0,2\pi)\sim S^1$,
the Hamiltonian $\Hh(\ph_z,\de)$ can be regarded as an effective 2D Hamiltonian
and it is in a topologically nontrivial phase with the Chern number $C=-1$ [\eq{C}].
According to the bulk-boundary correspondence \eqn{bbc}, it must have a chiral bound-state band with chirality $N=-1$.
The dependence $\Ec(\de-\nu)$ of the bound state energy on $\de$ (for a fixed BC angle $\nu$), shown in \figr{Ec},
has precisely these properties and now becomes particularly transparent.
Indeed, $\Ec(\de-\nu)$ can be regarded as a chiral edge-state band of an effective 2D QAH system with $\Hh(\ph_z,\de)$, with the chirality number
\[
	N[\Ec(\de-\nu)]=\x{sgn}(\de_{+_b}-\de_{-b})=-1.
\]
The bound states are indeed robust, as the band covers half $(\nu,\nu+\pi)$ of the $S^1$ parameter space of $\de$.

\subsection{Connection of 1D class-A system to 1D class-AIII chiral-symmetric system \lbl{sec:1DAIII}}

The second explanation is based on relating the 1D class-A model of interest to a 1D class-AIII model,
with the same dimension but of a different symmetry class (chiral symmetry).
In fact, we have already provided such type of explanation for the edge states of a nodal 2D semimetal in Refs.~\ocite{KharitonovLSM,KharitonovQAH},
which is a mathematically equivalent situation; we reproduce it in \secr{2D}.

Following the ideas of Ref.~\ocite{KharitonovLSM,KharitonovQAH}, this time it is instructive to consider the general 1D class-A model [\eqs{H}{bc}]
with no assumed symmetries as a deviation from the chiral-symmetric subsets in the parameter space, described by \eqss{HS}{bcgenS}{Sdenu} (\figr{EcS}).
As in \secr{2DA} above, we consider the variation of the bulk parameters, while the BC parameter $\nu$ in \eq{bcgenS} is assumed fixed.
To always have the BC chiral-symmetric, this fixes the chiral-symmetry operator to $\Sh(\nu)$ [\eq{S}] via \eq{Ssnu}.
After this, there are two discrete options \eqn{Sdenu} for the value of $\de$, at which the Hamiltonian \eqn{HS} is chiral-symmetric.
Starting at $\de=\nu+\f\pi2$, at which there is a zero-energy bound state, and varying $\de$,
the Hamiltonian $\Hh(\ph_z)$ [\eq{H}] deviates from chiral symmetry and the bound-state energy $\Ec(\de-\nu)$ deviates from zero.
The formation of the chiral bound-state band $\Ec(\de-\nu)$ as a function of $\de$
can thus be seen as a consequence of this mechanism: deviation from the chiral-symmetry point $\de=\nu+\f{\pi}2$,
at which the zero-energy bound state is guaranteed.
There necessarily exists a finite-size stability region in $\de$ around $\de=\nu+\f{\pi}2$, which happens to be $(\nu,\nu+\pi)$ of length $\pi$.
This region ends with the bound-state band merging with the bulk-band boundaries $E_{\pm_b}$ [\eqs{E}{Ecmerge}].
Similarly, starting at $\de=\nu-\f\pi2$, there exist a finite-size region $\de\in(\nu-\pi,\nu)$ of length $\pi$ around this point,
where bound states are absent.
Together, these two regions cover the whole parameter space $\de\in S^1$.

To summarize, we provide two (at first glance, seemingly independent) explanations for the existence of bound states [\eq{Ec}]
in a topologically trivial 1D class-A model \eqsn{H}{bc},
by relating it to two different topologically nontrivial models:
in one case (2D class-A model with no symmetries) -- by changing the dimension while preserving the symmetry class;
in the other (1D class-AIII with chiral symmetry) -- by changing the symmetry class while preserving the dimension.
We further notice that these are precisely the two cases related by Bott periodicity~\cite{Ryu,Chiu}.
In fact, it seems that the above construction can be viewed as an explicit manifestation of Bott periodicity.
The very 1D class-A system itself serves as the link that connects the 1D class-AIII and 2D class-A systems:
the angle parameter $\de$ of the 1D class-A system simultaneously describes the deviation from chiral symmetry
and is regarded as the second dimension of the effective 2D class-A QAH system.

These arguments identify a systematic ``propagation'' effect,
whereby bound states from the topologically nontrivial classes, where they are protected and guaranteed to exist,
propagate to the related, ``adjacent'' in dimension or symmetry, topologically trivial classes.
We discuss this in more detail in the concluding \secr{conclusion}.

\section{One node of a 3D Weyl semimetal \lbl{sec:3D}}
\subsection{General Hamiltonian, optimal current-diagonal form. \lbl{sec:H3}}

\begin{figure}
\includegraphics[width=.45\textwidth]{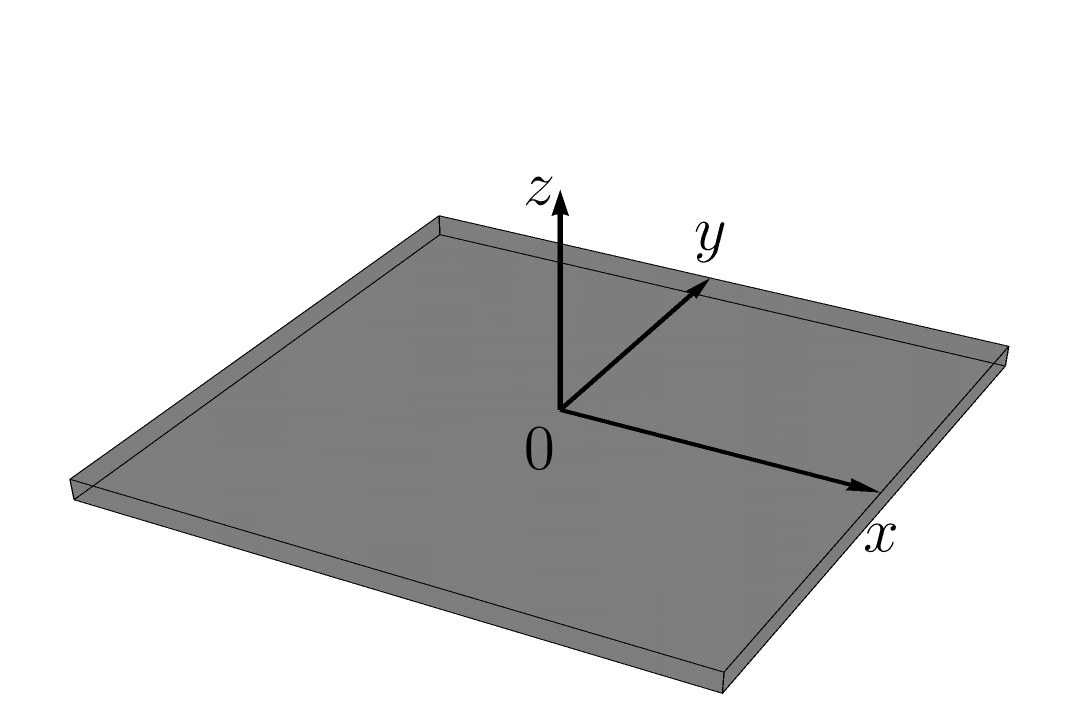}
\caption{3D half-infinite system occupying the $z\geq0$ half-space. %{\bf[Decide on orientation. Show momentum space as well?]}
}
\lbl{fig:3Dsystem}
\end{figure}

We now demonstrate the application of the formalism of general continuum models with BCs to the minimal model in 3D.
The two-component wave function $\psih(\rb)$ now depends on the radius vector $\rb=(x,y,z)$.
The most general form of the Hamiltonian, linear in the momentum operator
\[
	\pbh=(\ph_x,\ph_y,\ph_z)=-\ix(\pd_x,\pd_y,\pd_z),
\]
for this wave function reads
\beq
	\Hh_3(\pbh)=\sum_{\al=0,x,y,z} \tauh_\al(\eps_\al+v_{\al x} \ph_x+v_{\al y} \ph_y+v_{\al z} \ph_z).
\lbl{eq:H3init}
\eeq
Here, we also exploit the available basis degrees of freedom to recast this Hamiltonian in the form
most convenient for the analysis of the surface-state problem.
This is analogous to the steps performed in the Supplemental Material of Ref.~\ocite{KharitonovQAH} for the most general 2D model.
Such available degrees of freedom are:
the origin (three real translation parameters) and orientation
(three real parameter of the SO(3) rotation) of the affine real-space basis of $\rb$;
momentum origin (three real translation parameters);
orientation (three real parameters of the SU(2) rotation) of the basis in the vector space of the wave function $\psih$.
Note that we will utilize only the pure rotations SO(3) [and not O(3)] of the real-space basis to preserve its right-handedness;
this is important for the chirality of the Weyl node.

We assume that the sample occupies a half-space (\figr{3Dsystem}).
We use three origin and two of the three SO(3) rotation parameters of the $\rb$-basis
to orient it so that the sample occupies the $z\geq0$ half-space.
One parameter of the SO(3) rotation remains: the rotation about the $z$ axis.

Next, we deal with the Hamiltonian \eqn{H3init}.
Since we will only be using the final form of the Hamiltonian, we preserve the same notation in the intermediate formulas below;
this should not lead to confusion.
The effect of these basis changes can be viewed as elimination of certain terms in the general Hamiltonian \eqn{H3init}.

There are four energy parameters $\eps_\al$ ($\al=0,x,y,z$)
and twelve velocity parameters $v_{\al\be}$ ($\al=0,x,y,z$ and $\be=x,y,z$) in \eq{H3init}.
The energy parameter $\eps_0$ represents an inconsequential shift, which can be eliminated by changing the energy reference point;
this way, the energy of the Weyl point becomes $\e=0$.

The other three energy parameters $\eps_\al$ ($\al=x,y,z$) can be eliminated via a momentum shift.
In real space, this is realized by the wave-function change
\[
	\psih(\rb)\rarr \ex^{\ix(\qb\cd\rb)}\psih(\rb),
\]
where the momentum components $\qb=(q_x,q_y,q_z)$ are the solutions to the system of equations:
\[
	\eps_\al+v_{\al x} q_x+v_{\al y} q_y+v_{\al z} q_z=0,\spc \al=x,y,z.
\]
The Hamiltonian takes the form
\beq
	\Hh_3(\pbh)=\sum_{\al=0,x,y,z} \tauh_\al(v_{\al x} \ph_x+v_{\al y} \ph_y+v_{\al z} \ph_z).
\lbl{eq:H31}
\eeq

The three energy parameters have this way been eliminated by the shifts of the three momentum components, 
which means that there is a node in the spectrum.
As a manifestation of the known fact that 3D Weyl semimetals exist generically (without any symmetries or other conditions),
we obtain that the 3D linear-in-momentum Hamiltonian of the most general form \eqn{H3init} for a two-component wave function
describes the vicinity of a Weyl node.
The momentum operator in \eq{H31} now corresponds to the deviation from the Weyl node.

Similarly to 1D in \secr{1D}, we then use two of the three parameters of the SU(2) wave-function basis
to diagonalize the velocity matrix at the momentum component $\ph_z$ perpendicular to the surface:
\[
	\Hh_3(\pbh)
	=\sum_{\al=0,z} \tauh_\al(v_{\al x} \ph_x+v_{\al y} \ph_y)
	+\tauh_x(v_{xx} \ph_x+v_{xy} \ph_y)
	+\tauh_y(v_{yx} \ph_x+v_{yy} \ph_y)
	+(\tauh_0v_{0z}+\tauh_z v_{zz}) \ph_z
\]
(i.e., eliminate the $v_{xz}$ and $v_{yz}$ terms),
and in such a way that the necessarily nonzero velocity $v_{zz}>0$ is positive.
We use the remaining one parameter of the SU(2) wave-function basis rotation (rotation about the pseudospin $z$ axis)
and one real parameter of the SO(3) coordinate basis rotation (rotation about the real $z$ axis)
to also eliminate the $v_{xy}$ and $v_{yx}$ terms, and in such a way that $v_{xx}>0$ is positive.
We arrive at the desired form of the Hamiltonian:
\beq
	\Hh_3(\pbh)=\tauh_0 (v_{0x} \ph_x+v_{0y}\ph_y)
	+\tauh_z (v_{zx} \ph_x+v_{zy}\ph_y)
	+\tauh_x v_{xx} \ph_x+ \tauh_y v_{yy} \ph_y + (\tauh_0 v_{0z}+\tauh_z v_{zz})\ph_z.
\lbl{eq:H3}
\eeq
As a ``bookkeeping'' check: we have used all three parameters of the SU(2) rotation and one parameter of the SO(3) rotation
to reduce the number of velocity parameters in a desired way from twelve in \eq{H3init} to eight in \eq{H3}.
As in \secr{1D}, we emphasize again that any Hamiltonian of the most general form \eqn{H3init}
can always be expressed in the form \eqn{H3} by utilizing the available basis degrees of freedom.
For the velocities $v_{0z}$ and $v_{zz}$ at $\ph_z$, the same notation is preserved as in 1D [\eqs{vpm}{vr0z}].
As in 1D, the wave function in this basis will be denoted as
\[
	\psih(\rb)=\lt(\ba{c} \psi_{+}(\rb) \\ \psi_{-}(\rb) \ea\rt),
\]
with the signs denoting the propagation direction along the real $z$ axis [the signs of the contributions to the current \eqn{jz}].

\subsection{General local translation-symmetric boundary condition\lbl{sec:bc3}}

We now derive the BC of the most general form for the 3D Hamiltonian \eqn{H3}, under of a couple of reasonable assumptions.
Repeating the same procedure as in \secr{bc} for 3D,
we arrive at the expression similar to \eq{herm} for the hermiticy requirement for the Hamiltonian \eqn{H3}.
The bulk contribution imposes the constraint that $\Hh_3(\pb)$ must be a hermitian matrix when $\pb$ is a real vector,
i.e., that all velocity and energy parameters in \eqs{H3init}{H3} must be real.
The boundary contribution imposes the constraint that the {\em total} current through the $z=0$ surface vanishes:
\beq
	J_z=\int\dx x\dx y\, j_z(x,y,0)=0.
\lbl{eq:Jz=0}
\eeq
Here,
\beq
	j_z(\rb)=v_{+}\psi^*_{+}(\rb)\psi_{+}(\rb) - v_{-}\psi^*_{-}(\rb)\psi_{-}(\rb)
\lbl{eq:jz}
\eeq
is the density of the $z$ component of the current, normal to the $z=0$ surface, and the velocities $v_{\pm}$ are given by \eq{vpm}.

We notice a new effect that arises in higher dimensions, which is absent in 1D~\cite{1Dtwoboundaries}:
to ensure the norm conservation of the wave function, it is sufficient that only the total current, integral over the surface, vanishes.
Although mathematically correct,
this is clearly a way {\em too general} constraint for most real physical systems, where certain locality along the surface can be expected.
This ``nonlocality issue'' perhaps deserves a more detailed separate study, which we do not undertake here.
In this work, we will consider only a (natural) subset of the constraints \eqn{Jz=0},
where the current {\em density} vanishes individually at each point $(x,y)$ of the surface:
\beq
	j_z(x,y,0)=0.
\lbl{eq:jz=0}
\eeq

Similarly to 1D, the boundary can be introduced only when $\pm v_{\pm}\gtrless0$
[for which \eq{vrange} must be satisfied and $v_{zz}>0$ must be positive],
so that $\psi_{\pm}(\rb)$ are right- and left-movers,  respectively, providing positive and negative contributions to the current density \eqn{jz}.
In the context of Weyl semimetals, this also has to do with type-I or II band structure, which we discuss in more detail in \secr{Wtype}.

The second assumption we make is that the surface is translation-symmetric.
In this case the BC is translation-symmetric as well, i.e., the same BC resolves that constraint \eqn{jz=0} at every point $(x,y)$.
Under these two assumptions, we arrive at the translation-symmetric BC of the most general form
\beq
	\sq{v_{+}}\psi_{+}(x,y,0)=\ex^{-\ix\nu}\sq{v_{-}}\psi_{-}(x,y,0), \spc \nu\in[0,2\pi),
\lbl{eq:bc3}
\eeq
which have the same form at every point $\rb_\p=(x,y)$ along the surface, as the 1D BC \eqn{bc},
and is characterized by the same phase-shift angle parameter $\nu$.
An analogous general BC, but with broken translation symmetry along the edge, was explored for the 2D model in Ref.~\cite{Walter}.

\subsection{Bulk and surface-state spectra \lbl{sec:spectra}}

\subsubsection{Bulk spectrum}

%leave out for now
%\begin{figure}
%%\includegraphics[width=.45\textwidth]{Ec.pdf}
%\caption{{\bf[haven't decided on this fig yet, actually...]}
%Bulk-band spectrum $\eps_{3,\pm_b}(\pb)$ [\eq{e3}] of the general Hamiltonian $\Hh_3(\pbh)$ [\eq{H3}]
%of the 3D linear-in-momentum two-component model.
%}
%\lbl{fig:e3}
%\end{figure}

The bulk spectrum of the general 3D linear-in-momentum Hamiltonian \eqn{H3}, describing the vicinity of a Weyl node, reads
\beq
	\eps_{3,\pm_b}(\pb)=v_{0x} p_x+v_{0y}p_y+v_{0z} p_z \pm_b \sq{(v_{xx}p_x)^2+(v_{yy}p_y)^2+(v_{xz}p_x+v_{zy}p_y+v_{zz}p_z)^2}, \spc \pb=(p_x,p_y,p_z).
\lbl{eq:e3}
\eeq
%and is plotted in \figr{e3}.
The spectrum is necessarily gapless.
It consists of the lower (valence) $-_b$ and upper (conduction) $+_b$ band, which touch at the Weyl point $\pb=\nv$.
Due to the linear scaling in momentum, the spectrum is presentable as
\[
	\eps_{3,\pm_b}(\pb) = v_{zz} p\,\vr_{\pm_b}(\nb),
\]
where
\beq
	\vr_{\pm_b}(\nb)
	=\vr_{0x} n_x+\vr_{0y}n_y+\vr_{0z} n_z \pm_b \sq{(\vr_{xx}n_x)^2+(\vr_{yy}n_y)^2+(\vr_{xz}n_x+\vr_{zy}n_y+\vr_{zz}n_z)^2},
	\spc \vr_{\al\be}=\f{v_{\al\be}}{v_{zz}},
\lbl{eq:vr}
\eeq
are the dimensionless velocities of the bands, functions of the unit vector
\[
	\nb=(n_x,n_y,n_z)=(\sin\tht\cos\phi,\sin\tht\sin\phi,\cos\tht)
\]
of momentum $\pb=p\nb$, $p=|\pb|$.
The velocity $v_{zz}>0$ is chosen as the reference unit,
since it is the key parameter that cannot be zero, to have a well-defined boundary [\eq{vrange}], as explained in \secr{bc}.

\subsubsection{Surface-state spectrum}

\begin{figure}
\includegraphics[width=.45\textwidth]{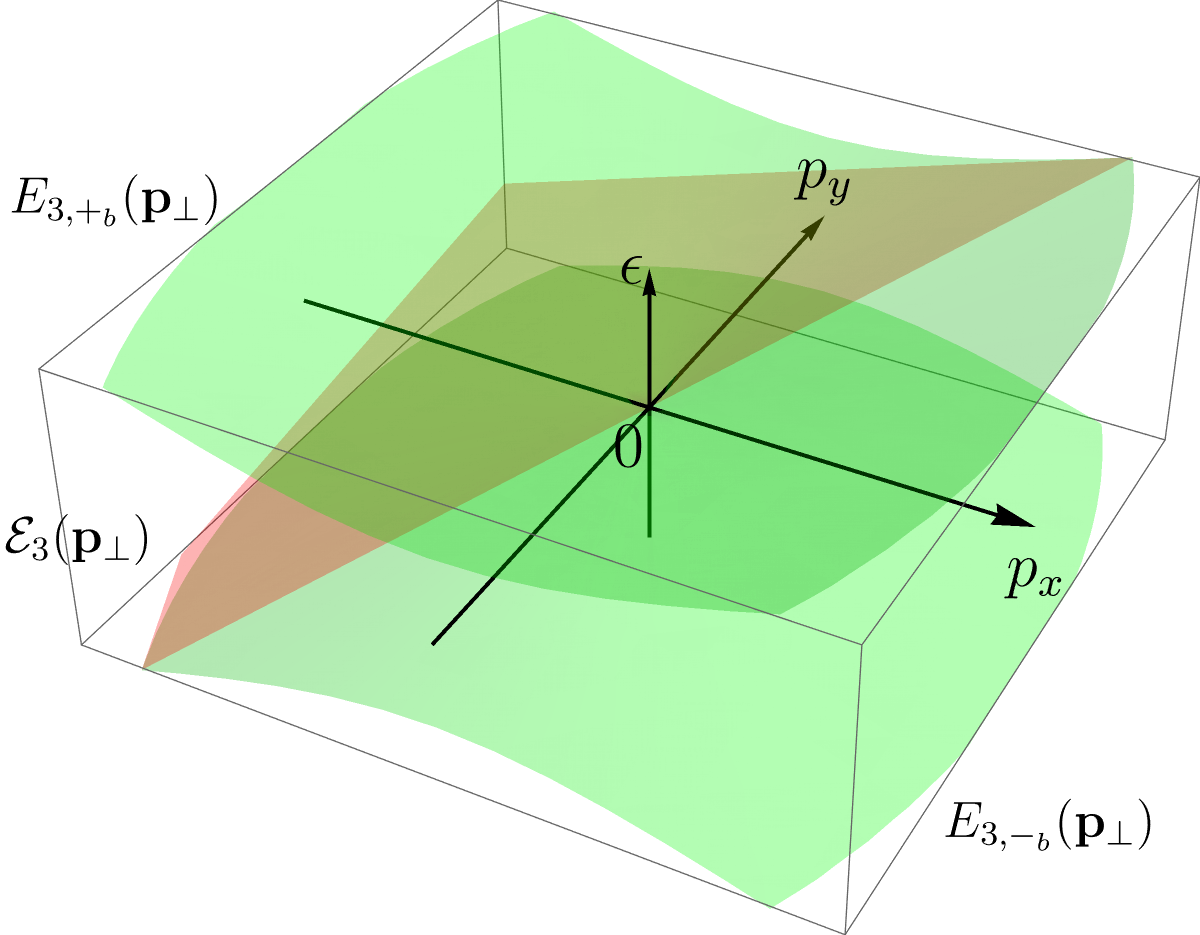}
\caption{
An example of the surface-state spectrum $\Ec_3(\pb_\p)$ [red, \eq{Ec3}]
of the 3D linear-in-momentum two-component model of the most general form,
with the Hamiltonian $\Hh_3(\pbh)$ [\eq{H3}] and BC \eqn{bc3}.
The bulk-band boundaries $E_{3,\pm_b}(\pb_\p)$ [\eq{E3}] are shown in green.
The same parameters as in \figr{Vcr}(b) and \figr{pp} were used: $\nu=0.2\pi$, $\Vr_{0x}=0$, $\Vr_{0y}=0$
(absent additive term), $\f{v_{yy}}{v_{xx}}=0.6$ (present $xy$ anisotropy).
}
\lbl{fig:Ec3}
\end{figure}

For the conserved momentum
\beq
	\pb_\p=(p_x,p_y)=p_\p(\cos\phi,\sin\phi), \spc p_\p=\sq{p_x^2+p_y^2},
\lbl{eq:pp}
\eeq
along the surface, the model and the surface-state problem reduce completely to the 1D case studied in \secr{1D}.
The eigenstates can be sought in the form
\beq
	\psih(\rb)=\ex^{\ix(p_x x+p_y y)}\psih(z),
\lbl{eq:psiz3}
\eeq
in which $\psih(z)$ becomes identical to the 1D wave function \eqn{psi}.
The general BC for $\psih(z)$, following from \eq{bc3}, becomes identical to \eq{bc}.
The Hamiltonian \eqn{H3}
\beq
	\Hh_3(\pb_\p,\ph_z)
	=\tauh_0 d_0(\pb_\p)
	+\tauh_z d_z(\pb_\p)
	+\tauh_x d_x(\pb_\p)+ \tauh_y d_y(\pb_\p) + (\tauh_0 v_{0z}+\tauh_z v_{zz})\ph_z
\lbl{eq:H3pp}
\eeq
for $\psih(z)$ and a given conserved $\pb_\p$ becomes identical to \eq{H}.
The effect of $\pb_\p$ is to generate the energy terms, which become its functions:
\beq
    d_0(\pb_\p)=v_{0x} p_x+v_{0y}p_y
,\spc
    d_z(\pb_\p)=v_{zx} p_x+v_{zy}p_y,
\lbl{eq:d0zphi}
\eeq
\beq
    d_x(\pb_\p)=d_\p(\phi)\cos\de(\phi)=v_{xx} p_x=v_{xx} p_\p\cos\phi
,\spc
    d_y(\pb_\p)=d_\p(\phi)\sin\de(\phi)=v_{yy} p_y=v_{yy} p_\p\sin\phi,
\lbl{eq:dxyphi}
\eeq
\beq
    d_\p(\phi)=\sq{v_{xx}^2 p_x^2+v_{yy}^2 p_y^2}= p_\p \sq{v_{xx}^2 \cos^2\phi+v_{yy}^2 \sin^2\phi}.
\lbl{eq:dpphi}
\eeq
The dependence of the phase $\de(\phi)$ of the gap terms (for the effective 1D motion along $z$)
on the polar angle $\phi$ of the surface momentum \eqn{pp} is given by
\beq
	(\cos\de(\phi),\sin\de(\phi))=\f1{\sq{v_{xx}^2\cos^2\phi+v_{yy}^2\sin^2\phi}}(v_{xx}\cos\phi,v_{yy}\sin\phi).
\lbl{eq:dephi}
\eeq
There is also a relation
\beq
	\tan\de(\phi)=\f{v_{yy}}{v_{xx}}\tan\phi,
\lbl{eq:tandephi}
\eeq
which can be used to determine $\de(\phi)$ modulo $\pi$, while \eq{dephi} can be used to determine it modulo $2\pi$.

Once the Hamiltonian has been presented in the effective 1D form \eqn{H3pp},
the solution of the surface-state problem is then obtained by simply taking the 1D expressions from \secr{bs}
and substituting the constant energy terms $d_{0,x,y,z}$ there with the above functions \eqssn{d0zphi}{dxyphi}{dpphi} of the surface momentum $\pb_\p$.
The surface-state spectrum of the general linear-in-momentum model \eqn{H3},
which describes the vicinity of an isolated node of a generic Weyl semimetal,
therefore consists of one band and reads [\eq{Ec}]
\beq
	\Ec_3(\pb_\p) = d_0(\pb_\p)-\vr_{0z} d_z(\pb_\p) +\sq{1-\vr_{0z}^2} [d_x(\pb_\p) \cos\nu + d_y(\pb_\p) \sin\nu].
\lbl{eq:Ec3}
\eeq
Its range in $\pb_\p$ and the merging lines with the bulk-band boundaries are discussed in \secr{Ec3range}.

Since every energy term is linear in $\pb_\p$, the surface-state spectrum
\[
    \Ec_3(\pb_\p) = v_{zz} p_\p \Vcr(\phi)
\]
is linear in the absolute value $p_\p$ [\eq{pp}] and can therefore be fully characterized by its dimensionless velocity
\beq
	\Vcr(\phi)=\dr_0(\phi)-\vr_{0z}\dr_z(\phi)+\sq{1-\vr_{0z}^2} \dr_\p(\phi) \cos[\de(\phi)-\nu]
	=\Vr_{0x}\cos\phi+\Vr_{0y}\sin\phi+\sq{1-\vr_{0z}^2} (\vr_{xx}\cos\phi\cos\nu+\vr_{yy}\sin\phi\cos\nu)
\lbl{eq:Vcr}
\eeq
along the direction $\pb_\p/p_\p=(\cos\phi,\sin\phi)$ set by the polar angle $\phi$.
Here, we introduce the dimensionless energy terms
\beq
    \dr_0(\phi)=\vr_{0x} \cos\phi +\vr_{0y}\sin\phi
,\spc
    \dr_z(\phi)=\vr_{zx} \cos\phi +\vr_{zy}\sin\phi,
\lbl{eq:dr0zphi}
\eeq
\beq
	\dr_x(\phi)=\vr_{xx}\cos\phi,\spc \dr_y(\phi)=\vr_{yy}\sin\phi,
\lbl{eq:drxyphi}
\eeq
\beq
	\dr_\p(\phi)=\sq{\vr_{xx}^2\cos^2\phi+\vr_{yy}^2\sin^2\phi},
\lbl{eq:drpphi}
\eeq
as
\beq
	d_\al(\pb_\p)=v_{zz} p_\p \dr_\al(\phi), \spc \al=0,x,y,z,
\lbl{eq:drdef}
\eeq
and the velocity combinations
\[
	\Vr_{0x}=\vr_{0x}-\vr_{0z}\vr_{zx}, \spc \Vr_{0y}=\vr_{0y}-\vr_{0z}\vr_{zy}.
\]

Moreover, due to the special property of the 1D model that the bound-state energy [\eqs{Ec}{Ecd}] depends linearly on the energy parameters $d_{0,x,y,z}$,
the surface-state spectrum \eqn{Ec3} is linear not only in $p_\p$, but also in each momentum component:
\[
	\Ec_3(\pb_\p)=\Vc_x p_x+\Vc_y p_y, %= p_\p(\Vc_x\cos\phi+\Vc_y\sin\phi)
\]
and, equivalently, the dimensionless velocity \eqn{Vcr}
\beq
	\Vcr(\phi)=\Vcr_x\cos\phi+\Vcr_y\sin\phi
\lbl{eq:Vcrlin}
\eeq
is linear in $p_x/p_\p=\cos\phi$ and $p_y/p_\p=\sin\phi$ individually.
Here,
\beq
	\Vc_{x,y}=v_{zz}\Vcr_{x,y}
,\spc
	\Vcr_x%=\vr_{0x}-\vr_{0z}\vr_{zx}+\sq{1-\vr_{0z}^2}\vr_{xx}\cos\nu
	=\Vr_{0x}+\sq{1-\vr_{0z}^2}\vr_{xx}\cos\nu
,\spc
	\Vcr_y%=\vr_{0y}-\vr_{0z}\vr_{zy}+\sq{1-\vr_{0z}^2}\vr_{yy}\sin\nu
	=\Vr_{0y}+\sq{1-\vr_{0z}^2}\vr_{yy}\sin\nu.
\lbl{eq:Vcxy}
\eeq
This means that the 3D plot of the surface-state band $\Ec_3(\pb_\p)$ is a segment of a plane (it is, actually, a half-plane, see below).
This was first demonstrated in Refs.~\ocite{Hashimoto2016,Hashimoto2019},
where the surface-state spectrum near one node of a Weyl semimetal
was calculated for the most general form of the BC (of a different, but equivalent parametrization),
but for specific, simpler forms of the linear-in-momentum Hamiltonian:
in Ref.~\ocite{Hashimoto2016} -- for the fully isotropic case, with equal velocities $v_{xx}=v_{yy}=v_{zz}$ and other velocities absent;
in Ref.~\ocite{Hashimoto2019} -- with additional $v_{0x}$, $v_{0y}$, $v_{0z}$ velocities, relevant for the type-II Weyl semimetal.
Here, we demonstrate this and other properties for the most general forms of both the Hamiltonian and BC.

The boundaries
\beq
	E_{3,+_b}(\pb_\p)=\min_{p_z}\eps_{3,+_b}(\pb), \spc
	E_{3,-_b}(\pb_\p)=\max_{p_z}\eps_{3,-_b}(\pb)
\lbl{eq:E3def}
\eeq
of the bulk-state bands \eqn{e3} are given by
\beq
	E_{3,\pm_b}(\pb_\p) = d_0(\pb_\p) - \vr_{0z} d_z(\pb_\p) \pm_b \sq{1-\vr_{0z}^2}d_\p(\pb_\p).
\lbl{eq:E3}
\eeq
They are also linear in $p_\p$:
\beq
	E_{3,\pm_b}(\pb_\p)=v_{zz} p_\p \Vr_{\pm_b}(\phi),
\lbl{eq:Er3def}
\eeq
and can be fully characterized by the dimensionless velocity functions
\beq
	\Vr_{\pm_b}(\phi)
	=\dr_0(\phi)-\vr_{0z}\dr_0(\phi) \pm_b \sq{1-\vr_{0z}^2} \dr_\p(\phi)
	=\Vr_{0x}\cos\phi+\Vr_{0y} \sin\phi \pm_b \sq{1-\vr_{0z}^2} \sq{\vr_{xx}^2\cos^2\phi+\vr_{yy}^2\sin^2\phi}
\lbl{eq:Vr}
\eeq
of just the polar angle $\phi$ of the surface momentum [\eq{pp}]. The local-in-$\phi$ bulk ``gap'' (at a fixed radius $p_\p$)
\beq
	\Vr_{+_b}(\phi)-\Vr_{-_b}(\phi)=2\sq{1-\vr_{0z}^2}\dr_\p(\phi)
\lbl{eq:dVr}
\eeq
is determined by \eq{drpphi}.

We recognize that the shape of the surface-state velocity function $\Vcr(\phi)$ [\eq{Vcr}]
is fully characterized by just {\em three} dimensionless combinations of the velocity parameters (in addition to the BC angle $\nu$);
the same concerns the velocities $\Vr_{\pm_b}(\phi)$ [\eq{Vr}] of the bulk-band boundaries.
For example, $(\Vr_{0x},\Vr_{0y})/(\sq{1-\vr_{0z}^2} \vr_{xx})$ can be used to characterize the relative effect of the additive term \eqn{Vr0}
and $v_{yy}/v_{xx}$ -- to characterize the $xy$ anisotropy, see the next \secr{mainW}.
The whole surface-state structure (both the surface-state band and bulk-band boundaries)
%The surface-state spectrum 
in the vicinity of an isolated Weyl node is therefore fully characterized by just five real parameters:
the BC angle $\nu$ and four velocity parameters, of which one is just an overall magnitude;
compare this to the twelve velocity parameters in the initial Hamiltonian \eqn{H3init}.
This could be particularly useful for analyzing experimental data,
as fits to this general dependence could be used to extract these parameters for real materials.

In \figsr{Vcrbase}{Vcr}, the surface-state spectrum in terms of the velocity dependencies on $\phi$ 
is plotted for several characteristic values of parameters.

\begin{figure}
\includegraphics[width=.40\textwidth]{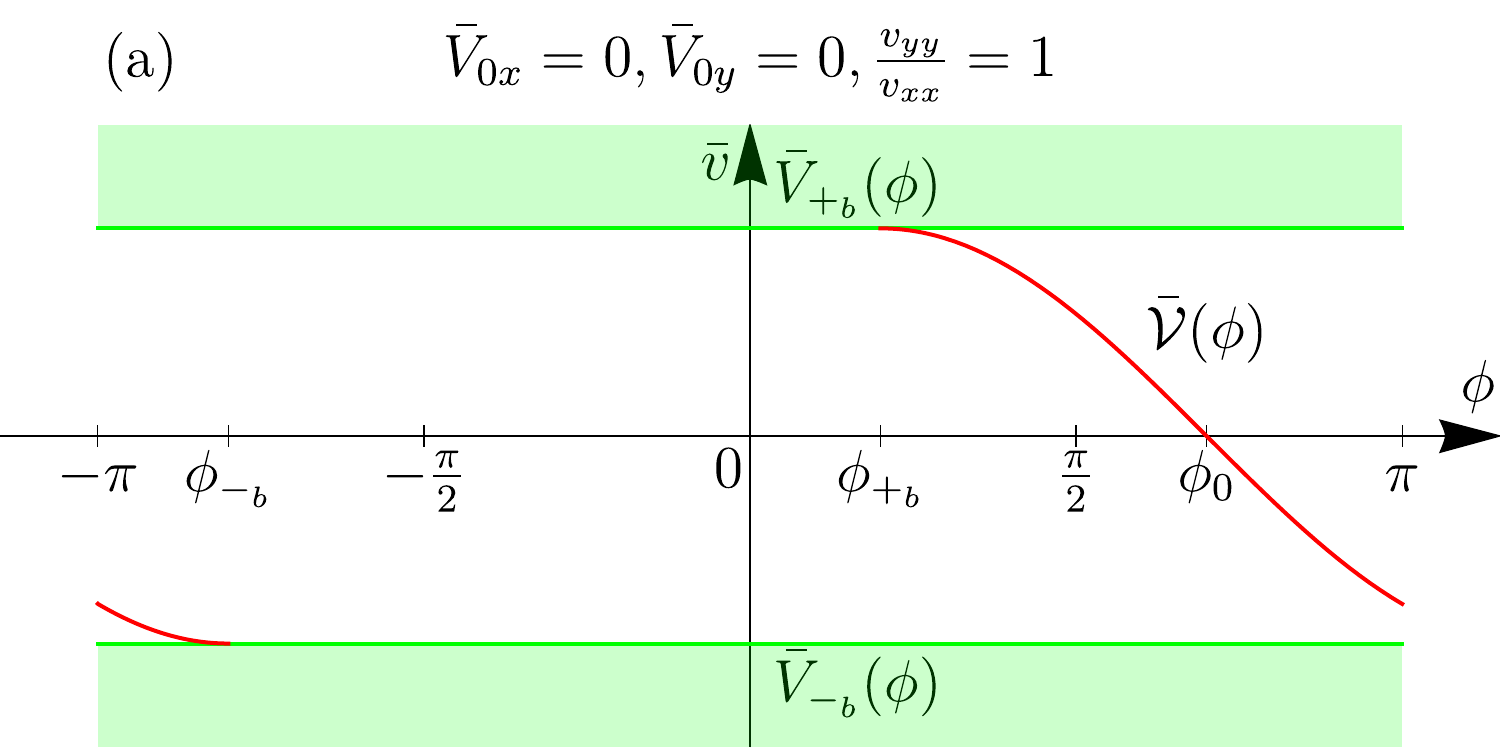}
\includegraphics[width=.40\textwidth]{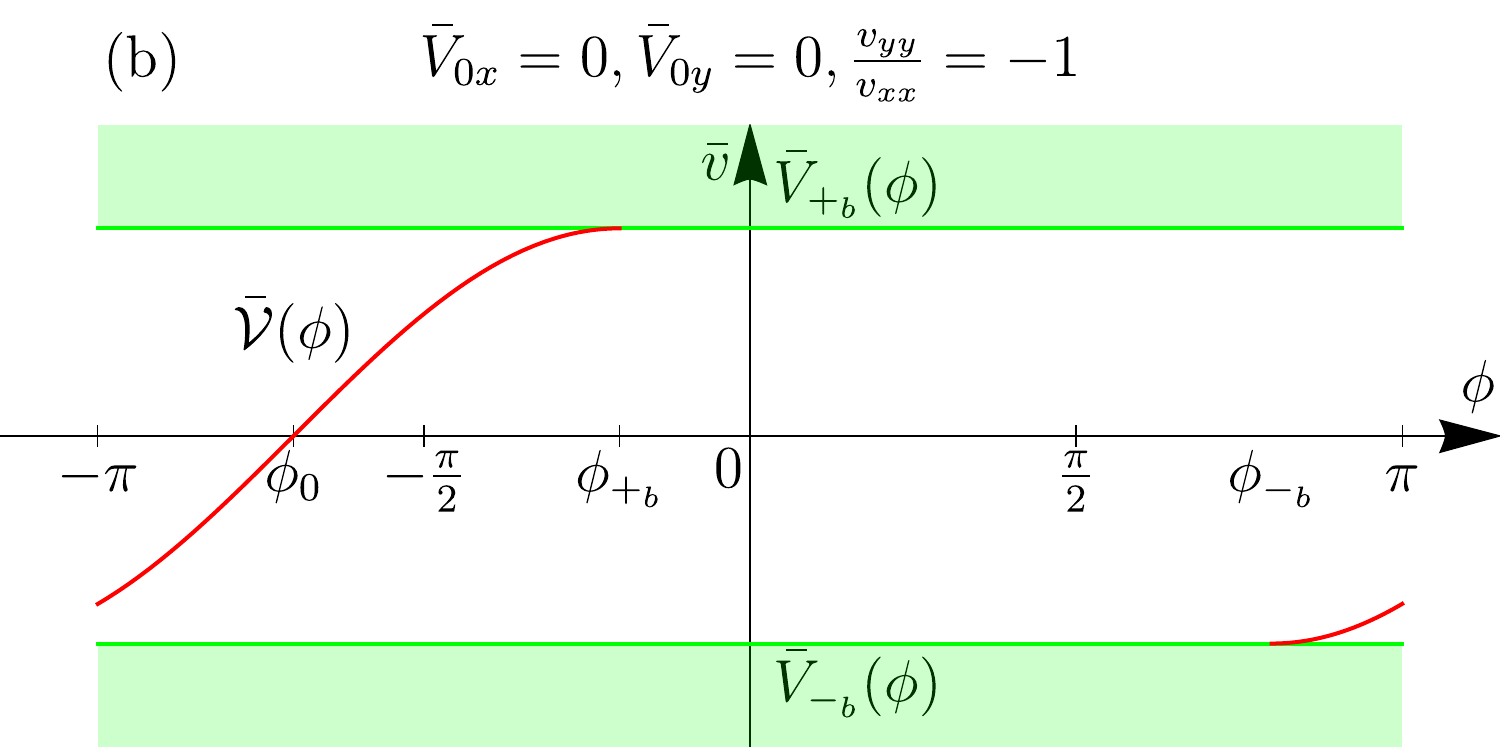}
\caption{
The dimensionless velocity $\Vcr(\phi)$ [\eq{Vcr}] of the surface-state spectrum $\Ec_3(\pb_\p)$ [\eq{Ec3}]
for absent additive term $\Vr_0(\phi)\equiv0$ [\eq{Vr0}], $\Vr_{0x}=\Vr_{0y}=0$, and absent $xy$ anisotropy, $\f{v_{yy}}{v_{xx}}=\pm1$,
shown in (a) and (b), respectively.
In this case, $\Vcr(\phi)$
%and the velocities $\Vr_{\pm_b}(\phi)$ [\eq{Vr}] of the bulk boundaries (independent of $\phi$)
is identical (up to an overall magnitude) to the dependence
$\Ec(\pm\phi-\nu)$ [\eqs{de=+phi}{de=-phi}] of the 1D bound-state energy on the angle $\de$ [\eq{d}] of the gap terms.
%See caption to \figr{Vcr} [and text] for more details.
The green regions depict the continuum of the bulk states.
The value $\nu=0.2\pi$ of the BC phase-shift angle was used in all \figssr{Vcrbase}{Vcr}{pp}.
The sign $v_{yy}\gtrless 0$ determines the chirality of $\Vcr(\phi)$, as well as the Chern number of the Weyl point,
which are related via the bulk-boundary correspondence \eqn{bbcW}. %that we formulate.
}
\lbl{fig:Vcrbase}
\end{figure}

\begin{figure}
\includegraphics[width=.40\textwidth]{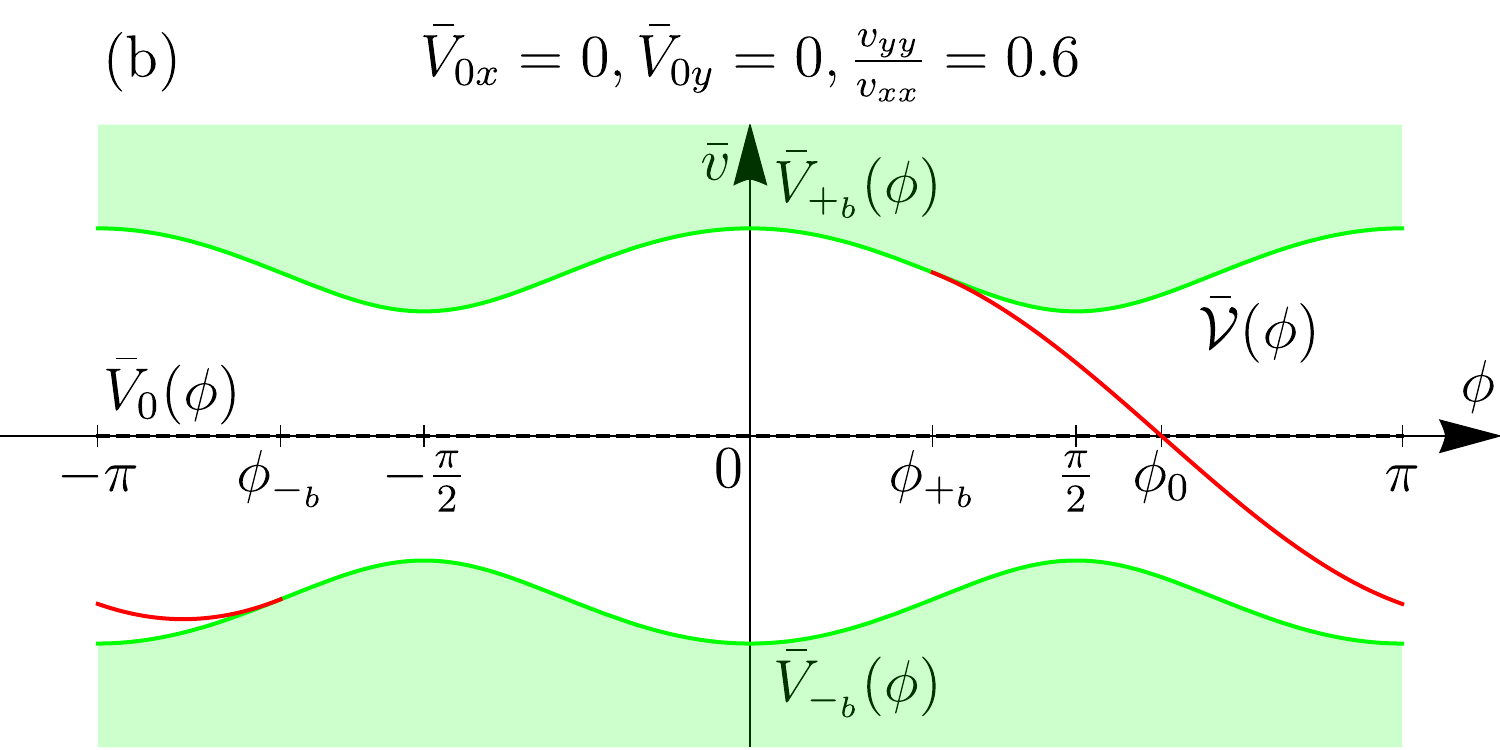}\\
\includegraphics[width=.40\textwidth]{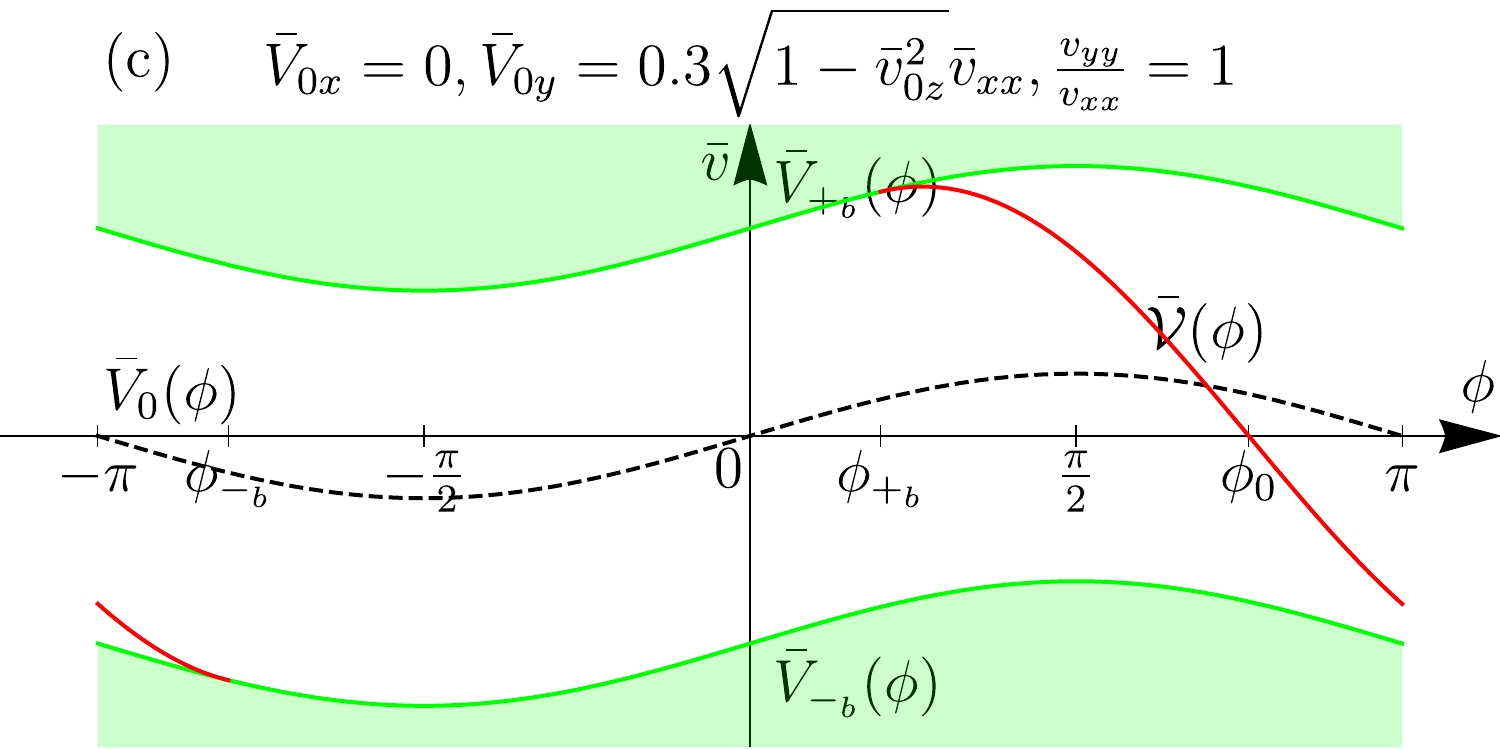}\\
\includegraphics[width=.40\textwidth]{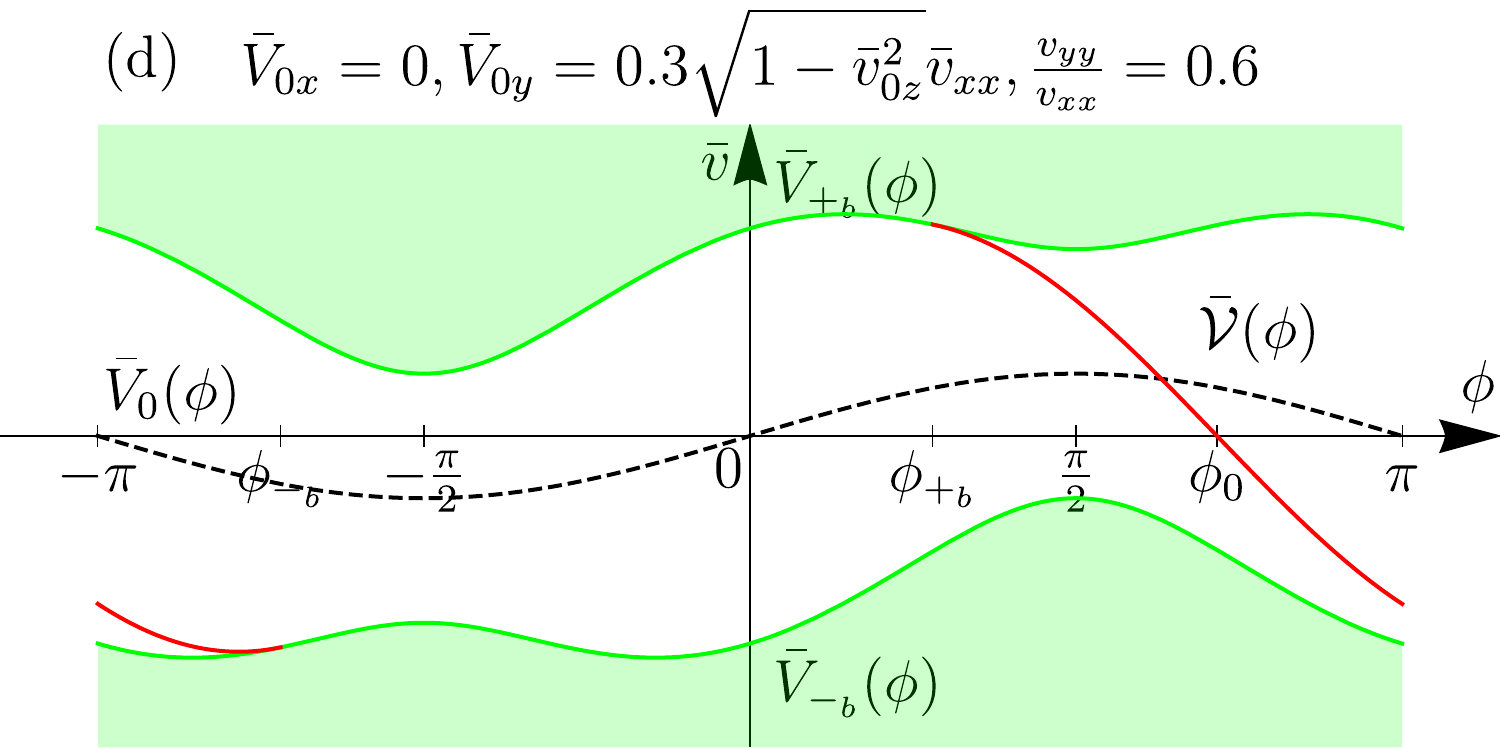}\\
\includegraphics[width=.40\textwidth]{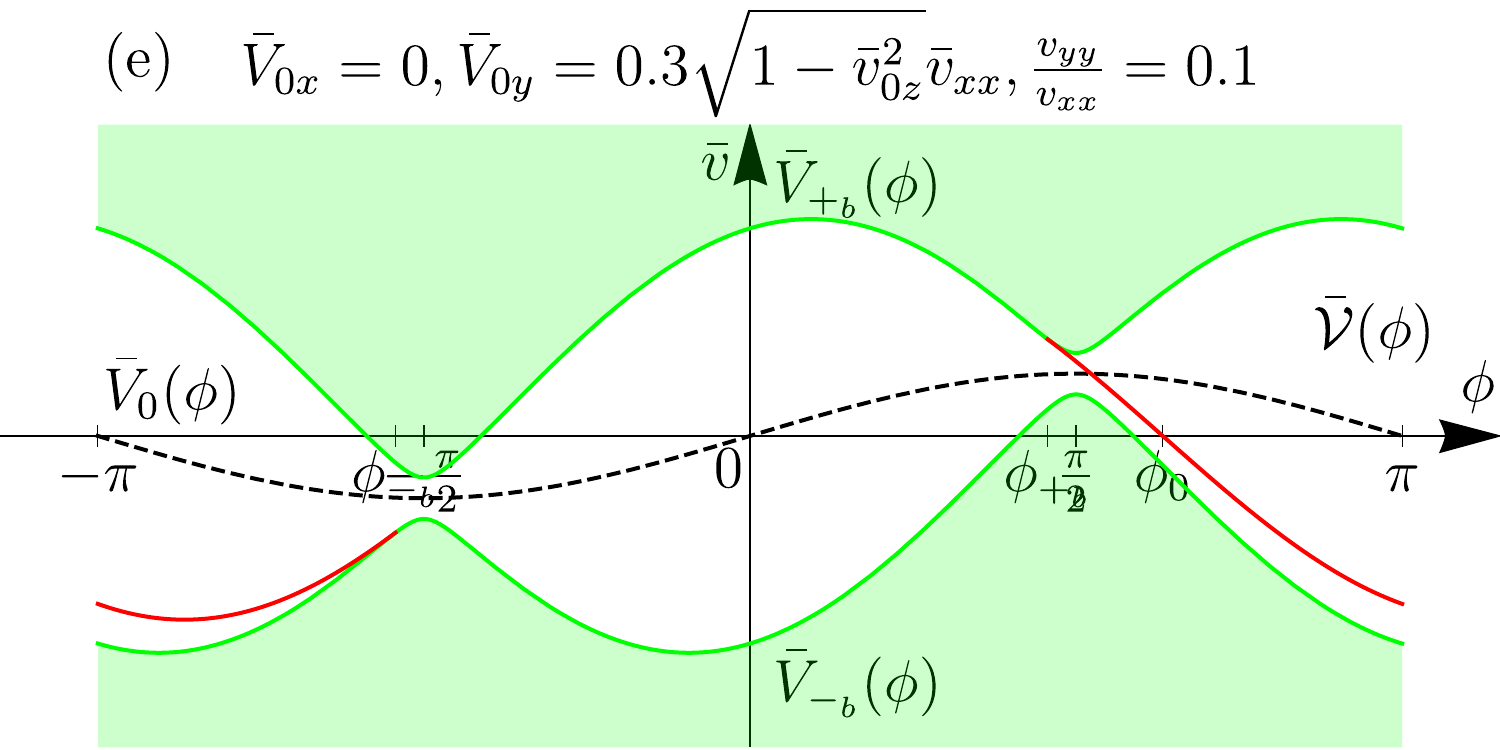}\\
\includegraphics[width=.40\textwidth]{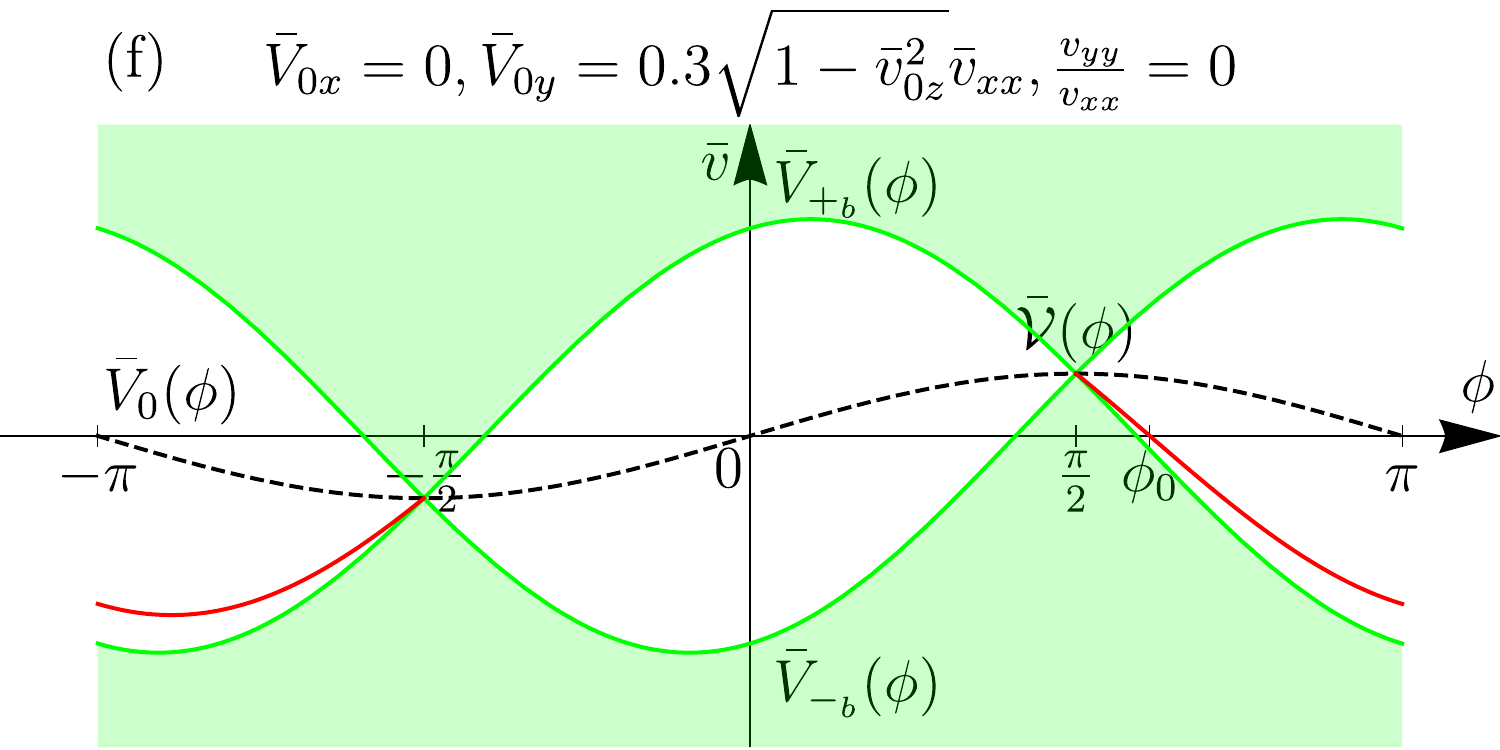}\\
\caption{
The dimensionless velocity $\Vcr(\phi)$ [\eq{Vcr}] of the surface-state band \eqn{Ec3}
as a function of the polar angle $\phi$ of the surface momentum $\pb_\p$ [\eq{pp}],
for various values of the velocity parameters, showing the effects of the additive term $\Vr_0(\phi)$ [\eq{Vr0}]
and $xy$ anisotropy $\f{|v_{yy}|}{v_{xx}}\neq 1$.
(b) Shows separately the effect of the $xy$ anisotropy $\f{v_{yy}}{v_{xx}}=0.6$ for absent additive term, $\Vr_{0x,0y}=0$.
(c) Shows separately the effect of the additive term for $\Vr_{0x}=0$, $\Vr_{0y}=0.3\sq{1-\vr_{0z}^2}\vr_{xx}$
and absent $xy$ anisotropy.
(d) Shows the combined effect of both the $xy$ anisotropy of (b) and the additive term of (c) present.
(e) Shows even larger anisotropy $\f{v_{yy}}{v_{xx}}=0.1$. Due to the present additive term with $\Vr_{0y}\neq 0$,
upon decreasing $\f{v_{yy}}{v_{xx}}$, the system eventually turns from type-I (as in all of the above cases)
to type-II (as here) Weyl semimetal. For small $\f{|v_{yy}|}{v_{xx}}\ll 1$, the Weyl semimetal is also strongly $xy$ anisotropic.
(f) The limiting case of the line-node semimetal with $\f{v_{yy}}{v_{xx}}=0$.
The ``gap'' in the velocity $\Vcr(\phi)$ closes at $\phi=\pm\f{\pi}2$.
The topological chirality properties of $\Vcr(\phi)$ remain the same in (b),(c),(d),(e),
in accord with the Chern number of the Weyl point via the bulk-boundary correspondence \eqn{bbcW} formulated in \secr{topoWgen}.
%The labeling denotes a modification/continuation of \figr{Vcrbase}(a), showing various added effects.
}
\lbl{fig:Vcr}
\end{figure}

\subsection{Main properties of the surface-state spectrum \lbl{sec:mainW}}

\begin{figure}
\includegraphics[width=.35\textwidth]{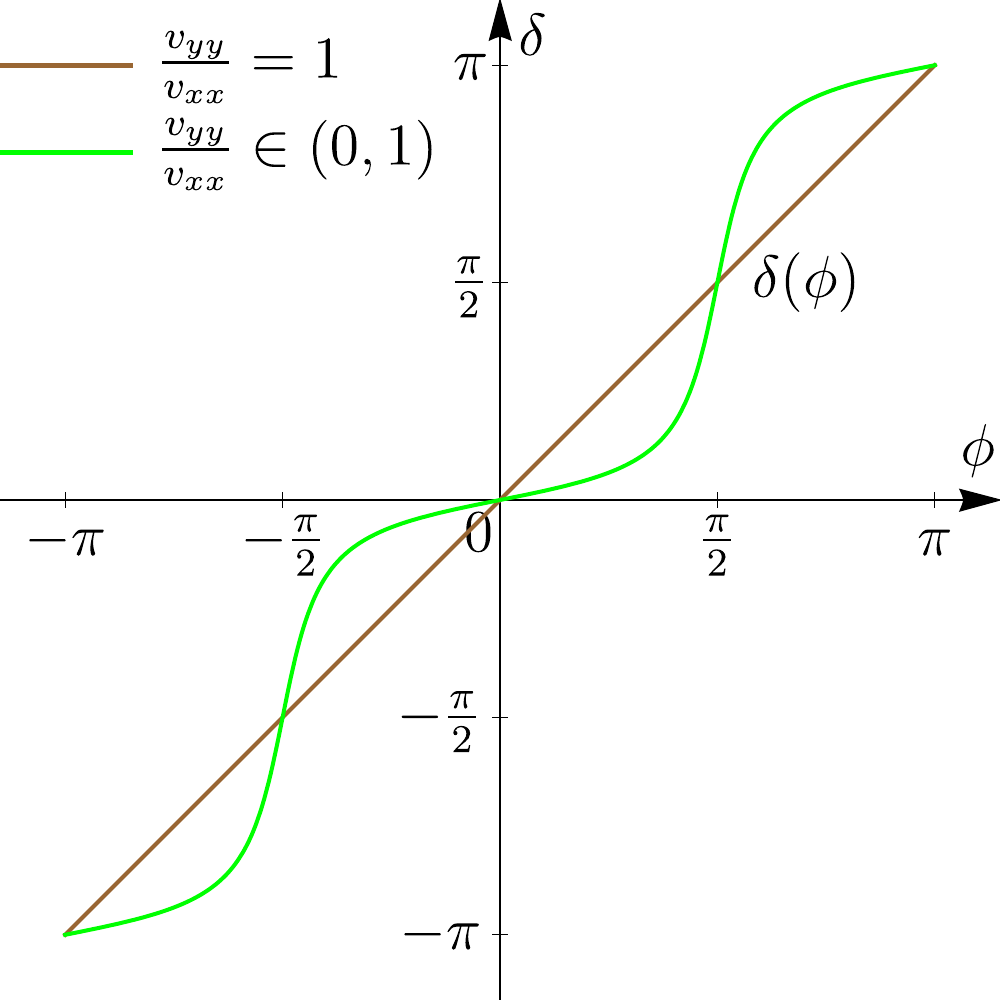}
\includegraphics[width=.35\textwidth]{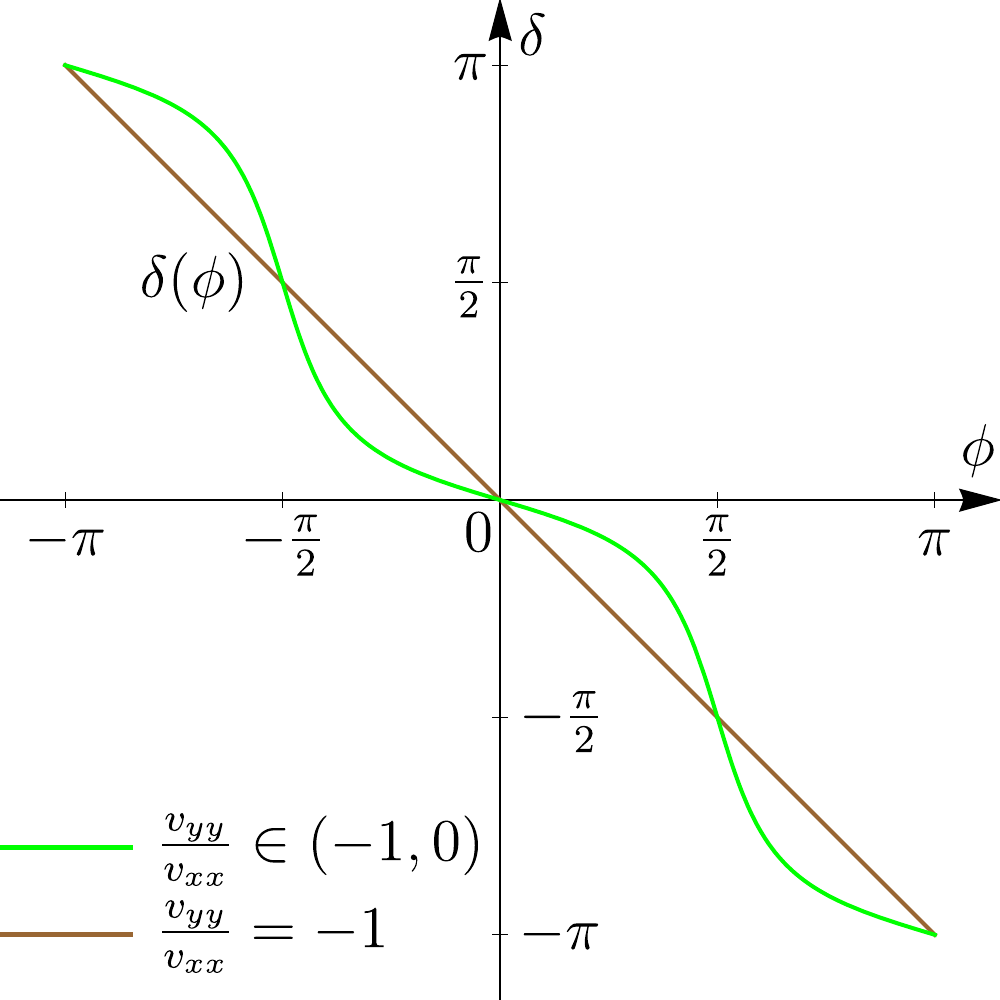}
\caption{The dependence $\de(\phi)$ [\eq{dephi}] of the angle of the gap terms [\eq{d}] on the polar angle $\phi$ of the surface momentum [\eq{pp}].
In our convention, the relative winding direction is determined by the sign of the velocity $v_{yy}$.
% (whereas $v_{zz}>0$ and $v_{xx}>0$ by construction/convention).
%This winding ultimately determines the sign of the Chern number of the Weyl points.
The values $\f{v_{yy}}{v_{xx}}=\pm0.3$ were used for the green curves.}
\lbl{fig:dephi}
\end{figure}

In the 1D case (\secr{1D}), $d_{0,x,y,z}$ are independent energy parameters.
The dependence of the bound-state energy $\Ec(\de-\nu)$ [\eq{Ec}] on the polar angle $\de$ of the gap parameters $(d_x,d_y)$ [\eq{d}]
(for a fixed BC angle $\nu$)
is of the main interest there, while the dependence on $d_{0,z,\p}$ is trivial and just determines the energy shift and gap size [\eq{Ec}].
In the 3D case, the dependence of the dimensionless surface-state velocity $\Vcr(\phi)$ [\eq{Vcr}]
on the polar angle $\phi$ of the surface momentum $\pb_\p$ [\eq{pp}] is of the main interest;
the dimensionless energy terms $\dr_{0,x,y,z}(\phi)$ [\eqss{dr0zphi}{drxyphi}{drpphi}] are no longer independent parameters,
but are all functions of $\phi$.
There are two main new effects in the dependence of $\Vcr(\phi)$ on $\phi$ versus the dependence of $\Ec(\de-\nu)$ on $\de$ that should be tracked.

The first new effect concerns the combination
\beq
	\Vr_0(\phi)=\dr_0(\phi)-\vr_{0z}\dr_z(\phi)=\Vr_{0x}\cos\phi+\Vr_{0y}\sin\phi,
\lbl{eq:Vr0}
\eeq
%{\bf[see if to intro above, notation used in plots already]}
which is present as an additive term in both the surface-state spectrum $\Vcr(\phi)$ [\eq{Vcr}] and bulk-band boundaries $\Vr_{\pm_b}(\phi)$ [\eq{Vr}]:
while it was regarded as a constant in the 1D bound-state solution, it now also has $\phi$-dependence.
Also note that this part is linear in both $p_x$ and $p_y$,
and hence gives a plane contribution to the 3D plots of $\Ec_3(\pb_\p)$ [\eq{Ec3}] and $E_{3,\pm_b}(\pb_\p)$ [\eq{E3}].
It also does not depend on the BC angle $\nu$ [which certainly cannot enter the bulk-band boundaries $E_{3,\pm_b}(\pb_\p)$].

The second new effect concerns the ``gap'' part $(\dr_x(\phi),\dr_y(\phi))$.
There are two further effects here, controlled by $v_{yy}$ relative to $v_{xx}$,
that affect both $\dr_\p(\phi)$ and $\de(\phi)$ dependencies on $\phi$:
(i) the sign of $v_{yy}$ and (ii) the anisotropy, when $v_{xx}\neq|v_{yy}|$ differ (we remind that $v_{xx}>0$ by construction).

The sign of $v_{yy}$ determines the relative winding direction of $\de(\phi)$ and $\phi$.
For $v_{yy}>0$, the winding direction is the same, while for $v_{yy}<0$ it is opposite.
The sign of $v_{yy}$ in our basis convention directly determines the sign of the Chern number of the Weyl node, as we discuss in \secr{topoW}.

If the anisotropy is absent, $\de(\phi)$ is equal to $\phi$ or opposite (\figr{dephi}):
\beq
	\de(\phi)=+\phi, \spc v_{yy}=+v_{xx},
\lbl{eq:de=+phi}
\eeq
\beq
	\de(\phi)=-\phi, \spc v_{yy}=-v_{xx}.
\lbl{eq:de=-phi}
\eeq
If the anisotropy $|v_{yy}|\neq v_{xx}$ is present, the dependence $\de(\phi)$ deviates from the linear ones above.
Also, in this case, $d_\p(\phi)$ itself acquires dependence \eqn{drpphi} on $\phi$.
This, along with the additive term \eqn{Vr0} above, is another source of the $\phi$-dependence of the bulk-band boundaries $\Vr_{\pm_b}(\phi)$.

\subsection{The range and merging contours of the surface-state band\lbl{sec:Ec3range}}

\begin{figure}
\includegraphics[width=.30\textwidth]{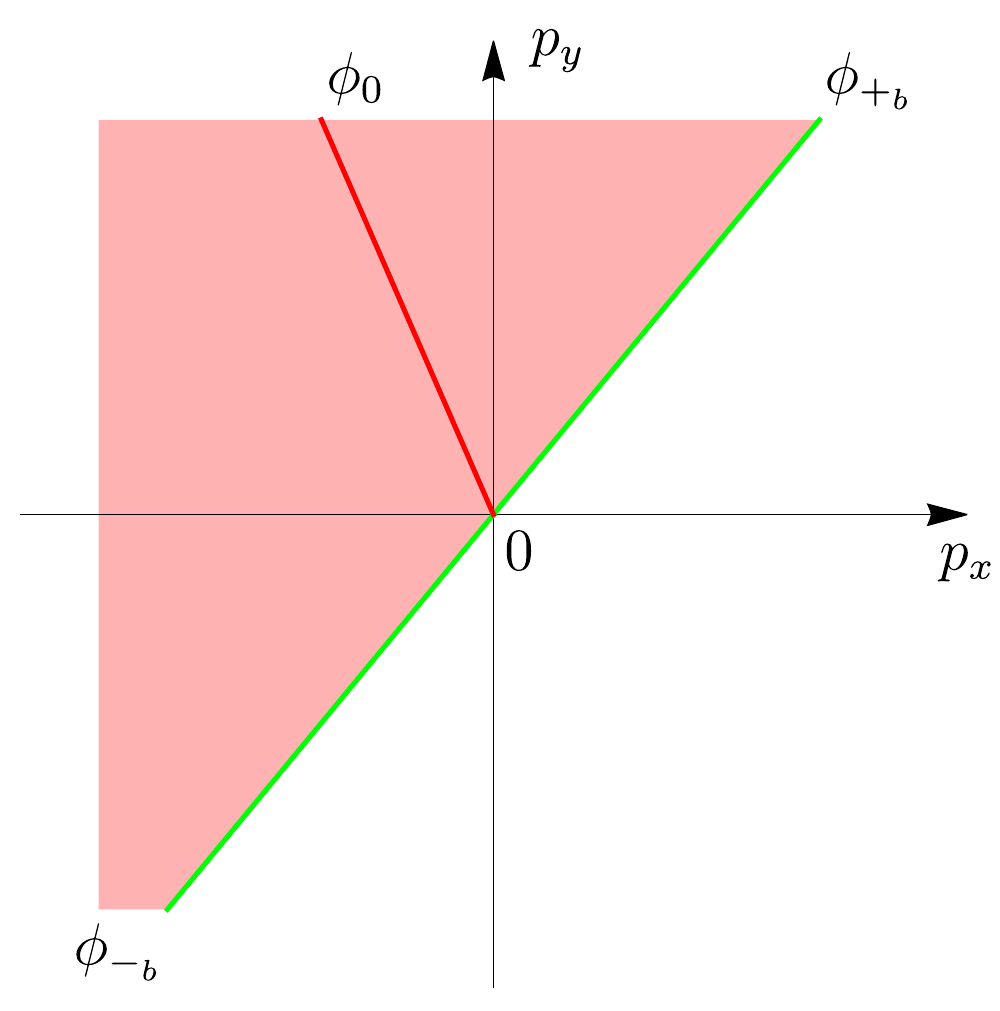}
\caption{The plane of the surface momentum $\pb_\p=(p_x,p_y)$ [\eq{pp}].
The region where the surface-state band $\Ec_3(\pb_\p)$ [\eq{Ec}] is present is shown in red.
It is bounded by the merging half-lines (green) of the surface-state band
%spectrum $\Ec_3(\pb_\p)$ [\eq{Ec}]
with the boundaries $E_{3,\pm_b}(\pb_\p)$ [\eq{E3}] of the upper $+_b$ and lower $-_b$ bulk bands.
The angle between the merging half-lines is always $\pi$ [\eq{phim}]; therefore, they always form one line.
The red half-line is the Fermi contour, determined by $\Ec_3(\pb_\p)=0$,
where the surface-state band crosses the zero energy level of the Weyl point.
Generally, the angle between the Fermi-contour half-line and the merging half-lines differs from $\f{\pi}2$.
The same parameters as in \figr{Vcr}(b) were used: $\nu=0.2\pi$, $\Vr_{0x}=0$, $\Vr_{0y}=0$
(absent additive term), $\f{v_{yy}}{v_{xx}}=0.6$ (present $xy$ anisotropy).
}
\lbl{fig:pp}
\end{figure}

Let us now find the merging angles $\phi_{\pm_b}$ of the surface-state velocity $\Vcr(\phi)$
with the velocities $\Vr_{\pm_b}(\phi)$ of the bulk-band boundaries, at which
\beq
    \Vcr(\phi_{\pm_b})=\Vr_{\pm_b}(\phi_{\pm_b}).
\lbl{eq:phimdef}
\eeq
In the surface-momentum $\pb_\p$ plane, the corresponding merging contours of $\Ec_3(\pb_\p)$ with $E_{3,\pm_b}(\pb_\p)$
are straight half-lines at these fixed angles $\phi_{\pm_b}$ with $p_\p>0$ spanning all positive values.

The merging angles $\phi_{\pm_b}$ are determined by the same values \eqn{demerge} in terms of $\de$,
found from the 1D bound-state solution in \secr{bs}, which now only need to be expressed in terms of $\phi$ via the dependence $\de(\phi)$:
\beq
	\de(\phi_{+_b})=\de_{+_b}=\nu, \spc \de(\phi_{-_b})=\de_{-_b}=\nu+\pi.
\lbl{eq:phim}
\eeq
The merging angles $\phi_{\pm_b}$ can be determined modulo $\pi$ from \eq{tandephi} as
\beq
	\tan\phi_{\pm_b}=\f{v_{xx}}{v_{yy}}\tan\nu.
\lbl{eq:phimtan}
\eeq
The merging angles $\phi_{\pm_b}$ therefore depend only on the velocity ratio $v_{yy}/v_{xx}$ and do not depend on any other velocity parameters.
It follows from \eqn{dephi} that for any two values $\phi$ and $\phi+\pi$ of the argument separated by $\pi$,
the values of the function are also separated by $\pi$:
\[
	\de(\phi+\pi)-\de(\phi)=\pi.
\]
And since this holds for the merging points $\de_{\pm_b}$ in $\de$, the same holds for the merging points in $\phi$:
\[
	\phi_{-_b}=\phi_{+_b}+\pi,
\]
as confirmed explicitly by \eq{phimtan}.
Importantly, in terms of the dependence on $\pb_\p$, this means that the two half-lines,
at which the surface-state band $\Ec_3(\pb_\p)$ merges with the lower $E_{3,-_b}(\pb_\p)$ and upper $E_{3,+_b}(\pb_\p)$
bulk-band boundaries, actually form {\em one} straight line.
Therefore the sector in $\pb_\p$, where the surface-state band is present, is always a half-plane,
and the 3D plot of the surface-state band is always a half-plane.
This result has previously been obtained in Ref.~\ocite{Hashimoto2016,Hashimoto2019}
for the most general form of the BC and specific forms of the Hamiltonian.
Now we have proven that this is a generic property, valid asymptotically in the vicinity of any isolated Weyl node.

The sign of $v_{yy}$ determines which half-plane is occupied by the surface-state band relative to the merging lines in $\pb_\p$ (points in $\phi$).
For $v_{yy}>0$ [\figr{Vcrbase}(a), \figr{Vcr}], when the winding directions of $\de(\phi)$ and $\phi$ are the same (see \secr{mainW}, \figr{dephi}),
the surface-state velocity $\Vcr(\phi)$ decreases as $\phi$ increases (\figr{Vcr}),
starting from the upper bulk band and ending at the lower bulk band [like $\Ec(\de-\nu)$ vs $\de$], occupying the sector
\beq	
	\Vr_{+_b}(\phi)>\Vcr(\phi)>\Vr_{-_b}(\phi),\spc
	\phi\in(\phi_{+_b},\phi_{-_b}), \spc v_{yy}>0.
\lbl{eq:phirange+}
\eeq
For $v_{yy}<0$ [\figr{Vcrbase}(b)], when the winding directions of $\de(\phi)$ and $\phi$ are opposite,
the surface-state band $\Vcr(\phi)$ increases as $\phi$ increases,
starting from the lower bulk band and ending at the upper bulk band [opposite to $\Ec(\de-\nu)$ vs $\de$], occupying the sector
\beq
	\Vr_{-_b}(\phi)<\Vcr(\phi)<\Vr_{+_b}(\phi),\spc
	\phi\in(\phi_{-_b},\phi_{+_b}), \spc v_{yy}<0.
\lbl{eq:phirange-}
\eeq

We will show in \secr{topoW} that the winding properties of the surface-state velocity $\Vcr(\phi)$
are directly related to the Chern number of the Weyl node, via the bulk-boundary correspondence that we formulate.

\subsection{Fermi contour}

A commonly used characteristic of the surface-state spectrum of the Weyl semimetal is the ``Fermi contour'',
which is an equal-energy contour in the surface-momentum $\pb_\p$ plane
along which the surface-state band crosses the energy level $\e=0$ of the Weyl point (named so assuming that this is the Fermi level).
For the considered model, the Fermi contour is determined from the equation
\[
	\Ec_3(\pb_\p)=0.
\]
Because of the linear scaling, it is, of course, a straight half-line in $\pb_\p$,  specified by the angle $\phi_0$ found as [\eq{Vcxy}]:
\beq
    \Vcr(\phi_0)=\Vcr_x \cos\phi_0+\Vcr_y\sin\phi_0=0 \Rarr
    \tan\phi_0=-\f{\Vcr_x}{\Vcr_y}=-\f{\Vr_{0x}+\sq{1-\vr_{0z}^2}\vr_{xx}\cos\nu}{\Vr_{0y}+\sq{1-\vr_{0z}^2}\vr_{yy}\sin\nu}.
\lbl{eq:phi0}
\eeq
The solution must be chosen within the half-plane sector occupied by the surface-state band [\eqs{phirange+}{phirange-}].
Unlike the merging angles $\phi_{\pm_b}$ [\eq{phimdef}],
the Fermi-contour angle $\phi_0$ does involve the velocity parameters of the $\dr_{0,z,\p}(\phi)$ functions.

We demonstrated above that the angle between the merging half-lines is always $\pi$, i.e., they form one straight line.
Comparing \eqs{phi0}{phimtan}, we see that when the velocity parameters are such that the additive term \eqn{Vr0} is absent
and there is no $xy$ anisotropy ($v_{yy}=v_{xx}$), the Fermi-contour half-line forms a $\f{\pi}2$ angle with the merging line.
However, this angle deviates from $\f{\pi}2$ when these conditions are not satisfied. An example is shown in \figr{pp}.

\section{Generalized topology and bulk-boundary correspondence of a Weyl node \lbl{sec:topoW}}

\subsection{Bulk topology of a Weyl semimetal}

\begin{figure}
\includegraphics[width=.30\textwidth]{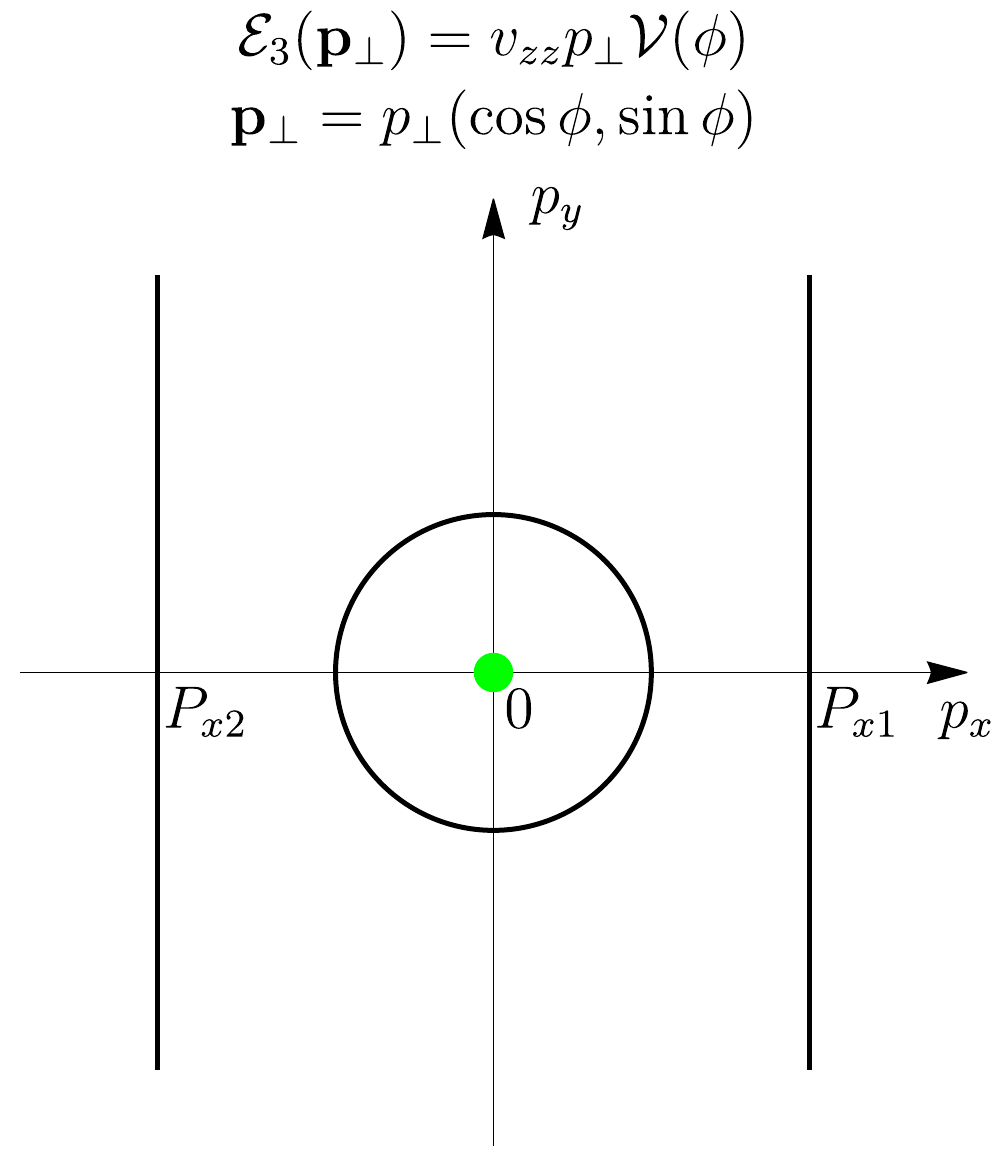}
\caption{
%{\bf[too much text?-now ok?]}
Generalized topology and bulk-boundary correspondence of a Weyl semimetal, as formulated in \secr{topoWgen}.
An arbitrary path $\ga$ in the 2D surface momentum $\pb_\p=(p_x,p_y)$ plane [\eq{pp}]
defines an effective 2D QAH system %with the Hamiltonian $\Hh(\pb_\p^\ga(t),\ph_z)$
on a generalized cylinder $\Sig^\ga$ in the 3D momentum space,
obtained by passing straight $p_z$ lines (perpendicular to the surface) through this path.
%The Chern number $C(\Sig^\ga)$ [\eqs{Cdef}{CSig}] of such system is given by the flux through this generalized cylinder.
Various such paths as shown.
This leads to the generalized bulk-boundary correspondence \eqn{bbcga}.
%where the chirality of the surface-state spectrum along such path is given by the Chern number of the respective cylinder $\Sig^\ga$.
%in the 3D momentum $\pb=(p_x,p_y,p_z)$ space.
As a special case of it,
we formulate the variant \eqn{bbcW} of bulk-boundary correspondence
most suitable for characterizing the vicinity of the projected Weyl point:
%the chirality of the surface-state spectrum along a path enclosing the projected Weyl point $\pb_\p=(0,0)$
%is given by the the Chern number of the Weyl point.
%The dimensionless velocity $\Vcr(\phi)$ [\eq{Vcr}] as a function of the polar angle $\phi$
%realizes precisely such a case, as it is the surface-state band [\eq{Er3def}] on a circle of any radius $p_\p$.
the chirality $N[\Vcr(\phi)]$ of the surface-state velocity \eqn{Vcr} as a function of the polar angle $\phi$ of the surface momentum
is equal to the Chern number of the Weyl point.
Both are determined by the sign of $v_{yy}$ in our basis convention
and \figsr{Vcrbase}{Vcr} fully confirm this statement.
%with the Chern number of the Weyl point via \eq{bbcW},
%both determined by the sign of $v_{yy}$ in our basis convention.
}
\lbl{fig:pptopo}
\end{figure}

The topological properties of 3D Weyl semimetal are directly tied to those of 2D QAH systems (Chern insulators)~\cite{Armitage2017}.
The Chern number $C(\Sig)$ can be defined for any surface $\Sig$ in the 3D momentum space
that is closed or extends to infinity and does not pass through any Weyl points.
It is given by the flux
\beq
	C(\Sig)=\f1{4\pi}\int_\Sig \dx^2 \sigb_\pb \, \Bb(\pb)
%\sin\tht_s(\pb(\ab))\pd_{a_1} \tht_s(\pb) \pd_{a_2} \phi_s(\pb())
%	=\int \dx (-?)\cos\tht_s \dx \phi_s
\lbl{eq:Cdef}
\eeq
($\sigb_\pb$ is the normal area vector at point $\pb\in\Sig$ on the surface)
of the Berry curvature through that surface.
The Berry curvature
\beq
	\Bb(\pb)=[\n_\pb\tm\Ab(\pb)]
\lbl{eq:B}
\eeq
%\[
%	[\n_\pb\tm\Ab]_z=\n_{p_x}A_y-\n_{p_y}A_x
%\]
is a wave-function-basis-independent vector field in momentum space, given by the rotor of
the wave-function-basis-dependent vector field, the Berry connection $\Ab(\pb)$.
For the considered model,
\beq
	\Ab(\pb)
%=\ix\lan \s(\pb)|\n_\pb|\s(\pb)\ran=\ix\psih^\dg(\pb) \n_\pb\psih(\pb)
	=\ix\wh^\dg(\s(\pb)) \n_\pb\wh(\s(\pb))
,\spc
	\n_\pb=(\pd_{p_x},\pd_{p_y},\pd_{p_z}),
\lbl{eq:A}
\eeq
derives from the bulk plane-wave eigenstates $\wh(\s(\pb))$ [\eq{w}] of the lower (occupied) band $\eps_{-_b}(\pb)$ [\eq{e3}],
with the pseudospin
\[
	\s(\pb)=-\f{\hb(\pb)}{|\hb(\pb)|},
\]
where $\hb(\pb)$ is the effective ``Zeeman field'' of the Hamiltonian \eqn{H3} in the same representation as \eq{HZeeman}.
The Berry curvature \eqn{B} is expressed in terms of the pseudospin $\s(\pb)$ as [\eq{s}]
%{\bf[figure the sign and coeff -could have $1/2$, r?]}
\[
	B_\al(\pb)=e_{\al\be\ga}\sin\tht_s(\pb) \pd_{p_\be}\tht_s(\pb)\pd_{p_\ga}\phi_s(\pb), \spc \al,\be,\ga=x,y,z.
\]
Geometrically, the Chern number
\beq
	C(\Sig)
%=\f1{4\pi}\int \dx p_z \dx \al \, \sin\tht_s(p_z,\al)\pd_{p_z} \tht_s \pd_{\al} \phi_s
	=\f1{4\pi}\int_\Sig \sin\tht_s \dx\tht_s \dx\phi_s
\lbl{eq:CSig}
\eeq
of a surface $\Sig$ is a skyrmion winding number of the mapping
\[
	\s(\pb):\spc \pb\in\Sig \rarr S^2
\]
of the surface $\Sig$ onto the pseudospin Bloch sphere $S^2$, realized by pseudospin.

Since the Berry curvature $\Bb(\pb)$ is a rotorless vector field, $[\n_\pb\tm\Bb(\pb)]=\nv$,
the Chern numbers of different surfaces obey conservations laws, which follow from the Gauss theorem.

Weyl points are singularities of the Berry curvature $\Bb(\pb)$.
The Chern number
\beq
	C^W=C(\Sig_\nv)
\lbl{eq:CWdef}
\eeq
of a Weyl node, say, at $\pb=\nv$ is defined as the Chern number over a surface $\Sig_\nv$ enclosing only that point,
e.g., a sphere of sufficiently small radius centered at the Weyl point.
Weyl points can be seen as sources and drains of the Berry curvature field, depending on the signs of their Chern numbers.

In the model \eqn{H3},
the Chern number $C^W$ of the (only) Weyl point $\pb=\nv$ is determined by the signs of the diagonal velocities $v_{xx}$, $v_{yy}$, and $v_{zz}$.
[The handedness of the coordinate (and hence, momentum) basis also affects the sign of the Chern number, as per \eq{Cdef};
we have taken care to preserve its right-handedness during the basis changes of \secr{H3}.]
In our basis convention, $v_{xx}>0$ and $v_{zz}>0$ by construction, and therefore,
the sign of $C^W$ is determined by the sign of the velocity $v_{yy}$.
We obtain
\beq
	C^W=\mp 1, \spc v_{yy}\gtrless 0.
\lbl{eq:CW}
\eeq

\subsection{Generalized bulk-boundary correspondence of a Weyl node \lbl{sec:topoWgen}}

Bulk topology has consequences for the bound/edge/surface states,
known as the concept of bulk-boundary correspondence~\cite{Chiu}.
Here, we formulate the bulk-boundary correspondence for Weyl semimetals that is more general than commonly considered.
As its special case, we formulate the bulk-boundary correspondence that directly
relates the dimensionless velocity $\Vcr(\phi)$ [\eq{Vcr}] of the surface-state spectrum \eqn{Ec3}
to the Chern number $C^W$ [\eqs{CWdef}{CW}] of the Weyl point.

This more general formulation arises when one considers effective 2D QAH systems in the generalized sense explained in \secr{topo},
when the additional parameter of the effective 1D system (besides $p_z$) is not necessarily a cartesian momentum component.
Consider an arbitrary path $\ga$: $\pb_\p^\ga(t)$, parameterized by $t$, in the surface-momentum plane
that is either closed or terminates with both ends at infinity $|\pb_\p|=+\iy$, and does not cross any projected Weyl points (\figr{pptopo}).
(For paths terminating at infinity, there is an important subtlety, particularly relevant for the linear-in-momentum Hamiltonian,
which we discuss in \secr{subtlety}.)
Each such path defines an effective 2D QAH system with the Hamiltonian $\Hh(\pb_\p^\ga(t),\ph_z)$
on a {\em generalized cylinder} $\Sig^\ga$ in the 3D momentum space,
obtained by passing straight $p_z$ lines (perpendicular to the surface) through this path.
The Chern number $C(\Sig^\ga)$ [\eqs{Cdef}{CSig}] of such system is given by the flux through this generalized cylinder.

The surface-state spectrum $\Ec_3(\pb_\p^\ga(t))$ [\eq{Ec3}] along this path becomes the ``edge-state'' spectrum as a function of $t$
of this effective 2D QAH system defined on the space $(p_z,t)$.
According to \secr{topo}, the bulk-boundary correspondence for this system means that the signed number
\beq
	N(\Sig^\ga)=N[\Ec_3(\pb_\p^\ga(t))] %=\x{sgn}(t_{+_b}-t_{-_b})
\lbl{eq:Nga}
\eeq
of the effective chiral edge-state bands is given by the Chern number $C(\Sig^\ga)$ of the cylinder $\Sig^\ga$:
\beq
	N(\Sig^\ga)=C(\Sig^\ga),
\lbl{eq:bbcga}
\eeq
with all the definitions introduced in \secr{topo}.

The bulk-boundary correspondence of Weyl semimetals is commonly expressed
in terms of ``conventional'' 2D QAH systems defined on infinite 2D planes in 3D momentum space.
This now becomes a special case of the above more general formulation, when the paths $\pb_\p^\ga(t)$ are straight lines,
and the generalized cylinders on which the QAH systems are defined are such planes.
For example, for a plane with the fixed $p_x=P_x=\x{const}$, to be denoted as $\Sig_{P_x}$,
the parameter $t=p_y$ is the momentum component itself.
This way, the Hamiltonian $\Hh_3(P_x,p_y,p_z)$ [\eq{H3}] is seen as that of a family of 2D QAH systems defined
on the momentum plane $\Sig_{P_x}$ with coordinates $(p_y,p_z)$ and parameterized by $P_x$.

Spanning the planes $\Sig_{P_x}$ with different $P_x$, the regions of $P_x$ between the Weyl points can be seen as topological phases,
within which the Chern number $C(\Sig_{P_x})$ is conserved, and the crossing of a Weyl point by the plane can be seen as a topological phase transition.
According to the Gauss theorem,
the difference in the Chern numbers of the two planes is equal to the Chern number of the Weyl point(s) enclosed between the planes, e.g.,
\beq
	C^W=C(\Sig_{P_{x1}})-C(\Sig_{P_{{x2}}})
\lbl{eq:dCPx}
\eeq
for $P_{x1}>0$ and $P_{x2}<0$ on the opposite sides of the node (\figr{pptopo}).

The surface-state spectrum $\Ec_3(P_x,p_y)$ is interpreted in this picture as effective edge-state spectra (functions of $p_y$)
of a family a 2D QAH systems defined on the planes $\Sig_{P_x}$.
According to the bulk-boundary correspondence [\eqs{Nga}{bbcga}], the signed number
\beq
	N(\Sig_{P_x})=N[\Ec_3(P_x,p_y)]=C(\Sig_{P_x})
\lbl{eq:NPx}
\eeq
of the chiral edge-state bands is determined by the Chern number $C(\Sig_{P_x})$ of the plane $\Sig_{P_x}$ (again, see the subtlety in \secr{subtlety}).
The Chern number $C^W$ [\eqs{CWdef}{CW}] of the Weyl point determines via \eqs{dCPx}{NPx}
the corresponding {\em changes} in the chiralities of the effective edge-state spectra upon crossing the Weyl point by the plane:
\[
	N(\Sig_{P_{x1}})-N(\Sig_{P_{x2}})=C^W.
\]

However, the above more general formulation \eqn{bbcga}
allows us to formulate, as its special case, a different type of bulk-boundary correspondence of Weyl semimetals
that is much better suited for the characterization of the vicinity of a Weyl node.

Consider now a closed path $\ga$ that encloses the projected Weyl point $\pb_\p=(0,0)$. The Chern number
\beq
	C(\Sig^\ga)=C^W
\lbl{eq:Cga=CW}
\eeq
of the 2D QAH system defined on the cylinder $\Sig^\ga$ generated by such path
is equal to the Chern number of the enclosed Weyl point [\eq{CWdef}],
since such cylinder and the sphere around the Weyl point can be continuously deformed into each other.
According to the bulk-boundary correspondence \eqn{bbcga},
the signed number $N(\Sig^\ga)$ [\eq{Nga}] of the chiral effective edge-state band $\Ec_3(\pb_\p^\ga(t))$ along the path $\ga$
is given directly by the Chern number of the Weyl point.

Hence, one path enclosing the projected Weyl point
already fully characterizes the topology of the Weyl point.
We notice that the dimensionless velocity $\Vcr(\phi)$ [\eq{Vcr}] of the surface-state spectrum \eqn{Ec3}
as a function of the surface-momentum polar angle $\phi$ [\eq{pp}] realizes precisely such paths:
circles of any radius $p_\p$ around the projected Weyl point, parameterized by the polar angle $t=\phi$ (\figr{pptopo}).
According to the generalized bulk-boundary correspondence \eqn{bbcga} and \eq{Cga=CW},
the chirality of dimensionless velocity $\Vcr(\phi)$, viewed as an effective ``edge-state'' band defined on a unit circle $\phi\in S^1$,
is given by the Chern number of the Weyl point:
\beq
	N[\Vcr(\phi)]=C^W.
\lbl{eq:bbcW}
\eeq
The chirality of $\Vcr(\phi)$ in \figr{Vcr} now becomes transparent.
%{\bf[probably emph but also later: This constitutes a variant/version of the bulk-boundary correspondence where ...]}

As already stated above, in our basis convention,
both the chirality $N[\Vcr(\phi)]$ of the surface-state velocity [\eqs{phirange+}{phirange-} in \secr{Ec3range}]
and the Chern number $C^W$ [\eq{CW}] of the Weyl point are controlled by the sign of the velocity $v_{yy}$.
From \eqs{phimdef}{phim} for the merging angles $\phi_{\pm_b}$, we obtain
\[
	N[\Vcr(\phi)]=\x{sgn}(\phi_{+_b}-\phi_{-_b})=\mp 1,\spc v_{yy}\gtrless 0,
\]
fully confirming \eq{bbcW}.

\subsection{Subtlety of topology and bulk-boundary correspondence in continuum models \lbl{sec:subtlety}}

For continuum models of 2D QAH systems (similar statements hold for other symmetry classes),
there is an important subtlety that, for topology to be well-defined,
the Hamiltonian has to be defined on a 2D manifold topologically equivalent to a sphere.
Only in that case the Chern number is guaranteed to be integer, being the skyrmion winding number of the mapping from a sphere onto a sphere.
To satisfy that, proper behavior of the Hamiltonian at momentum infinity must be satisfied.
For example, for the effective 2D QAH system introduced in \secr{topo}, defined on a cylinder,
with the Hamiltonian $\Hh(p_z,\al)$ and the parameter $\al\in S^1$ on a unit circle,
this means that the limits of the pseudospin $\s(p_z,\al)$ [\eq{stopo}] have to be well-defined at infinities $p_z\rarr \pm\iy$,
i.e., $\s(p_z=\pm\iy,\al)\equiv\s(p_z=\pm\iy)$ have to be independent of $\al$ (but not necessarily equal to each other).
This turns our particularly important for the linear-in-momentum Hamiltonians we consider.

We observe that the topological properties of the (effective) edge-state spectrum are in concert
with this requirement for the bulk topology:
the Chern number is a well-defined integer (as per above) if and only if the edge-state bands have merging points with the bulk bands
and hence their chirality numbers are also well-defined integers.
This is to be anticipated, given the bulk-boundary correspondence \eqn{bbc},
since the two sides of the relation can only simultaneously be well-defined or not.

The Chern number and the edge-state-band chirality number are well-defined
in the cases of the 1D Hamiltonian $\Hh(p_z,\de)$, with the polar angle $\de\in S^1$ of the gap terms,
and the 3D Hamiltonian $\Hh_3(p_\p\cos\phi,p_\p\sin\phi,p_z)$ at fixed $p_\p$ with the polar angle $\phi\in S^1$ of surface-momentum.
The Chern numbers over the cylinders are well-defined
since the pseudospin limits $\s(p_z=\pm\iy,\de)=(0,0,\mp1)$ are well-defined.
Accordingly, the effective edge-state bands $\Ec(\de-\nu)$ [\eq{Ec}] and $\Vcr(\phi)$ [\eq{Vcr}] have well-defined merging points.

One the other hand, the Chern number and the edge-state-band chirality number are not well-defined
in the cases of a 2D QAH system (\secr{2D}) defined on a plane
(and equivalent effective 2D QAH systems defined on planes in the 3D momentum space of a Weyl semimetal).
Due to the linear momentum dependence of the Hamiltonian,
these planes are not topologically equivalent to a sphere in terms of the Hamiltonian and pseudospin dependence,
since the pseudospin limit at momentum infinity depends on the direction and is therefore not well-defined.
Rather, the system is effectively defined on a half-sphere, which is not closed.
The pseudospin field $\s(p_x,p_z)$ forms a meron, rather than a skyrmion, configuration
and the Chern number is finite, but half-integer, rather than integer.
%-hm, also for anisotropic?
Accordingly, there are no merging points of the edge-state bands $\Ec_2(p_x,\De)$ [\eq{Ec2}, \figr{Ec2}] at large momenta $p_x$
and their chirality numbers are undefined.
At the same time, we note that the {\em changes} in non-integer Chern numbers of a 2D QAH system defined on a plane
and the corresponding changes in chirality of the edge-state bands remain integer and well-defined.
Hence, a sort of {\em ``relative''} bulk-boundary correspondence still holds,
relating the differences of the Chern numbers and chiralities, as in \eq{dCPx}.

\section{Type-I and type-II Weyl semimetals \lbl{sec:Wtype}}

The derived results for the most general model [\eqs{H3}{bc3}] of an isolated Weyl node include all possible situations, in particular,
the regimes of the so-called type-I and type-II Weyl semimetals.
In the velocities $\vr_{\pm_b}(\nb)$ [\eq{vr}] of the two linear bulk bands $\eps_{3,\pm_b}(\pb)$ [\eq{e3}],
there are opposite-sign contributions $\pm_b\sq{\ldots}$,
arising from the $\tauh_{x,y,z}$ terms in the Hamiltonian \eqn{H3}, and same-sign contributions, arising from the $\tauh_0$ terms;
the latter cause a ``tilt'' of $\eps_{3,\pm_b}(\pb)$.
If the velocities $\vr_{\pm_b}(\nb)$ [\eq{vr}] are of opposite signs for every direction $\nb$, one talks about a type-I Weyl semimetal.
If the tilt terms ($\tauh_0$) are so strong that for some range of $\nb$ the velocities $\vr_{\pm_b}(\nb)$ of both bands $\pm_b$
have the same sign, one talks about a type-II Weyl semimetal.

In the type-I case, the model of a single Weyl node with a boundary is well-defined for any boundary orientation.
In the type-II case, it matters for the surface-state problem how the directions with the same-sign velocities are aligned relative to the surface.
For the $p_z$ direction perpendicular to the surface, the velocities in our basis are $\pm v_\pm=v_{zz}\pm v_{0z}$ [\eq{vpm}]:
\[
	\eps_{3,\pm_b}(0,0,p_z)=v_{0z} p_z \pm_b v_{zz}|p_z|
	%= (v_{0z} \pm v_{zz})p_z
	=\pm v_{\pm} p_z.
\]
Same-sign velocities $\pm v_\pm$ for this direction would precisely correspond to the situation when it is impossible to introduce any BCs
(and hence, any boundary) in such a model with a {\em single node},
since the probability current would not vanish for a nonvanishing wave function, see \secsr{bc}{bc3}.
This means that for a type-II Weyl semimetal,
for those surface orientations for which the velocities for perpendicular motion are of the same sign for one node,
the minimal continuum model of a system with a boundary must include {\em two nodes},
and the velocities of the other node would both have the opposite sign.
Such model would have two right- and two left-movers, for which BCs do exist and the boundary is well-defined.

Therefore, for a type-II Weyl semimetal, a single-node model is applicable only for those surface orientations
for which the velocities for the motion perpendicular to the surface are of opposite signs.
For the motion along the surface, on the other hand,
it appears from the obtained solution (\secr{spectra}) that there are no restrictions on the signs of the velocities.

Consider the velocities $\Vr_{\pm_b}(\phi)$ [\eq{Vr}] of the bulk-band boundaries in more detail.
The $\pm_b$ part gives opposite-sign contributions, while the additive part \eqn{Vr0} gives the same-sign shift.
Starting from small (say, positive) values of $\Vr_{0x}$ and $\Vr_{0y}$ in the additive part relative to $v_{xx}$ and $v_{yy}$,
there is a global-in-$\phi$ gap between $\Vr_{\pm_b}(\phi)$:
the minimum $\x{min}_\phi \Vr_{+_b}(\phi)>0$ of the upper band is positive
and the maximum $\x{max}_\phi \Vr_{-_b}(\phi)<0$ of the lower band is negative.
\figr{Vcrbase} and \figr{Vcr}(b),(c),(d) are the examples.
However, upon increasing $\Vr_{0x}$ or $\Vr_{0y}$, eventually the global gap will close: in some range of the surface-momentum directions $\phi$,
both $\Vr_{\pm_b}(\phi)>0$ will become positive, and, in a complementary range, both will become negative. \figr{Vcr}(e) is an example.
This also results in changes in the $\e=0$ contours: regions of bulk states will emerge.

At the same time, we note that topologically, in terms of the surface-state-band dependence $\Vc(\phi)$, there are no qualitative changes,
since the local-in-$\phi$ gap, defined as the difference \eqn{dVr}, remains also in the type-II Weyl semimetal (as long as $v_{yy}\neq0$, see \secr{LN}).

\section{Line-node semimetal and strongly anisotropic Weyl semimetal \lbl{sec:LN}}

%\begin{figure}
%%\includegraphics[width=.30\textwidth]{ChiralSymmetryInQHFigs/velocity-newmodel.pdf}
%\caption{Line-node semimetal and strongly anisotropic Weyl semimetal.
%{\bf[decide if here or there]}
%}
%\lbl{fig:LN}
%\end{figure}

Consider the case of the vanishing velocity $v_{yy}=0$ (the velocities $v_{zz}>0$ and $v_{xx}>0$ are nonzero by construction).
This is the case of a line-node semimetal.
As follows from the bulk spectrum \eqn{e3}, at $v_{yy}=0$, the argument of the square root vanishes
for the direction $\nb$ with $n_x=0$ and $v_{zy}n_y+v_{zz}n_z=0$;
hence, the two bulk bands touch along the straight line $\pb=(0,p_y,-\f{v_{zy}}{v_{zz}} p_y)$: $\eps_{3,+_b}(\pb)=\eps_{3,-_b}(\pb)$.

The general surface-state solution \eqn{Ec3}, of course, still applies to this limiting case. At $v_{yy}=0$, we have
\[
	\dr_x(\phi)=\vr_{xx}\cos\phi,\spc \dr_y(\phi)=0,\spc \dr_\p(\phi)=\vr_{xx}|\cos\phi|.
\]
Hence, the gap phase \eqn{dephi} becomes a step function: $\de(\phi)=0$ for $\phi\in(-\f\pi2,\f\pi2)$ and $\de(\phi)=\pi$ for $\phi\in(\f\pi2,\f{3\pi}2)$.
The velocities \eqn{Vr} of the bulk-band boundaries \eqn{E3} read
\[
	\Vr_{\pm_b}(\phi)=\Vr_{0x}\cos\phi+\Vr_{0y}\sin\phi \pm_b \sq{1-\vr_{0z}^2} \vr_{xx}|\cos\phi|,
\]
an example is shown in \figr{Vcr}(f).
The gap \eqn{dVr} between the bulk-band boundaries, determined by $\dr_\p(\phi)$,
closes at $\phi=\f{\pi}2,\f{3\pi}2$, in accord with the above.
The velocity \eqn{Vcr} of the surface-state band \eqn{Ec3} reads
\beq
	\Vcr(\phi)=\Vr_{0x}\cos\phi+\Vr_{0y}\sin\phi+\sq{1-\vr_{0z}^2} \vr_{xx}\cos\phi\cos\nu.
\lbl{eq:VcrLN}
\eeq
For the BC parameter $\nu\in(0,\pi)$, the surface-state band is present for $\phi\in(\tf\pi2,\tf{3\pi}2)$;
for $\nu\in(\pi,2\pi)$, it is present for $\phi\in(-\tf\pi2,\tf\pi2)$.
%{\bf[-double confirm in M -seems correct]}
In the line-node semimetal, the surface-state band thus exists in one of the regions separated by the line node,
depending on the range of $\nu$, but regardless of its specific value.

By continuity, a similar behavior of the surface-state band still persists in the case of a Weyl semimetal with strong $xy$ anisotropy [\figr{Vcr}(e)],
$|v_{yy}|\ll v_{xx}$, when $v_{yy}$ is not zero, but is much smaller than $v_{xx}$,
when mini-gaps open up around the line-node directions $\phi=\tf\pi2,\tf{3\pi}2$.
The surface state band is still located in approximately the same regions, depending on the range of $\nu$.
However, the sign of $v_{yy}$ determines the signs of the mini-gaps, and hence,
determines with which bulk band the surface-state band merges.
This crucially determines the chirality of the surface-state band $\Vcr(\phi)$ [\eq{Vcr}],
to correspond to the Chern number of the Weyl node, as per the bulk-boundary correspondence \eqn{bbcW} we formulated in \secr{topoW}.
The line-node semimetal at $v_{yy}=0$ can also be regarded as a phase-transition point
between the Weyl point-node semimetal phases $v_{yy}\gtrless 0$ with opposite Chern numbers $\mp1$
of the node, respectively [\eq{CW}].
The case of strong anisotropy $|v_{yy}|\ll v_{xx}$ is relevant, e.g., for the Weyl semimetal that emerges
in the Luttinger semimetal model under strain and bulk inversion asymmetry, which describes materials such as HgTe under strain~\cite{Ruan}.

This behavior of the velocity $\Vcr(\phi)$ [\eqs{Vcr}{VcrLN}] of the surface-state spectrum
of a 3D line-node semimetal and strongly anisotropic Weyl semimetal in the vicinity of the line node
is in full accord with the edge-state spectrum of the 2D QAH system we present in the next \secr{2D}:
the velocity $v_{yy}$ corresponds to the gap parameter $\De$;
the 3D line-node semimetal with $v_{yy}=0$ corresponds to the $\De=0$ case of a 2D point-node semimetal, realized at the phase transition.

This 3D line-node semimetal case is also directly related to chiral symmetry.
Applying the same symmetry analysis as in \secr{S} to the 3D model [\eqs{H3}{bc3}],
since $v_{zz}>0$ and $v_{xx}>0$ are assumed to be present, the most general form of the chiral symmetry operator is
\beq
	\Sh=\tauh_y.
\lbl{eq:SW}
\eeq
This prohibits all $\tauh_0$ and $\tauh_y$ terms; hence, $v_{0x},v_{0y},v_{0z}=0$ and, most importantly, $v_{yy}=0$.
The most general form of the chiral-symmetric Hamiltonian then reads
\beq
	\Hh_3^\Sc(\pbh)=
	+\tauh_z (v_{zx} \ph_x+v_{zy}\ph_y)
	+\tauh_x v_{xx} \ph_x+ \tauh_z v_{zz}\ph_z.
\lbl{eq:H3S}
\eeq
and represents a line-node semimetal.
The bulk-band boundaries are
\[
	\Vr_{\pm_b}^\Sc(\phi)=\pm_b \vr_{xx}|\cos\phi|.
\]
There are only two forms of the chiral-symmetric BC:
\[
	\sq{v_{zz}}\psi_{+}(x,y,0)=\sq{v_{zz}}\ex^{-\ix\nu^\Sc}\psi_{-}(x,y,0),
\]
with $\nu^\Sc=\tf\pi2,\tf{3\pi}2$. The surface-state band [\eq{VcrLN}] is flat:
\beq
	\Vcr^\Sc(\phi)=0,
\lbl{eq:VcrS}
\eeq
and occupies $\phi\in(\tf\pi2,\tf{3\pi}2)$ for $\nu=\tf{3\pi}2$ and $\phi\in(-\tf\pi2,\tf\pi2)$ for $\nu=\tf\pi2$.

We conclude that for two bands, a 3D chiral-symmetric system is necessarily a line-node semimetal.
One can also consider a general two-band Hamiltonian $\tauh_x h_x(\pb)+\tauh_z h_z(\pb)$, chiral-symmetric under \eq{SW}.
Its spectrum $\pm\sq{h_x^2(\pb)+h_z^2(\pb)}$ will generally have line nodes,
since one has {\em three} available momentum components to nullify {\em two} functions $h_{x,z}(\pb)$.
This means that constructing a 3D gapped chiral-symmetric system requires at least four bands.
According to the general topological classification~\cite{Ryu,Chiu}, a class-AIII gapped system is topologically nontrivial in 3D.

\section{2D quantum anomalous Hall system in the vicinity of a topological phase transition\lbl{sec:2D}}

\begin{figure}
\includegraphics[width=.24\textwidth]{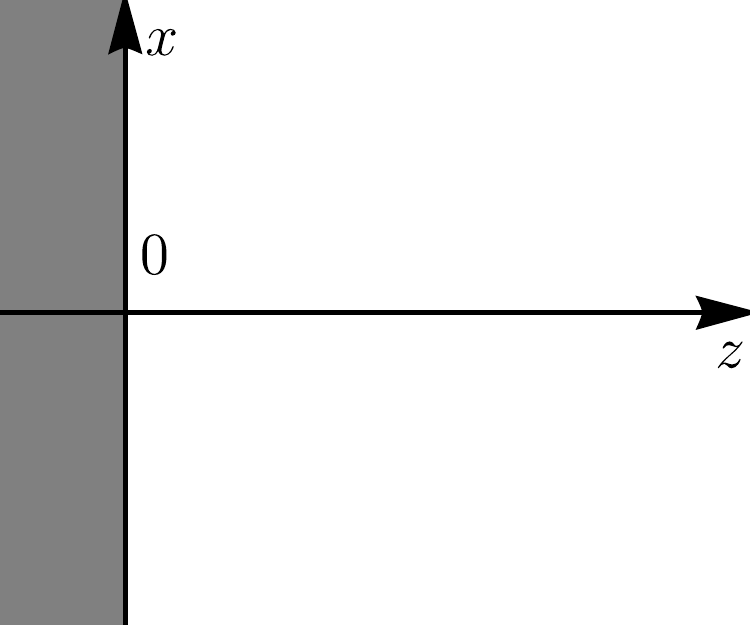}
\caption{
2D system occupying the half-plane $z\geq 0$ in $\rb_2=(x,z)$.
}
\lbl{fig:2Dsystem}
\end{figure}

\begin{figure}
\includegraphics[width=.70\textwidth]{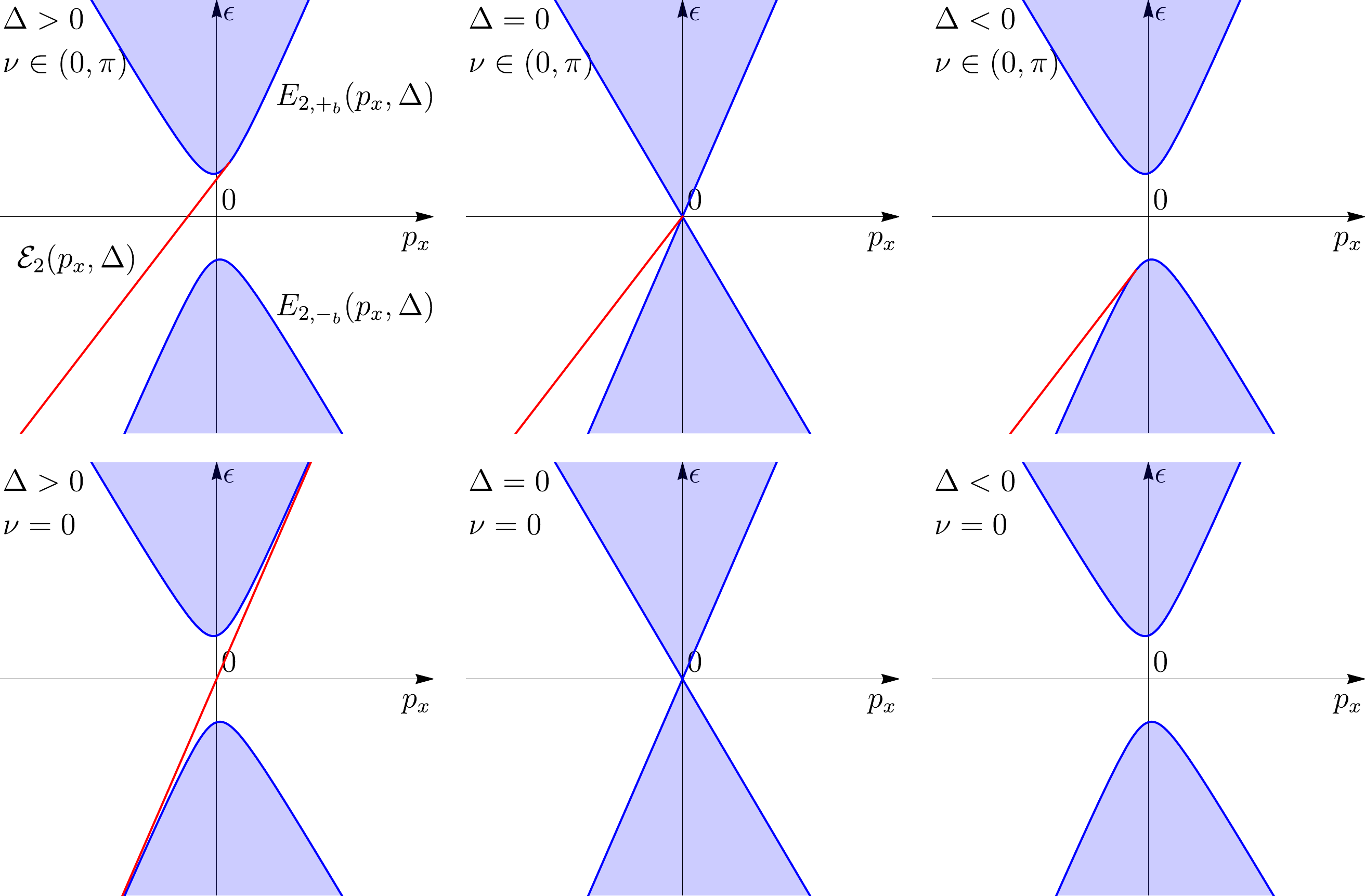}
\caption{The edge-state spectrum $\Ec_2(p_x,\De)$ [red, \eq{Ec2}] of the 2D linear-in-momentum two-component model of the most general form,
with the Hamiltonian \eqn{H2} and BC \eqn{bc2}. The blue regions depict the continuum of the bulk states.
%The key feature is the existence of the
}
\lbl{fig:Ec2}
\end{figure}

\begin{figure}
\includegraphics[width=.35\textwidth]{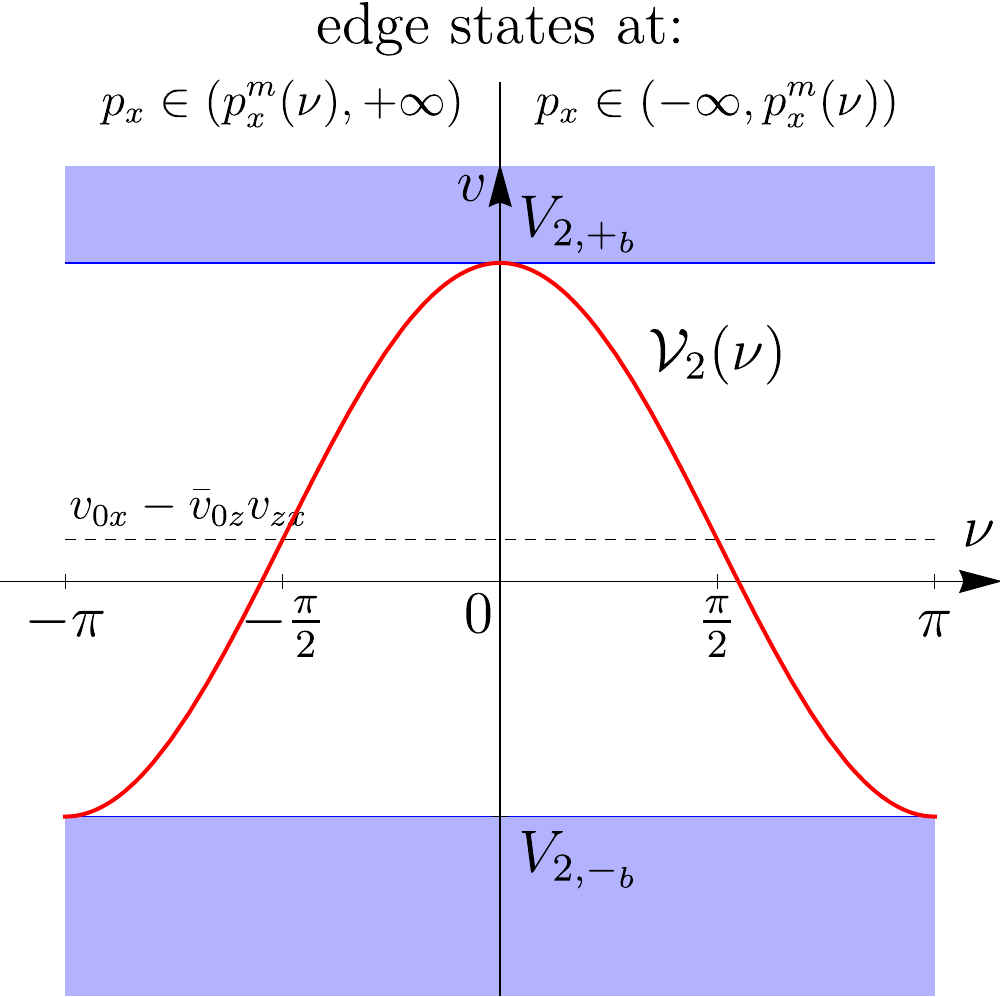}
\caption{The velocity $\Vc_2(\nu)$ of the the edge-state spectrum $\Ec_2(p_x,\De)$ [\eq{Ec2}]
as a function of the phase-shift angle parameter $\nu$ of the general BC \eqn{bc2}.
%of the 2D linear-in-momentum two-component model of the most general form, with the Hamiltonian \eqn{H2} and BC \eqn{bc2}.
}
\lbl{fig:Vc2}
\end{figure}

We now demonstrate the application of the formalism of general continuum models with BCs to the minimal model in 2D.
The most general 2D model with the linear-in-momentum Hamiltonian and a two-component wave function
has already been derived and studied in the Supplemental Material of Ref.~\ocite{KharitonovQAH}.
Here, we reproduce this result, with the main goal of demonstrating the connection to the 1D (\secr{1D}) and 3D (\secr{3D}) models.

As explained therein (an analogous procedure has been carried out in \secr{3D}),
one can always exploit the basis degrees of freedom to express the Hamiltonian of the most general form as
\beq
	\Hh_2(\ph_x,\ph_z) = (\tauh_0 v_{0x}+\tauh_z v_{zx})\ph_x +\tauh_x v_{xx} \ph_x +\tauh_y \De +(\tauh_0v_{0z}+v_{zz}\tauh_z) \ph_z
\lbl{eq:H2}
\eeq
and have the sample occupy the $z\geq 0$ half-plane in the coordinate plane $\rb_2=(x,z)$ (\figr{2Dsystem}).
Also, both $v_{zz}>0$ and $v_{xx}>0$ can be made positive by this basis choice.
There is an energy parameter $\De$ in 2D, which determines the gap in the bulk spectrum.
This is the key difference between the models in 3D and 2D:
In 3D (\secr{3D}), all three energy parameters describe momentum shifts and can be this way eliminated;
as a result, the system is necessarily a point-node semimetal.
In 2D, only two can be eliminated, while one energy parameter remains and determines the gap.

The wave function in this basis reads
\[
	\psih(x,z)=\lt(\ba{c} \psi_+(x,z) \\ \psi_-(x,z) \ea\rt).
\]
As in the 3D case (\secr{bc3}), we consider the constraint that the current density
\[
	j_z(\rb_2)=v_{+}\psi^*_{+}(\rb_2)\psi_{+}(\rb_2) - v_{-}\psi^*_{-}(\rb_2)\psi_{-}(\rb_2)
\]
perpendicular to the edge vanishes at every point $x$ of the edge:
\[
	j_z(x,0)=0.
\]
The family of general BCs, translation-symmetric along the edge, that resolve this constraint reads
\beq
	\sq{v_+}\psi_+(x,0)=\ex^{-\ix\nu}\sq{v_-}\psi_-(x,0), \spc \nu\in[0,2\pi),
\lbl{eq:bc2}
\eeq
with the phase-shift angle $\nu$ parameterizing all possible BCs.

The model \eqsn{H2}{bc2} describes a generic 2D QAH system (Chern insulator) with an edge in the vicinity of a topological phase transition.
The model belongs to class A of gapped systems, which is topologically nontrivial in 2D.
The phase transition between the two topological phases with $\De\gtrless 0$
occurs at $\De=0$, where the system becomes a nodal semimetal.

For a conserved momentum $p_x$ along the edge, the problem again reduces completely to the 1D model (\secr{1D}) for $\psih(z)$ in the wave function
\[
	\psih(x,z)=\ex^{\ix p_x x}\psih(z).
\]
The energy parameters of the 1D model become functions of $p_x$ and $\De$:
\[
	d_0(p_x)= v_{0x} p_x,\spc
	d_z(p_x)= v_{zx} p_x,
\]
\beq
	d_x(p_x)=d_\p(p_x,\De)\cos\de(p_x,\De)=v_{xx} p_x, \spc
	d_y(\De)=d_\p(p_x,\De)\sin\de(p_x,\De)=\De.
\lbl{eq:d2}
\eeq
The edge-state band reads
\beq
	\Ec_2(p_x,\De)= (v_{0x}-\vr_{0z}v_{zx})p_x+\sq{1-\vr_{0z}^2}(v_{xx} p_x \cos\nu + \De\sin\nu).
\lbl{eq:Ec2}
\eeq
The bulk-band boundaries read
\[
	E_{2,\pm_b}(p_x,\De)= (v_{0x}-\vr_{0z}v_{zx})p_x \pm_b \sq{1-\vr_{0z}^2} \sq{v_{xx}^2 p_x^2+\De^2}.
\]
%The gap phase $\de(p_x,\De)$ spans a half circle as $p_x\in(-\iy,+\iy)$. For $\De>0$, it spans the upper half, for $\De<0$, the lower half.
The edge-state band merges with either the upper or lower bulk-band boundary at momentum
\[
	p_x^m(\nu)=\f{\De}{v_{xx}}\cot\nu
\]
and is located in $p_x\in(p_x^m(\nu),+\iy)$ for $\nu\in(\pi,2\pi)$, and in $p_x\in(-\iy,p_x^m(\nu))$ for $\nu\in(0,\pi)$.
The edge-state spectrum is shown in \figr{Ec2}.

The main interesting observation about this result made in Refs.~\ocite{KharitonovLSM,KharitonovQAH}
is that at absent gap $\De=0$, when the system is a nodal semimetal, the edge-state band
\beq
	\Ec_2(p_x,\De=0) %= (v_{0x}-\vr_{0z}v_{zx} +\sq{1-\vr_{0z}^2} v_{xx} \cos\nu)p_x
	=\Vc_2(\nu) p_x
\lbl{eq:Ec2De=0}
\eeq
with the velocity
\beq
	\Vc_2(\nu)=v_{0x}-\vr_{0z}v_{zx} +\sq{1-\vr_{0z}^2} v_{xx} \cos\nu
\eeq
exists ``nearly always'', i.e., for every value of the velocity parameters and every value of the BC parameter $\nu$ except for two points $\nu=0,\pi$.
The edge-state band \eqn{Ec2De=0} exists in $p_x\in(0,+\iy)$ and $\nu\in(\pi,2\pi)$
and in $p_x\in(-\iy,0)$ for $\nu\in(0,\pi)$. The velocity $\Vc_2(\nu)$ is plotted in \figr{Vc2}.

Again, as in \secr{2reasons} for 1D system, this presence of robust edge states in a topologically trivial system
(2D point-node semimetal with no assumed symmetries) is not accidental.
This behavior was originally explained in Ref.~\ocite{KharitonovLSM} in terms of chiral symmetry and the ``stability argument'',
analogous to \secr{1DAIII} for the 1D model.
We reproduce this explanation here.
The application of chiral symmetry to the 2D model \eqsn{H2}{bc2} is as follows.
Since in addition to $\tauh_z v_{zz} \ph_z$, the term $\tauh_x v_{xx} \ph_x$ must also be present,
the only option for the chiral-symmetry operation allowing both terms is
\[
	\Sh=\tauh_y.
\]
This prohibits all $\tauh_0$ and $\tauh_y$ terms; hence, $v_{0x}=0$, $v_{0z}=0$, $\De=0$.
The most general form of the 2D chiral-symmetric Hamiltonian therefore reads
\beq
    \Hh_2^\Sc(\ph_x,\ph_z)=(\tauh_z v_{zx} +\tauh_x v_{xx})\ph_x + v_{zz}\tauh_z \ph_z,
\lbl{eq:H2S}
\eeq
which describes a point-node semimetal, confirming the common understanding.
The bulk-band boundaries of \eq{H2S} read
\[
	E_{2,\pm_b}^\Sc(p_x)= \pm_b v_{xx}|p_x|.
\]
There are only two points $\nu^\Sc=\tf\pi2,\tf{3\pi}2$ in the $S^1$ parameter space of the general BCs,
at which the BC is chiral-symmetric:
\beq
	\sq{v_{zz}}\psi_+(x,y,0)=\sq{v_{zz}}\ex^{-\ix\nu^\Sc}\psi_-(x,y,0).
\lbl{eq:bc2S}
\eeq
The edge-state band is flat and exists on one side of the node, depending on $\nu^\Sc$
\beq
	\Ec_2^\Sc(p_x)=0, \x{ in $p_x\in(0,+\iy)$ for $\nu^\Sc=\tf{3\pi}2$ and in $p_x\in(-\iy,0)$ for $\nu^\Sc=\tf\pi2$}.
\lbl{eq:Ec2S}
\eeq

Equations \eqsn{H2S}{bc2S} is the model of the most general form of a 2D chiral-symmetric system with a boundary,
which describes the topologically nontrivial class AIII of 2D chiral-symmetric semimetals.
The edge-state bands \eqn{Ec2S} are topologically protected by chiral symmetry.
The existence of the edge-state band \eqn{Ec2De=0} for the general semimetal system at $\De=0$,
with both the velocities and BC parameters deviating from the chiral-symmetric values, can then be explained by the stability argument:
when chiral-symmetry is broken, the edge states will still persist in some finite ranges of $\nu$ around $\nu^\Sc=\tf\pi2,\tf{3\pi}2$ points.
The existence of such ``stability regions'' around symmetric points is a standard and generic situation.
What {\em is} particularly remarkable is that these regions $(0,\pi)$ and $(\pi,2\pi)$ are of length $\pi$
and thus together occupy the whole $\Ux(1)\sim S^1$ unit-circle parameter space, except for the two points $\nu=0,\pi$.
Tracing back to the 1D bound-state solution of \secr{bs},
this property is equivalent to the fact that the bound-state ``band'' $\Ec(\de-\nu)$ [\eq{Ec}]
at a fixed $\de$ covers {\em exactly half} of the BC parameter space $\nu$: the interval of length $\pi$
with the end points excluded (which are the merging points).
For the 2D semimetal, there are two sides $p_x\gtrless 0$ of the node corresponding to $\de(p_x,\De=0)=0,\pi$ [\eq{d2}], respectively,
and hence, the edge-state band \eqn{Ec2De=0} exists in the {\em whole} BC parameter space except for the two merging points $\nu=0,\pi$.
That the extent of the bound-state band $\Ec(\de-\nu)$ [\eq{Ec}] is exactly $\pi$ and not less
seems to be a special mathematical property of the considered minimal model.

This effect, of course, persists when the gap $\De$ is finite, in either of the gapped phases,
since the gap only modifies the edge-state spectrum in the region $|p_x|\sim |\De|/v_{xx}$ determined by it:
the edge state-band \eqn{Ec2} still exists nearly always (for every $\nu\in S^1$ expect $\nu=0,\pi$)  at larger momenta $|p_x|\gtrless |\De|/v_{xx}$.
As explained in Ref.~\ocite{KharitonovQAH},
this property can be regarded as the ``extension'' of the edge-state structure of a 2D QAH system beyond the minimal one required to
satisfy the Chern numbers via bulk-boundary correspondence.

\section{Conclusion \lbl{sec:conclusion}}

Although the development of the presented {\em formalism of general continuum models with BCs}
was initially motivated mainly by the desire to establish and explore the general bound-state structures of topologically nontrivial systems,
the first findings following from its application, presented here (with the groundwork laid in Refs.~\ocite{KharitonovLSM,KharitonovQAH}),
are quite unanticipated and all the more exciting.
The main emerging overarching conclusion of our analysis is that bound states appear to be more widespread
and persistent than the mere strict topological classification suggests.
Moreover, these persistent bound-state structures in topologically trivial classes
are not some accidental mathematical properties of the considered model 
(and even if they were, this is a {\em general} low-energy model) %for a multitude of systems)
but have sound topological explanations.
Namely (\secr{2reasons}), we identify a systematic {\em ``propagation'' effect},
whereby bound states from the topologically nontrivial classes, where they are protected and guaranteed to exist,
propagate to the related, ``adjacent'' in dimension or symmetry, topologically trivial classes.

It is quite common in the theoretical studies of topological systems
to treat the bound-state features that are not strictly topologically protected
as fringe cases that deserve little attention. %that may be disregarded.
The generality of the model derived from the developed formalism allows us to establish that
bound states in topologically trivial systems
are neither accidental (as per above) nor rare, but are a generic persistent effect.
Since the model is of the most general form, a precise volume fraction of its parameter space where the bound states exist can be determined.
In 1D, this fraction is equal to the half of the parameter space (\secr{bs}). %of either $\de$ or $\nu$.
In a 2D semimetal (and, as a consequence, in a 2D QAH system),
the persistence effect turns out to be particularly extreme (\secr{2D}):
so much so that the edge-state band, not topologically protected, exists ``nearly always'',
i.e., in the whole parameter space except for the two discrete points.
One ends up with a rather bizarre situation, completely contrary to the common expectation,
where one would actually have to fine-tune the system to these special points
to make the supposedly ``non-protected'' edge states disappear.

Since the considered minimal model is a universal general low-energy model
(in the vicinity of the gap closing in 1D and 2D and in the vicinity of a Weyl node in 3D), describing numerous physical systems,
the established bound-state structures with their interesting properties can be expected to be widespread in real materials.

We conclude by pointing out that it is quite remarkable
how many insights about the general behavior of the bound/edge/surface states
can be extracted with the help of the presented formalism from such a mathematically simple model.
%[application of the formalism should continue].

\begin{acknowledgments}
%\section*{Acknowledgements}
M. K. acknowledges the financial support by the DFG Grant No. KH 461/1-1.
\end{acknowledgments}

\end{document}